\documentclass[aps,pre,twocolumn,groupedaddress]{revtex4}
\usepackage{graphicx}
\usepackage{epsfig}

\usepackage{amsmath, amsthm, amssymb}
\numberwithin{equation}{section} 

\begin{document}

\newcommand {\be}{\begin{equation}}
\newcommand {\ee} {\end{equation}}
\newcommand {\bea}{\begin{eqnarray}}
\newcommand {\eea} {\end{eqnarray}}
\newcommand {\non}{\nonumber}
\newcommand {\Eq}[1] {Eq.~(\ref{#1})}
\newcommand {\Fig}[1] {Fig.~\ref{#1}}
\def \bp {\mbox{\boldmath $\partial$}}
\def \q {{\bf q}}
\def \k {{\bf k}}
\def \zero {{\bf 0}}
\def \bfw {{\bf w}}
\def \bfx {{\bf x}}
\def \bfy {{\bf y}}
\def \bfz {{\bf z}}
\def \F2 {FPL${}^2$ }
\def \Rs {\sf I\hskip-1.5pt R}
\def \Zs {\mbox{\sf Z\hskip-5pt Z}}
\def \Cs {\rm C\!\!\!I\:}
\def \rb {\rm b}
\def \rg {\rm g}
\def \OMIT #1{}
\def \rem #1 {{\it #1}}

\def \p {\ell}
\def \pt {p_0}
\def \pc {p_c}
\def \bt {{\bf T}}
\def \ct {{\mathcal T}}

\def \bigO {\mathcal{O}}
\def \indic {\mathbb{I}}
\def \olgam {\overline{\gamma}}
\def \olg {\overline{\Gamma}}
\def \gxy {G_{\bfx, \bfy}}
\def \gtxy {\widetilde{G}_{\bfx,\bfy}}
\def \gamxy {\Gamma_{\bfx,\bfy}}
\def \gxyz {G_{\bfx, \bfy, \bfz}}
\def \gamxyz {\Gamma^{\bfz}_{\bfx,\bfy}}
\def \bargamxy {\olg_{\bfx,\bfy}}
\def \bargamxyz {\olg^{\bfz}_{\bfx,\bfy}}
\def \gama {\Gamma_\alpha}
\def \gamb {\Gamma_\beta}
\def \gamab {\Gamma_{\alpha \beta}}
\def \dmsf {d_{\text{MSF}}}
\def \Omsf {\mathcal{O}_{\text{MSF}}}
\def \bOmsf {\overline{\mathcal{O}}_{\text{MSF}}}
\def \gmst {\gamma_{\text{MST}}}
\def \gmsf {\gamma_{\text{MSF}}}
\def \calC {\mathcal{C}}
\def \calM {\mathcal{M}}
\def \ta {\widetilde{a}}
\def \tb {\widetilde{b}}
\def \tPhi {\widetilde{\Phi}}
\def \cE {\mathcal{E}}
\def \cG {\mathcal{G}}
\def \cH {\mathcal{H}}
\def \cT {\mathcal{T}}
\def \cxyz {\widetilde{C}^{\bfz}_{\bfx, \bfy}}
\def \icxy {\indic_{c} (\bfx,\bfy) \vert_{p}}
\def \icxyz {\indic_{c} (\bfx,\bfy; \bfz) \vert_{p}}
\def \icz {\indic_{c}^{(\bfz)}}
\def \order {\pi}
\def \embed {\lambda}
\def \ie {\embed(\epsilon)}
\def \ve {\varepsilon}
\def \buo {u_0}
\def \beg {E_{G-g}}
\def \bog {\order_{G-g}}
\def \mst {\text{MST}}


\title{Theory of minimum spanning trees II: exact graphical methods
and perturbation expansion at the percolation threshold}

\author{T. S. Jackson} \email[]{thomas.s.jackson@yale.edu}
\author{N. Read}
\email[]{nicholas.read@yale.edu} \affiliation{Department of
Physics, Yale University, P.O. Box 208120, New Haven, CT
06520-8120, USA}
\date{\today}

\begin{abstract}
Continuing the program begun by the authors in a previous paper, we develop an exact low-density expansion for the random minimum spanning tree (MST) on a finite graph, and use it to develop a continuum perturbation expansion for the MST on critical percolation clusters in space dimension $d$. The perturbation expansion is proved to be renormalizable in $d=6$ dimensions. We consider the fractal dimension $D_{\rm p}$ of paths on the latter MST; our previous results lead us to predict that $D_{\rm p}=2$ for $d>d_c=6$. Using a renormalization-group approach, we confirm the result for $d>6$, and calculate $D_{\rm p}$ to first order in $\varepsilon=6-d$ for $d\leq 6$ using the connection with critical percolation, with the result $D_{\rm p} = 2 - \varepsilon/7 + \bigO (\varepsilon^2)$.

\end{abstract}

\maketitle

\section{Introduction}
\label{s_intro}

In this paper, we continue our treatment of the statistics of random minimum spanning trees, which was begun in Ref.\ \cite{jr} (to be referred to as I).
We first recall the definitions (see any of Refs.\ \cite{lawler,papst,tarj,ccps}): we consider an undirected, connected graph $\widehat{G}$ with vertex set $V$, edge set $E$ and a real-valued cost $\ell_e$ assigned to each edge $e \in E$. A spanning tree is then defined as a subset of the edges of $\widehat{G}$ that connects all the vertices and contains no cycles: in other words, it is a tree and it spans $V$. Such a tree must exist because the graph is assumed connected. A \emph{minimum} spanning tree $\bt$ is a spanning tree
such that the sum of the costs of its edges,
\be
\label{eq_costfn} \ell(\bt) = \sum_{e
\in \bt} \ell_e, \ee
is minimized over the set of all spanning trees on $\widehat{G}$. (If the costs $\ell_e$ are strictly positive, then any spanning subset of the edges that has minimum cost is automatically a tree.) In this paper,
we again consider the random problem, in which the edge costs are assumed to be independent and identically-distributed (iid) random variables, with a continuous probability distribution for the cost of each edge, and we are interested in the statistical geometry of the tree. This random model will be referred to simply as the MST. We also consider a generalization, introduced in Ref.\ \cite{jr}, in which the costs are iid, the probability distribution for the cost of any edge
is uniform on the interval $[0,1]$, and one finds the minimum spanning forest (a forest is a collection of vertex-disjoint trees) on the (not necessarily connected) graph formed by the subset of edges that have cost less than $p\in [0,1]$; this object is called MSF$(p)$. It is connected with the use of Kruskal's greedy algorithm for the MST \cite{krus,lawler,papst,tarj,ccps}. In Kruskal's algorithm, the edges are tested one by one, in order of increasing cost. Each edge is ``accepted'' as belonging to the MST unless doing so would form a cycle in combination with edges already accepted. If this algorithm (or ``Kruskal process'') is stopped when all edges with cost $<p$ have been tested, one obtains the MSF$(p)$. MSF$(p)$ is closely related to bond percolation \cite{stah}, which is obtained if the process is modified so that it  simply accepts every edge, in which case each edge is independently accepted (occupied) with probability $p$. This relationship plays a central role in the work to be reported here.

The background and motivation for studying this problem were discussed in the previous paper. Of particular interest there was the strongly-disordered spin-glass model of Newman and Stein (NS) \cite{ns}. Our results strongly suggested that when the graph is a lattice in Euclidean system of dimension $d$, the critical dimension for that model and for the MST is $6$. This is the dimension above which there are many large connected components of the MST that are visible within a window of size $W$; we found that the number is of order $W^{d-6}$ . The strongly disordered spin glass maps onto the MST, and the logarithm of the number of ground states of the spin glass that can be distinguished within a window is the same as the number of large connected components that are visible. For the MSF$(p)$, the properties such as the number of large connected components are expected to be similar (scale with the same exponents) for all $p>p_c$, where $p_c$ is the threshold for bond percolation. These results were strongly motivated by the solution of MSF$(p)$ on the Bethe lattice (Cayley tree) \cite{jr}.

Other problems related to the MST that have been studied in the physics (and related) literature include the fractal dimension of the paths on the MST in Euclidean space, and the relation with problems of optimal paths and transport in random media \cite{optpath,wu_superhwy_2}. In this paper we begin to address some of these issues. On a finite graph, there is a unique self-avoiding path on the MST between any two vertices. Eventually, we aim to study the fractal dimension $D_{\rm p}'$ of the paths on the MST. In general, such a dimension can be defined by a box-counting technique, that is counting the number of steps on the path inside a window. One would hope also to obtain the fractal dimension of the path from the expectation value of the total number of steps on the path, if this is of order $R^{D_{\rm p}'}$ for two vertices separated by a large Euclidean distance $R$. [In the Kruskal process or MSF$(p)$, this would again be expected to be the same (on large length scales) for any $p>p_c$, for two vertices that are connected on MSF$(p)$.]
Several earlier works \cite{cieplak,dd,middle,ww} have provided numerical estimates of $D_{\rm p}'$ for the MST at $p=1$ in various spatial dimensions. In $d=2$ dimensions, the value is around $1.23$.

However, in the limit of an infinite system in $d>6$, there are many connected components of the MST, and for a large finite system (say, a hypercube of side $L$) this means that for many pairs of vertices, the path connecting them makes a large excursion (of order the system size). To see this, first note that if one makes the system larger by adding additional edges at the boundary, chosen from the same probability distribution, keeping those in the interior the same, then one can study the behavior of the MST as $L$ increases. Locally, it will converge to a definite limit that has the number of components as described above. This implies that the path connecting two fixed vertices will be deformed until it goes off to infinity as $L\to\infty$, and so have infinite length, unless the two vertices are on the same component in the limit. For finite $L$, it will make an excursion typically of linear size of order $L$. These considerations, together with the probability of order $1/R^{d-6}$ that the two given vertices are on the same connected component in the limit, suggest heuristically that the expected length of the path scales as
\be
\bigO(R^{D_{\rm p}'}/R^{d-6})+ \bigO\left((1-1/R^{d-6})L^{D_{\rm p}'}\right)
\ee
where $D_{\rm p}'$ is the fractal dimension of the path on the MST, for which the Bethe lattice results in Ref.\ \cite{jr} suggest $D_{\rm p}'=2$ for $d>6$. For $L\gg R$, the last term $\bigO(L^2)$ dominates, and again the result is similar for MSF$(p)$ for all $p>p_c$. But if we could restrict attention to (or condition on) paths not going to infinity, the conditional expectation for the path length would scale as $R^{D_{\rm p}'}$. For $d\leq 6$, this problem does not arise, and we expect that, at least for $d<6$, where there is presumably only one connected component in the infinite system limit, the length of the path will indeed scale as $R^{D_{\rm p}'}$, independent of $L$ for large $L$. (In the numerical work cited above, cases in which $d>6$ were apparently not considered.)

For this and for other related technical reasons, we will restrict our calculation of exponents in this paper to the Kruskal process with $p\leq p_c$. In this case the probability that two points are connected by either the percolation or the Kruskal processes decays with increasing distance $R$. For $p=p_c$, where scaling will again apply, we will denote the fractal dimension of the path on the MSF$(p_c)$ by $D_{\rm p}$. The naive scaling for the total length of a path $R^{D_{\rm p}}$ still works in this case.

There are claims (the ``superhighways'' argument) \cite{wu_superhwy_1,wu_superhwy_2} that in the Kruskal process (on various families of graphs) the properties of the paths are mainly determined by the percolation threshold. Numerically, the fraction of steps on a path on the MST that are already present at $p=p_c$, averaged over all pairs of end vertices for the path, goes to a constant \cite{wu_superhwy_1,wu_superhwy_2}. This suggests that $D_{\rm p}'=D_{\rm p}$. For the Bethe lattice and for Euclidean systems with $d>6$ we can argue that both dimensions equal $2$ \cite{jr}, so equality holds, however, it is less clear whether it holds for $d<6$. Some support for it can be obtained from scaling arguments in percolation. Consider a hypercube of side $L$, and for bond percolation ask for the probability that two given opposite faces are connected when the occupation probability for each edge is $p$. The percolation threshold is the value of $p$ above which the connection occurs with probability one as the size $L\to\infty$. For $L$ finite, connection occurs at $p=p_c$ with low probability, but it occurs with probability approaching 1 at $p-p_c$ of order $1/L^{1/\nu}$, where $\nu$ is the correlation length exponent in percolation \cite{stah}. At this value of $p$, the correlation length $\xi$ is of order $(p_c-p)^{-\nu}=\bigO(L)$, and scaling properties at scales less than $\xi$ should be the same as those at threshold $p=p_c$. Hence we expect the fractal dimension $D_{\rm p}'$ of the path on the MST connecting the two faces (which is one of the paths on the percolation cluster that do so when the faces are first connected as $p$ increases) to be $D_{\rm p}$. However, this path does not have the boundary conditions we wanted, as we only asked for the connection of the two faces, not of two given vertices in the interior of a system. When the separation $R$ of the vertices is large, we may surround each by a nested sequence of concentric spheres of radii say $2^{-j} R/3$ for $j=0$, $1$, \ldots and ask the same question for each annulus bounded by two of these spheres. Then the relevant $p$ will be different for each sphere (higher $p$ is required to make the connections to the vertices on smaller scales), but also the scaling holds for each one. In addition, the clusters that connect each pair of spheres must also become connected together to form a single cluster and an MSF path. Similar arguments apply to all of these. This does suggest, heuristically, that $D_{\rm p}'=D_{\rm p}$.

On examining this argument, a key part of it can be seen to be the idea that there is a unique candidate superhighway (critical percolation cluster) that is used to make connections over large distances on each scale, and so it is clear which ones must be connected by higher-cost ``roads''. This is a property of critical percolation clusters that holds for $d<6$, but not for $d>6$, where the number of large clusters visible in a window of size $W$ is $W^{d-6}$ \cite{stah,aizen_perc}. This behavior may itself underlie the result \cite{jr} that the number of connected components of the MST (or of MSF$(p)$ for $p>p_c$) has this same form $W^{d-6}$ for $d\geq6$ (but order $1$ for $d<6$).

In this paper we do not assume the equality of $D_{\rm p}'$ and $D_{\rm p}$, but will study $D_{\rm p}$. We first construct in section \ref{s_latt} an exact series expansion for any finite graph that gives the probability that the path on MSF$(p)$ from vertex $\bfx$ to vertex $\bfy$ passes through vertex $\bfz$, by analogy with expansions in percolation. The expansion takes the form of a weighted sum of subgraphs. This expansion may be of general interest within various approaches to MSTs not considered further in this paper, such as high-temperature series.

In section \ref{s_cont_ft}, we then turn this expansion into an asymptotic or perturbation expansion in the continuum (with cutoff) by neglecting excluded volume requirements that were present in the exact expansion, and taking the graph to be the infinite lattice. (From here on, our results are not fully rigorous mathematically, though they will satisfy most theoretical physicists.) In this way, we obtain a Feynman diagram expansion. It is related to that for percolation, but with modified Feynman rules. In order to avoid dealing with the appearance of the order parameter for percolation (non-zero probability of connection to infinity), from this point forward we consider only $p\leq p_c$.

The perturbation expansion contains ultraviolet (short-distance) divergences as the wavevector cutoff goes to infinity (or as the lattice spacing goes to zero) with the separation of the vertices held fixed. We prove that the expansion is renormalizable in the field-theoretical sense to all orders in perturbation theory.
We then use standard techniques to formulate a renormalization group (RG) approach which gives the scaling behavior of the correlation functions (probabilities) already mentioned. The exponents or fractal dimensions are then calculated for $d<6$ as an asymptotic series in powers of $\varepsilon=6-d$, with the result
\be
D_{\rm p} = 2-\varepsilon/7 + \bigO(\varepsilon^2)
\ee
as $\varepsilon\to0$.
For $d>6$, the analysis of the cut-off expansion confirms that $D_{\rm p}=2$ to all orders in perturbation theory, and the path behaves as a Brownian random walk on large length scales. Further discussion is contained in the Conclusion. To improve readability of the paper, many detailed derivations have been relegated to the Appendices.

Our calculations can in principle be extended to other exponents, such as those defined in \cite{ABNW}, or carried to higher orders in $\varepsilon$. They can also be extended to include statistical properties that involve the cost of the MST. There do not appear to be any scaling relations that relate the geometric exponents for MSTs to those for percolation, unlike those found for the costs in \cite{read1}, even though the critical dimension $d_c=6$ is the same.


\section{Low-density expansion on a finite graph}
\label{s_latt}

In this section we describe how to set up an exact low-density expansion which enables one to calculate connectedness functions for the (random) MSF on any finite graph $\widehat{G}$. The expansion takes the form of a low-density expansion, similar to a high-temperature expansion familiar from statistical mechanics, with small $p$ playing the role of high temperature, and is modeled on a corresponding expansion for bond percolation. Although from the point of view of this paper the primary utility of the result is to provide a basis for the RG calculations of the following section (which determine the fractal dimension of paths on the MSF), the expansion is of interest in its own right, and might form the starting point for further mathematically-rigorous calculations, using e.g., lace expansion methods \cite{lace}. As the construction is somewhat involved, we concentrate here on describing the structure of the expansion, and relegate most details of its derivation to the Appendices.

\subsection{Graphical expansions for percolation}
\label{s_essam_inclexcl}

In view of the correspondence between bond percolation and Kruskal's algorithm discussed in I, and the relation of the MSF$(p)$ process with percolation, we will first briefly review the low-density expansion for bond percolation. Afterwards, we set up a corresponding expansion for MSF$(p)$ by generalizing the arguments using Kruskal's algorithm.

Bond percolation on a finite-dimensional lattice is conventionally treated as the $Q \to 1$ limit of the low-density expansion of the $Q$-state Potts model \cite{fk,wu_potts,zia_potts,luben_review}. The partition function of the Potts model, which is a polynomial in $Q$, is also known as the Tutte polynomial \cite{tutte}, and is a generating function for $Q$-colorings of the vertices of the graph $\widehat{G}$ with weights that depend on whether adjacent vertices are given the same or different colors. Although the $Q \to 1$ limit of this function can be taken, it lacks a mathematical definition in terms of state variables (colors), so in order to establish a correspondence with percolation we instead use a method originally due to Essam \cite{essam_dombgreen, essam_perc_review}. This has the advantage of being phrased explicitly in terms of geometric quantities.

The basic object of interest is the two-point connectedness function $C_{\bfx, \bfy} (p)$, the probability that two vertices $\bfx$ and $\bfy$ on the lattice are connected by a percolation cluster (connected component) when the probability that an edge is occupied is $p$. In Appendix \ref{s_essam_2pt}, we review the graphical expansion for this probability, which has the form
\be
\label{eq_perc_c2_copy}
C_{\bfx, \bfy} (p) = \sum_{G \in \gxy} d(G) \Pr[G \leq p],
\ee
where the sum is over all graphs $G$ in the set $\gxy$, defined as graphs on (i.e.\ subsets of) the lattice containing the endpoints $\bfx$, $\bfy$. The factor of $\Pr[ G \leq p]$ is the probability that all the edges of $G$ have cost less than $p$ (i.e.\ are occupied in the percolation process), which in the iid model (Bernouilli model of bond percolation) is simply $p^{|E(G)|}$ (where $E(G)$ is the set of edges of $G$). The function $d(G)$ can be defined as
\be
\label{eq_pottsweight_2pt_copy}
d(G) = \sum_{E' \subseteq E(G)} (-1)^{|E(G)|-|E'|} \indic [E' \text{ connects } \bfx, \bfy];
\ee
where the sum is over subsets $E'$ of the edge set $E(G)$. We use the notation $\indic[X]$ for the indicator function on events $X$, which takes the value $1$ when $X$ is true and $0$ when $X$ is false. $d(G)=0$ if $x$ and $y$ are not connected by $G$.

The preceding graphical expansion may be generalized straightforwardly to $n$-point connectedness functions: now we must sum over $G_{\bfx_1, \cdots, \bfx_n }$, the set of subgraphs of the lattice that contain all $n$ vertices, and in Appendix  \ref{s_essam_npt} we show that the correct generalization of \eqref{eq_pottsweight_2pt_copy} is \cite{essam_coniglio}
\begin{multline}
\label{eq_pottsweight_npt_copy}
d(G \in G_{\bfx_1, \cdots \bfx_n}) = \sum_{E' \subseteq E(G)} (-1)^{|E(G)|-|E'|} \\ \times \indic \left[E' \text{ connects }
\bfx_1, \cdots \bfx_n \right].
\end{multline}

The $d$-weights defined above have several properties worth remarking on. First, in Appendix \ref{s_essam_potts}, we prove that the expansions \eqref{eq_pottsweight_2pt_copy}, \eqref{eq_pottsweight_npt_copy} assign the same weights to the same diagrams as does the $Q \to 1$ limit of the low-density expansion for the $Q$-state Potts model, so the above expressions are entirely equivalent to the conventional description of bond percolation.

Second, the $d$-weight clearly only depends on the connectivity of the set $E(G)$ of the edges of $G$, not on the geometry of $G$, and further as shown in Appendices \ref{s_essam_2pt}, \ref{s_essam_npt}, the $d$-weight is invariant under replacing edges of $G$ with chains of edges (i.e., inserting vertices of degree two). Thus $d(G)$ is a topological invariant of graphs $G$ with two marked vertices. This means that we can consider the expansion in terms of topological graphs $\cG$, which are simply graphs (without any embedding in the lattice), with two distinct vertices labeled $\bfx$, $\bfy$ (and the others unlabeled), and which may be assumed to contain no vertices of degree two other than possibly $\bfx$, $\bfy$. Using this property, we may rewrite the expansion \eqref{eq_perc_c2_copy} as
\be
\label{eq_perc_c3}
C_{\bfx, \bfy} (p) = \sum_{\cG \in \cG_{\bfx, \bfy} } \frac{d(\cG)}{{\cal A}(\cG)} \sum_{\embed: \cG \to G}  \Pr [ G \leq p],
\ee
with analogous expressions for the $n$-point functions. Here $\cG_{\bfx, \bfy}$ is the set of all (topological equivalence classes of) topological graphs with two distinct labeled vertices $\bfx$, $\bfy$, and the inner sum is over all possible embeddings $\embed: \cG \to G$ which map the edges of $\cG$ into self-avoiding chains of edges on the lattice, producing the set of lattice graphs summed over in \eqref{eq_perc_c2_copy}. Note that these chains must not only be self-avoiding walks, but also must avoid intersection with chains arising from different edges of $\cG$. If the topological graph $\cG$ has any non-trivial automorphisms (leaving the root points $\bfx$, $\bfy$ fixed), then there is more than one way to produce the same embedded graph $G$. Consequently, we must divide by the number ${\cal A}(\cG)$, which is the number of elements in the automorphism group of $\cG$. This number is often called a ``symmetry factor'', especially in the context of Feynman diagrams, and will play such a role later.

This topological property of the $d$-weights is crucial for extending the lattice expansion \eqref{eq_perc_c3} to a continuum theory, a point to which we will return in Section \ref{s_exclvol}. It also simplifies lattice calculations, since it greatly reduces the number of different graphs for which $d(G)$ must be calculated. The $d$-weights have further properties that, when the graph $G$ has a connected subgraph, allow them to be factorized into pieces given by the $d$-weight of the subgraph and that of the ``quotient'' graph in which the subgraph is replaced by a single vertex. These are discussed in the Appendices.

There is a further function will be useful in making comparisons with the MSF theory. This is the derivative of $C_{\bfx, \bfy} (p)$ with respect to the value $p_e$ of $p$ on a particular edge $e$ (the generalization of the formulas to cases in which the occupation probabilities such as $p_e$ for edges differ should be obvious). The probability that $\bfx$ and $\bfy$ are connected by a cluster at $p$ changes with $p_e$ only if $\bfx$ and $\bfy$ are not connected when $p_e=0$, and are connected when $p_e=p$. This implies that for $p_e=p$, any path from $\bfx$ to $\bfy$ on the cluster must traverse $e$. An edge with this property is called a singly-connected edge. So we define
\be
C_{\bfx, \bfy}^e (p)=\int_0^p dp_e\frac{\partial}{\partial p_e} C_{\bfx, \bfy} (p,p_e),
\ee
which is the probability that at parameter $p$, $\bfx$ and $\bfy$ are connected, and $e$ is a singly-connected edge on the same cluster. Note that $C_{\bfx, \bfy}^e(p)$ is not the same as the $3$-point connectedness function that was defined above, and its lattice expansion (which may be obtained directly from the preceding definition) still contains the same $d$ weights as for $C_{\bfx, \bfy}(p)$, and in fact is given by the same expansion \eqref{eq_perc_c3}, but with the additional condition that $e$ be an edge of the embedded graph $G$ (note that for the graphs $G$ in the expansion, $e$ does {\em not} have to have the singly-connected property). We may choose to view the topological graphs as having the inverse image of $e$ as a marked edge (either of the ends of which may be degree-two vertices), so that the embeddings $\lambda$ map it to the {\em single} edge $e$; in this case the relevant automorphisms of $\cG$ must fix this edge as well as $\bfx$, $\bfy$, and we denote the number of these by ${\cal A}'(\cG)$. Clearly ${\cal A}'(\cG)\leq{\cal A}(\cG)$ (one group of automorphisms is a subgroup of the other). These different ways of writing the function are equivalent.

Although we have formulated the expansion here in terms of the infinite lattice, it proceeds in exactly the same way if the lattice is replaced by any finite graph $\widehat{G}$. Indeed it is best viewed as derived from some finite graph such as a portion of the lattice, followed by an infinite volume limit. The sums over all embedded graphs make sense for a finite graph because only a finite number of terms contribute. For the infinite lattice, the sum converges for $p<p_c$ (like a high-temperature expansion), but not for $p>p_c$. In the latter case, it needs to be re-summed, but we will not discuss this here.

In the next two subsections, we proceed to develop a graphical expansion for MSF connectedness functions, analogous to \eqref{eq_perc_c3} in that it takes the form of a weighted sum over topological graphs and their lattice embeddings. This will be the basis for the continuum theory analyzed in section \ref{s_cont_ft}.


\subsection{A low-density expansion for MSF paths}
\label{s_spec_vert}

We will study the random geometry of paths on the MSF by introducing a connectedness function $\cxyz (p)$, which is the probability that two vertices $\bfx, \bfy$ on the MSF are connected \emph{and} the connecting path passes through a third vertex $\bfz$. In the following, a vertex $\bfz$ satisfying this definition will be called a ``MSF path vertex'' (with respect to two other vertices $\bfx$, $\bfy$). (We could equally well define a similar function in terms of the probability that the MSF path passes through an edge $e$ instead of a vertex $\bfz$, which makes the analogy with the percolation function $C_{\bfx, \bfy}^e (p)$ closer; in the continuum formulation developed afterwards, there is no difference between the vertex and edge cases for MSF.) We obtain a diagrammatic expansion for these connectedness functions by relating the Kruskal process defining MSF$(p)$ to the expansion for bond percolation obtained in the previous section. The difference between the percolation and Kruskal processes is that, in the latter, as $p$ increases an edge with cost $\p=p$ is not accepted if, together with edges already accepted by $\p<p$, it forms a cycle.



In Appendices \ref{app_pathcompare} and \ref{s_more_dmst}, we develop the tools needed to obtain a diagrammatic expansion for the MSF connectedness functions involving the MSF path vertex. Appendix \ref{app_pathcompare} contains a careful discussion of properties of the MSF paths that are used, while Appendix \ref{s_more_dmst} uses the method of inclusion and exclusion to obtain the expansion itself. Here we will begin by defining notation. The ordering of the costs of the edges on a subgraph $G$ of the given graph $\widehat{G}$ plays an important role. We know that in fact the MST on a graph depends only on the ordering of the costs \cite{dd,jr}. We define an ordering on the set of edge costs of $\widehat{G}$ as a permutation $\order \in S_{|E|}$ on the set of $|E|$ elements. We index the edges of $\widehat{G}$ arbitrarily, and take the ordering of their costs to be defined by $\order$ via
\be
\label{eq_order_def}
\p_{\order(i)} < \p_{\order(j)} \iff i< j,
\ee
for the edges indexed $i$, $j$ (writing $\p_i$ for $\p_{e_i}$). For subsets of $E$, such as $E'$ or $E(G)$, we define the induced ordering by restriction, and denote it by $\order_{E'}$ or $\order_{E(G)}$. Thus $\order_{E'}$ is a permutation of the subset $E'$. We can obtain the probability for each possibility ordering from the iid probability distributions on the edge costs in the obvious way, and clearly $\Pr[\order_{E'}]=1/|E'|!$, in particular for the special case $E'=E$.


With these definitions, the result we obtain in Appendix \ref{s_more_dmst} for the path vertex connectedness function can be written
\begin{multline}
\label{eq_mst_clatt_copy}
\cxyz(p) = \sum_{G \in \gxy}  \Pr[ G \leq p]\\ \times\sum_{\order_{E(G)} \in S_{|E(G)|}}\dmsf(G | \order_{E(G)}) \Pr[\order_{E(G)}],
\end{multline}
which should be compared with the corresponding result \eqref{eq_perc_c2_copy} for percolation, or its analog for $C_{\bfx, \bfy}^e (p)$. Here $G_{\bfx,\bfy;\bfz}$ is the set of graphs $G$ that contain the root points $\bfx$ and $\bfy$, and a (self-avoiding) path from $\bfx$ to $\bfy$ passes through $\bfz$.
The diagrammatic weight $\dmsf(G | \order)$ implicitly depends on $\bfx$, $\bfy$, $\bfz$, and is ordering-dependent. It can be defined as
\begin{multline}
\label{eq_dmst_def}
\dmsf(G  |\order) = \\
\sum_{E' \subseteq E(G)} (-1)^{|E(G)| - |E'|} \indic[E' \text{ connects } \bfx, \bfy] \\
\times \indic[\gmst(G_{E'} | \order_{E'}) \text{ passes through } \bfz],
\end{multline}
which should be compared with \eqref{eq_pottsweight_2pt_copy}. Here $G_{E'}$ is the subgraph of $\widehat{G}$ with vertex set $V$ and edges $E'$, and $\gmst(G_{E'} | \order_{E'})$ is the path connecting $\bfx, \bfy$ on the MST on the graph $G_{E'}$ with costs on $E'$ induced by restriction from $\widehat{G}$. This MST can be assumed to exist because $G_{E'}$ can be assumed to be connected, in view of the indicator function $\indic[E' \text{ connects } \bfx, \bfy]$.

Next we wish to express this connectedness function as a sum over topological graphs $\cG$.  We consider a given lattice graph $G$ which is the image of some topological graph $\cG$ under some embedding $\embed$. For clarity in what follows, we denote elements of the edge sets of $G$, $\cG$ by different symbols: we have $e \in E(G)$ and $\epsilon \in E(\cG)$. We find it useful to extend our notation and let $\ie \subset E(G)$ denote the chain (self-avoiding path on the lattice) of $N_\epsilon$ edges on the lattice that the topological edge $\epsilon$ is mapped to.

For both percolation and MSTs, the only edge cost information relevant for the connectedness functions is the cost of the most expensive edge on $\ie$, which we denote by
\be
L_\epsilon = \max_{e\in \ie} \p_e.
\ee
Then in bond percolation at occupation probability $p$, to determine whether $\ie$ connects its endpoints, we only need to check whether $L_\epsilon \leq p$. Likewise, $\ie$ is a subset of the MST on $G$ only if $L_\epsilon$ is less than the maximal edge cost encountered on all other paths on $\cG$ connecting the same vertices. We may think of $L_\epsilon$ as the cost induced on the edge $\epsilon$ of $\cG$ by the embedding $\embed$. A corresponding order $\order'_{E(\cG)}$ is induced on these costs. Given an embedding of $\cG$, both $\Pr[L_\epsilon]$ and $\Pr[\order'_{E(\cG)}]$ depend on the embedding. The probabilities can be readily calculated, as we will see shortly.

Because $\dmsf(G|\order)$ is computed in terms of connectedness properties (whether or not the MST path between the root points of the diluted graph $G_{E'}$ goes through the MSF path vertex at $\bfz$), it may be computed using only the relative ordering $\order'_{E(\cG)}$ obtained from the $\{ L_\epsilon \}$. Therefore we have
\be
\dmsf(G|\order_{E(G)}) = \dmsf(\cG|\order'_{E(\cG)})
\ee
(and depends on the marked vertices of $\cG$ that are the inverse images of $\bfx$, $\bfy$, and $\bfz$ under $\lambda$).
In the expansion \eqref{eq_mst_clatt_copy}, only the ordering and its probability is required, and it is possible to fix an ordering before choosing an embedding (and finally summing over both). Hence it may be written as
\begin{multline}
\label{eq_mst_ctop0}
\cxyz(p) = \sum_{\cG \in \cG_{\bfx,\bfy;\bfz}}  \sum_{\order'_{E(\cG)} \in S_{|E(\cG)|}}  \left(\dmsf(\cG|\order'_{E(\cG)})/{\cal A}'(\cG) \right) \\
\times \sum_{\embed: \cG \to G}  \Pr[ \order'_{E(\cG)}\wedge( G \leq p) ].
\end{multline}
Here again ${\cal A}'(\cG)$ is the relevant symmetry factor as defined in section \ref{s_essam_inclexcl} (strictly, it is defined here for automorphisms fixing the inverse image of the vertex $\bfz$ rather than $e$; the cases relevant to the continuum expansion later are those in which  the vertex $\bfz$ has degree two, and should be compared with those in which, when $e$ is shrunk to a single vertex, then it has degree two, and for these the numbers are the same).
Eq.\ \eqref{eq_mst_ctop0} is the main result of this Section, and should be compared with the percolation result \eqref{eq_perc_c3}. It remains to find an expression for $\Pr[ \order'_{E(\cG)}\wedge( G \leq p) ]$. This is done in the following Subsection.

\subsection{Expression for $\Pr[ \order'_{E(\cG)}\wedge( G \leq p) ]$}

To put the quantity $\Pr[ \order'_{E(\cG)}\wedge( G \leq p) ]$ in a more tractable form, we return to basic considerations. As the simplest example, we take the case where $\cG$ consists of two root vertices connected in parallel by two edges $\epsilon_1, \epsilon_2$. We consider an embedding $\embed$ where $\embed(\epsilon_1), \embed(\epsilon_2)$ are chains of $N_1$, $N_2$ lattice edges, the most expensive edges of which have costs $L_1, L_2$, respectively. We can define and evaluate the probability for $L_1<\p_1$ and $L_2<\p_2$:
\be
\begin{split}
P_G (\p_1, \p_2) & \equiv \Pr[(L_1 \leq \p_1) \wedge (L_2 \leq \p_2)] \\
&= \p_1^{N_1} \p_2^{N_2}.
\end{split}
\ee
The probability density for $(L_1,L_2)$ is found by differentiating on both variables, and then the probability that $L_1<L_2<p$ is found by integration:
\begin{multline}
\label{eq_mst_p2edge}
 \Pr[L_1<L_2 \leq p] \\
= \int^{p}_{0} \! d \p_2 \int_{0}^{\p_2} \!  d \p_1 \, \frac{d}{d \p_1}
\frac{d}{d \p_2 } P_G( \p_1, \p_2)\\
= \frac{N_2}{N_1+N_2} p^{N_1+N_2}.
\end{multline}
This is clear from the iid behavior of the edge costs: the most costly edge could be any of the $N_2$ edges among the total number $N_1+N_2$. [The result is more general than the model we have been using, in which costs are uniformly distributed on $[0,1]$. For a general iid distribution of costs, the probability that $L_1<L_2<\ell_0$ is $N_2p^{N_1+N_2}/(N_1+N_2)$, where $p$ is the probability that a given edge is less than $\ell_0$. Similar statements apply to the following generalization.]

The result generalizes to any embedded graph $G$, with edges $E(\cG) = \{ \epsilon_1, \ldots, \epsilon_n\}$ which have lengths $N_1, \ldots, N_n$ and maximum costs $L_1, \ldots, L_n$. We let the ordering be $\order'_{E(\cG)}=\order'$, such that $L_{\order'(1)}<L_{\order'(2)}<\cdots<L_{\order'(n)}$.
The generalization of \eqref{eq_mst_p2edge} is
\begin{multline}
\label{eq_multi_l_int0}
\Pr[\order' \wedge (G \leq p)] = \\
\int^{p}_{0} \! d \p_{\order'(n)} \int_{0}^{\p_{\order'(n)}} \!  d \p_{\order'(n-1)} \cdots  \qquad \qquad \qquad \\
\cdots \int_{0}^{\p_{\order'(2)}} \!  d \p_{\order'(1)} \prod_{i=1}^n \, \frac{d}{d \p_i}\Pr \biggl[ \bigwedge_{i =1}^n (L_i \leq \p_i) \biggr], \\
\end{multline}
which can be evaluated as
\be
\begin{split}
\Pr[\order' \wedge (G \leq p)]
&=p^{\sum_i N_i}\prod_{i=1}^{n} \frac{N_{\order'(i)}}{\sum_{j=1}^i N_{\order'(j)}} \\
&=\Pr[ G \leq p ] \prod_{i=1}^{n}  \frac{N_{\order'(i)}}{\sum_{j=1}^i N_{\order'(j)}} .\label{eq_multi_l_int0'}
\end{split}
\ee

Although the preceding result is completely explicit, it will be useful in the following to utilize the expression as an integral of multiple derivatives. Accordingly, we will define the integro-differential operator $\Omsf (\order'_{E(\cG)},p)$ which depends on the set $E(\cG)$, the ordering $\order'_{E(\cG)}$, and the limit $p$, and which acts on a set of variables $\p_\epsilon$ indexed by the elements of $E(\cG)$. For notational simplicity, we again write it for $E(\cG)=\{1,\ldots,n\}$ with $\order'_{E(\cG)}=\order'\in S_n$:
\be
\label{eq_msfop_def}
\Omsf (\order',p) =   \idotsint\limits_{0 \leq \p_{\order'(1)} \leq \cdots \leq \p_{\order'(n)} \leq p} \prod_{i=1}^n d \p_i \frac{d}{d\p_i}.
\ee
We have
\be
\label{eq_sum_all_orderings}
\sum_{\order' \in S_n} \Omsf (\order', p) f( \p_1, \ldots, \p_n) = f( p, \ldots, p),
\ee
for any nonsingular function $f$ of $n$ parameters, because the domain of integration becomes the cube $\p_i<p$, all $i$.

Using eq.\ \eqref{eq_multi_l_int0'} in \eqref{eq_mst_ctop0}, and the definition of $\Omsf (\order'_{E(\cG)},p)$, gives the desired expansion in terms of topological graphs,
\begin{multline}
\label{eq_mst_ctop}
\cxyz(p) = \sum_{\cG \in \cG_{\bfx,\bfy;\bfz}} \sum_{\order'_{E(\cG)} \in S_{|E(\cG)|}} \left(\dmsf(\cG | \order'_{E(\cG)})/{\cal A}'(\cG)\right) \\
\times  \Omsf(\order'_{E(\cG)},p) \sum_{\embed: \cG \to G} \Pr \biggl[ \bigwedge_{\epsilon \in E(\cG)} (L_\epsilon \leq \p_\epsilon) \biggr].
\end{multline}
This is the final result of this section, which again should be compared with the percolation result \eqref{eq_perc_c3}.

The expansion for MSF$(p)$ that we have obtained is naturally organized as a low-density expansion, that is as an expansion in powers of $p$. For large graphs $\widehat{G}$, it becomes unwieldy, especially for $p$ greater than around the percolation threshold (to the extent that a threshold can be associated with a finite graph, for example for a portion of a hypercubic lattice we can consider the threshold $p_c$ of the infinite lattice). The corresponding expansion for the connectedness function $C_{\bfx,\bfy}(p)$ in percolation must produce the answer $1$ when $p\to 1$, but in a very complicated way, as a sum of a large number of terms. For the MST, obtained from MSF$(p)$ as $p\to1$, the probability that the path on the MST from $\bfx$ to $\bfy$ passes through $\bfz$ remains non-trivial in the limit, and is again given by a complicated set of terms. A general analysis of this sum on a large graph for $p>p_c$ would require a resummation of terms to allow for the presence of the ``giant cluster'' in the corresponding percolation. In the Potts model formulation of percolation, this is done by giving an expectation value to the Potts spin. A formulation of such a resummation suitable for the MSF$(p)$ problem will not be given in the present paper, which is consequently restricted to the region $p\leq p_c$ on large lattices from here on.

\section{Renormalized perturbation expansion for MSF paths}
\label{s_cont_ft}

In this section we describe how the exact low-density expansion obtained in the previous section may, for $p\leq p_c$, be turned into a continuum theory (with a cut-off) to which renormalization-group methods may be applied. This continuum theory is obtained from a naive-looking procedure of replacing the lattice paths representing edges of a topological graph by continuum random walks, and neglecting the excluded volume requirement that the graph be {\em embedded} in the lattice without using a vertex or edge more than once. The resulting perturbation expansion is expected to be asymptotic rather than convergent. The expansion is closely related to that for percolation, so that the resulting Feynman diagrams and corresponding integrals can be compared with those of the latter. However, the expansion does not arise from the path integral of an action functional, so we do not technically have a ``field theory of MSF paths,'' although we will show that many of the standard techniques of field theory remain applicable. In particular, we show in Section \ref{s_rg_justify} and Appendix \ref{app_RGibility} that the perturbation expansion may be renormalized and RG methods applied. This enables us to calculate in Section \ref{s_rg_calc} the fractal dimension of paths on MSF$(p_c)$ as an asymptotic expansion in $\ve = 6 - d$, which we perform to leading order.


\subsection{The excluded volume constraint}
\label{s_exclvol}

A lattice expansion in terms of topological graphs, such as \eqref{eq_perc_c3}, is very close to describing a continuum theory. The only remaining roadblock lies in the sum over embeddings $\embed$, which carries an effective excluded volume constraint: edges of $\cG$ must be mapped to self-avoiding chains of lattice edges, and which must all be edge-disjoint (and hence also vertex-disjoint): no edge on the lattice may be used more than once. This is technically more difficult to incorporate. If we drop this constraint, we have a sum over ``free embeddings'' $\bar{\embed}$, which map edges of $\cG$ to random walks in the lattice, which are allowed to intersect, and then Fourier transform techniques may be freely used. This is the starting point for a continuum theory: as is well known, the generating function for random walks may be thought of as the propagator of a free scalar field.

For percolation, the excluded volume constraint may be avoided by further modifications to the expansions given above. This is Essam's ``$\rho$ expansion'', given in Refs.\ \cite{essam_perc_review, essam_coniglio}. There certainly seems to be no obstacle to extending the $\rho$ expansion to the expansion for MSF paths derived in the following subsection, but we do not pursue this line of inquiry (which may be relevant for a mathematically rigorous reformulation of the results given here).
Instead, in what follows, we assume we may drop the excluded volume constraint without difficulty or modification of our lattice expansion. This is because our ultimate aim is an RG calculation around $d_c =6$. In the diagrammatic perturbation expansion, the important graphs for calculation of the exponents have vertices of degree 3 only. If we attempted to incorporate the excluded-volume constraints, perhaps following Essam's technique, additional diagrams (topological graphs) with vertices of degree four or more would enter, but these will be irrelevant close to six dimensions, as could be demonstrated by extending the techniques given below.

\subsection{Continuum perturbation expansion for MSF paths}
\label{s_msf_ft}

We first consider the percolation connectedness functions, defined via \eqref{eq_perc_c3} (and its generalization to $n$-point functions), which must be reproduced by the MSF$(p)$ process. Because we neglect the excluded volume constraint present in the sum over embeddings, for a given topological graph (which may now be referred to as a Feynman diagram) $\cG$, we may take the chains of edges produced by the lattice embedding to be independent random walks. In the sum over such embeddings, it is natural to consider the Fourier transform with respect to the positions $\bfx_i$ ($i=1$, \ldots, $n$), and to use the Fourier representation for the probability of a walk between two of the vertices; the latter takes the form $1/(\q^2+t_0)\equiv G_0(\q,t_0)$, the same as the propagator of a scalar field \cite{itzykdrouf}.  Here $t_0\geq0$ is the ``mass-squared'' parameter, which depends monotonically on $p$; naively, $t_0$ decreases to zero as $p$ increases to $p_c$ (however, this statement will be modified by perturbative corrections). Thus we make the substitution
\be
\sum_{\bar{\embed}:\cG \to G} \Pr[G \leq p] \to I(\cG,t_0),
\ee
where
\begin{multline}
\label{eq_cont_perc}
I(\cG,t_0) =g_0^{|V(\cG)|-n}\int \! \left(\prod_{\epsilon \in E} \frac{d^d \q_\epsilon}
{(2 \pi)^d}\right)\cdot \left(\prod_{\epsilon \in E}\frac{1}{\q^2_\epsilon + t_\epsilon} \right) \\
\times \prod_{v \in V} (2 \pi)^d \delta^d \left( (\k_\text{ext})_v - \sum_{\epsilon \in E} {\cal N}_{\epsilon,v} \q_e \right).
\end{multline}
Here $\cal N$ is the incidence matrix of $\cG$ under an arbitrary orientation of the edges:
\begin{align}
\begin{split}
{\cal N}_{\epsilon,v} &= 1 \text{ if $v$ is the head of $\epsilon$,} \\
{} &= -1 \text{ if $v$ is the tail of $\epsilon$, and } \\
{} & = 0 \text{ otherwise.}
\end{split}
\end{align}
The external momenta $\{ \k_{\text{ext},i} \}$ are the Fourier conjugates of the positions $\{ \bfx_i \}$ of the graph's root vertices, and $ (\k_\text{ext})_v$ is the net external momentum flowing into vertex $v$. The momentum (i.e.\ wavevector) integrals are subject to a cutoff: each variable $\q_\epsilon$ must obey $|\q_\epsilon|<\Lambda$. This cut off replaces the restriction of the integrals to a single Brillouin zone that is due to working on the lattice (note that in the latter case the propagators would be invariant under addition of a reciprocal lattice vector to any $\q_\epsilon$). Thus $\Lambda$ is initially taken to be of order $1/a$, where $a$ is the spacing of the lattice points. Finally, a factor $g_0^{|V(\cG)|-n}$ has been inserted, to absorb other numerical factors that are omitted, and because this parameter will be renormalized later. At this stage, $g_0$ is strictly speaking of order one, but will be viewed as small in the perturbation expansion. At the same time, we will restrict the sum to topological graphs with vertices of degree three, except for the marked points $\bfx_i$, which are of degree one (thus we have $g_0$ for each cubic vertex). Both of these simplifying assumptions can be justified because other contributions can be shown to be irrelevant (in a RG sense) near six dimensions, using the RG technology to be discussed later.

Thus, the continuum expansion for the percolation connectedness functions becomes
\be
C_{\{\bfx_i\}} (p) = \sum_{\cG \in \cG_{\bfx,\bfy,e} }d(\cG) I(\cG, t_0)/{\cal A}(\cG),
\ee
where the sum is over diagrams with trivalent vertices and $n$ ``external'' marked points on which $d$ depends, and ${\cal A}(\cG)$ is the appropriate symmetry factor. For the function $C_{\bfx,\bfy}^e(p)$, there is a similar expansion, in which we sum over graphs with two marked degree one external points at $\bfx$, $\bfy$, and with a single degree two vertex marked $\bfz$ (to replace $e$, as mentioned before); the symmetry factor becomes ${\cal A}'(\cG)$, and the power of $g_0$ is $|V(\cG)|-3$. In the same way that this function on the lattice was obtained by differentiating with respect to $p$, this function in the continuum can be obtained by differentiating the Feynman diagram expression for $C_{\bfx,\bfy}(p)$ with respect to $t_0$ (and no subsequent integration in the present case). In particular, this produces the correct symmetry factors. This operation gives the additional vertex with zero wavevector, but can be generalized to allow some momentum to enter at $\bfz$, as given above. In field theory it is referred to as insertion of a mass- or $\phi^2/2$ operator, where $\phi$ would be the field corresponding to the degree one external points \cite{amit_percFT}.

Because of the equivalence of the $d$ weights with those of the $Q=1$ Potts model (at least for $n=1$, $2$), this perturbation expansion reproduces the standard one for the field theory of the $Q\to 1$ Potts model, which is usually obtained via the Hubbard-Stratonovich technique \cite{luben_review,zia_potts}. We emphasize that for the purposes of what follows we are prohibited from making any reference to an action functional due to the fact that the MSF path vertex cannot be expressed in terms of any local operator: instead we must phrase our argument entirely in terms of diagrammatic expansions.

Now we turn to the path connectedness function $\cxyz (p)$ for MSF$(p)$, which is the probability that at parameter value $p$ there is a path on MSF from $\bfx$ to $\bfy$ passing through $\bfz$. We treat the continuum version of the expansion in exactly the same way as we did for percolation, with the function $C_{\bfx,\bfy}^e$ being the closest analogue. Then compared with percolation, we make the following modifications of the expansion: the diagrams are considered for each ordering $\order'$ of the costs $L_\epsilon$ of the edges $\epsilon$ of the diagram, the $d$-weights are modified as they depend on the ordering through the requirement that the path on the MSF passes through $\bfz$ (and depend on marked points $\bfx$, $\bfy$, $\bfz$), and the probability is modified to give the probability for the ordering. Further, for the MSF, when we pass to the continuum perturbation expansion, the probability for a given ordering of costs $L_\epsilon$ on the edges of the diagram must depend on squared-masses $t_\epsilon$ in place of $\ell_\epsilon$. These parameters are acted on by the integro-differential operator  $\Omsf(\order', p)$, which finally sets all $\ell_\epsilon$ to $p$. After the change of variable to $t_0$,
the operator $\Omsf(\order', p)$ may trivially be rewritten in terms of the $\{ t_\epsilon \}$ variables: because every derivative is paired with an integral, the Jacobians involved in changing from the $\{\p_\epsilon \} $ to the $\{ t_\epsilon \}$ cancel. The only difference is that the ordering $\order'$ applies to the $t_\epsilon$ in \emph{reverse}: small $L_\epsilon$ corresponds to large $t_\epsilon$. Therefore, \eqref{eq_msfop_def} becomes
\be
\label{eq_mst_op}
\Omsf(\order',t_0) =  \idotsint\limits_{\infty >  t_{\order'(1)} \geq  t_{\order'(2)} \geq \cdots \geq  t_{\order'(n)} \geq t_0} \prod_{\epsilon \in E(\cG)} dt_\epsilon \frac{d}{dt_\epsilon}.
\ee
The perturbation expansion for the MSF path vertex functions in the continuum is now
\begin{multline}
\label{eq_pathvert_rule}
\cxyz(p) = \\
\sum_{\cG \in \cG_{\bfx,\bfy}} \sum_{\order' \in S_{|E(\cG)|}}  \left(\dmsf(\cG| \order')/{\cal A}'(\cG)\right) \\
{}\times\Omsf (\order',t_0) I(\cG,\{t_\epsilon\}).
\end{multline}
The set of graphs involved are the same as those used in the differentiated two-point connectedness function of bond percolation (as described above), so we may compute this by starting with the expansion for $C_{\bfx,\bfy}^e (t_0)$ and making the substitution,
\begin{multline}
\label{eq_msf_from_perc}
d(\cG) I(\cG,t_0) \mapsto \\
\sum_{\order' \in S_{|E(\cG)|}} \dmsf(\cG| \order') \Omsf (\order',t_0) I(\cG,\{t_\epsilon\})
\end{multline}
on a diagram-by-diagram basis.

To summarize, the Feynman diagram rules for the MSF path correlation function, as specified in \eqref{eq_msf_from_perc}, are as follows:

\noindent
1) For each diagram $\cG \in \cG_{\bfx,\bfy,\bfz}$ contributing to the two-point correlation function between $\bfx$ and $\bfy$ of a cubic scalar field theory with a mass-insertion at $\bfz$, we associate a mass-squared $t_0 \leq t_\epsilon < \infty$ to each edge $\epsilon\in E(\cG)$.  For each ordering $\order$ of these mass parameters, we act on the integrand with the operator $\Omsf(\order',t_0)$ defined in \eqref{eq_mst_op}.

\noindent
2) After integrating over wavevectors, the contribution from each ordering is multiplied by the diagrammatic weight $\dmsf ( \cG | \order')/{\cal A}'(\cG)$, with $\dmsf$ defined in \eqref{eq_dmst_def}, and the sum of these over all mass parameter orderings is the contribution to the MSF path connectedness function.

Note that we must act with $\Omsf$ before any momentum integrations are performed, since the latter may produce expressions that diverge at the upper limit of the integrations in $\Omsf$. This situation could be remedied by cutting off the domain of integration in $\Omsf$ to $\{ t_\epsilon \} < \Lambda^2$, at the expense of complicating our RG calculation.

\subsection{Extraction of fractal dimensions and lowest order results}
\label{s_fracdim}

The perturbation expansion that we have now obtained can be organized as a loop expansion: the lowest order contribution to $\cxyz$ is order ${\cal O}(g_0^0)$, and is simply the diagram that takes the form of a path from $\bfx$ to $\bfz$ to $\bfy$, which possesses no loops (cycles), while higher orders in $g_0$ contain additional loops, one for each factor of $g_0^2$. The lowest order result, in position space and at $t_0=0$, takes the form (in this and the following, all separations like $|\bfx-\bfy|$ are assumed large, $\gg a$)
\be
\cxyz\propto \frac{1}{|\bfx-\bfz|^{d-2}|\bfz-\bfy|^{d-2}}.\ee
By contrast, the 2-point connectedness function at criticality, obtained as a single scalar propagator, is proportional in this order to $1/|\bfx-\bfy|^{d-2}$. [At zero-loop order, these results are the same for MSF$(p_c)$ and for critical percolation.] Dividing the two gives the conditional probability that there is a path from $\bfx$ to $\bfy$ passing through $\bfz$, given that there is a path from $\bfx$ to $\bfy$. Integrating over $\bfz$ gives $\propto |\bfx-\bfy|^2$. This is viewed as proportional to the total number of steps on the (lattice) path, even through the events of the path passing through the various $z$ are not disjoint. The exponent $2$ indicates that the fractal dimension of the path is 2, which is the correct result for a random walk. Thus we have shown that
\be
D_{\rm p}=2
\ee
at zero-loop order. This will be found to be correct for $d>6$, and also for $d=6$ up to logarithmic corrections. The same dimension is believed to hold for paths on critical percolation clusters for $d>6$, by similar field-theoretic arguments. Geometrically, it is because on large scales these clusters are trees, with no loops \cite{stah}, and hence are the same in the MSF$(p_c)$ process.

In general, and specifically for $d<6$ as we will show, the scaling exponents will be different. At $p_c$, quite generally $\cxyz$ will have the scaling behavior
\be
\widetilde{C}_{b\bfx,b\bfy}^{b\bfz}=b^{-(d-2+\eta)-(d-D_{\rm p})}\cxyz
\ee
for any $\bfx$, $\bfy$, and $\bfz$, and scalar $b$, while the 2-point connectedness behaves as
\be
C_{b\bfx,b\bfy}=b^{-(d-2+\eta)}C_{\bfx,\bfy}.
\ee
Thus these two functions determine two exponents $\eta$ and $D_{\rm p}$ for MSF$(p_c)$, and $\eta$ will be the same as for percolation, as we will explain shortly ($\eta=0$ for $d>6$). Then in the same way as at zero loops, we infer the fractal dimension $D_{\rm p}$ for the MSF path. From the geometric point of view, $d-D_{\rm p}$ is the co-dimension of the path.

The exponent $\eta$ describes the decay with distance $r$ of the probability that two points are connected by a critical percolation cluster, namely $\sim r^{-(d-2+\eta)}$. In a field-theoretic point of view, $(d-2+\eta)/2=x_\phi$ is the dimension of the Potts field operator $\phi$, while $d-D_{\rm p}=x_{\rm p}$ is the scaling dimension for the path-vertex ``operator''. $x_\phi$ is related to the fractal dimension $D_{\rm perc}$ of the critical percolation clusters as the codimension $x_\phi=d-D_{\rm perc}$, so $D_{\rm perc}=(d+2-\eta)/2$.

For the percolation function $C_{\bfx,\bfy}^e$, the corresponding $\phi^2$ operator at $e$ has dimension $x_{\phi^2}=d-D_{\rm sc}$, and $D_{\rm sc}$ is the fractal dimension of the set of singly-connected edges on the path from $\bfx$ to $\bfy$ on the critical percolation cluster \cite{stah}. For $d<6$ this set does not usually form a connected path. We see that the MSF path must include the singly connected edges, which leads to the inequality
\be
D_{\rm p}\geq D_{\rm sc}.
\ee
Because the function $C_{\bfx,\bfy}^e$ is connected via differentiation with the change in connectivity with $p$ (or $t_0$),
the scaling dimension $x_{\phi^2}$ of the $\phi^2$ insertion controls the length scale produced by taking $p<p_c$; this length is the correlation length $\xi$, and we can define the exponent $\nu$ by $\xi\sim (p_c-p)^{-\nu}$ as $p\to p_c$. It follows that $D_{\rm sc}=\nu^{-1}$ \cite{stah}. This discussion shows how the fractal dimension $D_{\rm p}$, and others, can be extracted from the renormalized perturbation calculations.

\subsection{Beyond lowest order: breakdown of perturbation theory for $d<6$}
\label{s_ginzburg}

The perturbation expansion for the MSF path connectedness functions can be treated in a similar manner as that for standard field theories. A first step is to introduce one-particle irreducible (1PI) functions. A diagram is defined to be 1PI if it does not become disconnected when a single edge is removed. Now for the MSF path connectedness function $\cxyz$, the (dominant) diagrams that contribute have a single edge emerging from $\bfx$ and $\bfy$. For terms of order $\bigO(g_0^2)$ (as $g_0\to0$), the diagram possesses at least one loop (cycle). It can then be decomposed into a chain of one or more disjoint 1PI 2-point graphs, connected by single edges. The vertex labeled $\bfz$ is either inside one of the subdiagrams (subgraphs), or on one of the single edges. The 1PI subdiagrams not containing $\bfz$ will be called self-energy diagrams.

For the $\dmsf$-weight of such a diagram, it is easy to see that the weights associated with each 1PI subdiagram factor. This is because the MSF path must pass through each of the 1PI subdiagrams in turn. For the self-energy diagrams, all paths through the subdiagram (which must be considered when evaluating the $\dmsf$-weight) contribute a non-zero amount (all diagrams we consider are connected to $\bfx$, $\bfy$, $\bfz$). Consequently, the factor in the $\dmsf$-weight for the subdiagram is independent of the ordering $\order'_E$ restricted to the subdiagram, and then the weight reduces to the same expression as in percolation. (This is not true, however, for the 1PI subdiagram that contains the path vertex at $\bfz$.) The application of the $\Omsf$ operator and the sum over orderings $\order'_E$ can now be carried out using \eqref{eq_sum_all_orderings}. Then the contribution of such a self-energy diagram is the same as in percolation. The self-energy diagrams can be formally summed to all orders in perturbation theory to yield the self-energy $\Sigma(\q,t_0)$, and then each of the two series of alternating $G_0(\q,t_0)$'s and $\Sigma(\q,t_0)$'s can be summed as a geometric series, giving the full Green's function $G(\q,t_0)$,
\be
G(\q,t_0)^{-1}=G_0(\q,t_0)^{-1}-\Sigma(\q,t_0)
\ee
(Dyson's equation). We pause to point out that the 2-point connectedness function for MSF$(p)$, in which we do not require the path on the MSF to pass through any particular point $\bfz$, is similarly shown in this diagrammatic point of view to be the same as in percolation, and is given by $G(\q,t_0)$. Consequently, the exponent $\eta$ defined above for $p=p_c$ must be the same as the similarly-defined exponent in percolation.

\begin{figure}[h]
\includegraphics[width=3.1in]{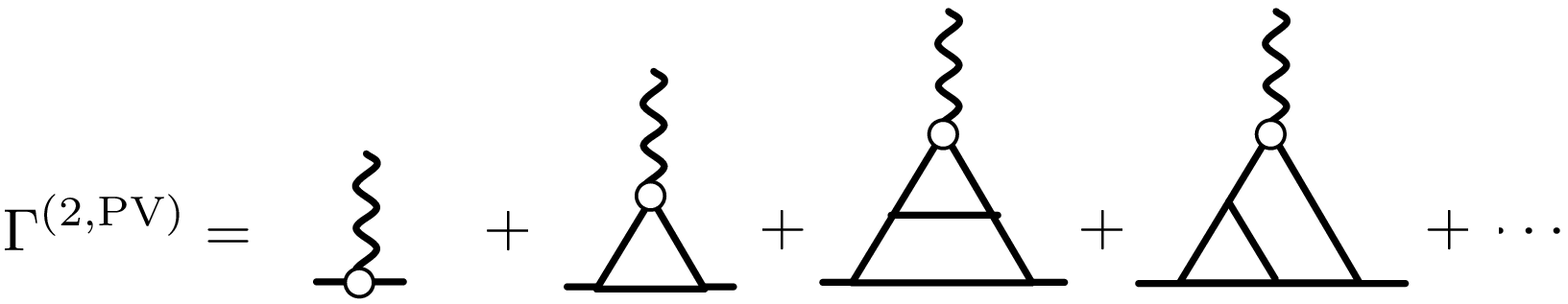}
\caption{Perturbation expansion for the path vertex function given in \eqref{eq_pathvert_rule}. We represent the path vertex by the open circle connected to a wavy line. Note that these diagrams do not include any notation corresponding to the operation of $\Omsf$. The lowest non-trivial term is the second one on the right-hand side. \label{f_pathvert}}
\end{figure}

The MSF path connectedness function, with zero momentum entering at the path vertex, can now be written formally as
\begin{multline}
\int \! d^d \bfz \, \cxyz(t_0) = \\
\int \! \frac{d^d \q}{(2 \pi)^d} e^{-i \q (\bfx - \bfy)}\, G(\q, t_0) \Gamma^{(2,\rm PV)} (\q, \zero; t_0) G(\q, t_0),
\end{multline}
where $\Gamma^{(2,\rm PV)}(\q,0;t_0)$, which we call the path vertex function, is the Fourier transform of the sum of all 1PI diagrams with two external points (connected to $\bfx$, $\bfy$), plus the path vertex at $\bfz$, which has here been assigned zero momentum. (The generalization to $\Gamma^{2,\rm PV}(\q_1,\q_2;t_0)$ should be obvious.) Diagrams contributing to $ \Gamma^{(2,\rm PV)} $ are depicted in figure \ref{f_pathvert}. Similarly, we also define, for the $N$-point connectedness functions $G^{(N)}(\q_1,\ldots,\q_N)$ without the path vertex ($N=2$, $3$; $G^{(2)}=G$), and the 2-point connectedness function with a mass (or $\phi^2$) insertion, which are the Fourier transforms of the percolation functions $G^{(2)}=C_{\bfx,\bfy}$, $G^{(2,1)}=C_{\bfx,\bfy}^\bfz$:
\be
G^{(N)}(\{\q_i\};t_0)= \\
\prod_{i=1}^NG(\q_i,t_0)\cdot \Gamma^{(N)}(\{\q_i\};t_0);
\ee
\begin{multline}
G^{(N,1)}(\{\q_i\};\q;t_0)= \\
\prod_{i=1}^NG(\q_i,t_0)\cdot \Gamma^{(N,1)}(\{\q_i\};\q;t_0).
\end{multline}
In these functions, a $\delta$-function that sets the total wavevector to zero has been removed, and $\{\q_i\}$ stands for the ordered set $\q_1$, \ldots, $\q_N$. This causes a minor difference in notation from that for the path vertex function above: in the functions $G$ or $\Gamma$ containing $N$ or $N+1$ wavevector arguments, one of the wavevectors could be eliminated, which is what was done in $\Gamma^{(2,\rm PV)}$ above, and we occasionally do this for the others also without further comment. The functions $\Gamma^{(N)}$ and $\Gamma^{(N,1)}$ are called the 1PI vertex functions (of the types indicated).
We identify $\Gamma^{(2)}(\q_1,-\q_1;t_0)=G(\q_1;t_0)^{-1}$.

The problem of calculating the path exponent $D_{\rm p}$ has now been reduced to the calculation of the 1PI path vertex function $\Gamma^{(2,\rm PV)}$. The external lines $G$ are the same as in percolation, because of the factorization and ordering independence of the weights for the self-energy diagrams. (Similar, but more subtle, factorizations play an important role in the later part of the argument also.) The path vertex, on the other hand,  is not the same as the mass-insertion vertex in percolation which it resembles.

To illustrate the perturbation expansion, let us now evaluate the first correction, of order $g_0^2$, to $\Gamma^{(2,\rm PV)}$ (see Fig.\ \ref{f_pathvert}) at zero external momentum. From the rules given above, this correction comes from the graph with three propagators connected to form a triangle, and the contribution is (note that ${\cal A}'=1$ for this graph)
\begin{multline}
I_{\rm MSF}(\bigtriangleup,t_0) = \\
\sum_{\order' \in S_3} \dmsf(\bigtriangleup|\order)  \Omsf(\order, t_0)  I_\Lambda(\bigtriangleup, \{ t_1, t_2, t_3 \}),
\end{multline}
where the last integral is
\begin{multline}
 I_\Lambda(\bigtriangleup, \{ t_1, t_2, t_3 \}) = \\
 g_0^2 \int^\Lambda \! \frac{d^d \k}{(2 \pi)^d} \, \frac{1}{(\k^2 +t_1) (\k^2+t_2)(\k^2+t_3)}.
\end{multline}
Here $t_3$ is the mass-squared on the edge not adjacent to the path vertex.
Our momentum-space rules required us to integrate up to radius $\Lambda$. Since the diagram is evaluated at zero external momenta, it is symmetric under permutations of the $\{ t_\epsilon \}$ and
\be
\Omsf(\order, t_0) I_\Lambda(\bigtriangleup, \{ t_1, t_2, t_3 \}) = \frac{1}{3!}  I_\Lambda(\bigtriangleup, \{ t_0, t_0, t_0 \}).
\ee
Applying \eqref{eq_dmst_def}, we have $ \dmsf(\bigtriangleup|\order) = 0$ for the two orderings in which $t_1$ and $t_2 > t_3$, and $ \dmsf(\bigtriangleup|\order) = -1$ for the other four orderings. Thus the result is
\be
I_{\rm MSF}(\bigtriangleup,t_0)=-\frac{2}{3}I_\Lambda(\bigtriangleup,
\{ t_0, t_0, t_0 \}).\label{eq_pathvertcorr}
\ee
For the corresponding simple mass-insertion vertex in percolation, the result would be instead $-I_\Lambda(\bigtriangleup,\{ t_0, t_0, t_0 \})$.

Dropping $g_0^2$ and numerical factors, the contribution to the path vertex function behaves like the integral
\be
\label{eq_ginz_int}
I'_\Lambda = \int_0^\Lambda \! \frac{k^{d-1}dk}{(k^2+t_0)^3}.
\ee
At present, we are interested in this for fixed $\Lambda$ as $t_0\to 0$, so as to reach $p=p_c$. We see that for $d>6$, \eqref{eq_ginz_int} behaves as $\Lambda^{d-6}$ as $\Lambda\to\infty$, and is finite (for any $\Lambda$) as $t_0\to0$. But for $d<6$, the reverse is the case: the integral converges as $\Lambda\to\infty$, but diverges (for any $\Lambda$) as $t_0^{(d-6)/2}$ as $t_0\to0$. In the borderline case $d=6$ the integral diverges logarithmically at both ends. For non-zero external wavevectors, the dependence of the integral on $\Lambda$ is the same in all cases. Note that similar statements apply for percolation; only the numerical prefactor is different.

There are similar results for diagrammatic contributions with more loops. Simply counting the number of propagators and integrations gives the ``superficial degree of divergence'', which for the path vertex function is always $\sim g_0^2\Lambda^{d-6}$ or $g_0^2t_0^{(d-6)/2}$ for $d>6$ and $d<6$ respectively, raised to the power of the number of loops (independent cycles) in the diagram, as in the one-loop example above. Note that this is the same as if the $\Omsf$ operator were absent, because $\Omsf$ leaves the overall degree (in $\k$, at $t_0\simeq0$) of the integrand unchanged. The consequence for perturbation theory at fixed $\Lambda$ and $d>6$ is simple: each term in the perturbation expansion of $\Gamma^{\rm PV}$ for MSF$(p)$ ($p\simeq p_c$) is finite as $t_0\to0$. This is true for the self-energy diagrams on the external lines also, and the value of $\Sigma(\q,t_0)$ at $\q=0$ determines an effective shift in the value of $t_0$ that corresponds to $p=p_c$: $p_c$ must correspond to the value of $t_0$ such that $t_0-\Sigma({\bf 0},t_0)=0$, and there are also other finite changes in the normalization of $G$. (In this case one would wish to sum up self-energy insertions in the lines inside of $\Gamma^{\rm PV}$ also. However we have not shown that these take the same form as on the external lines. This will be addressed below.) But the consequence is that in each order, $\Gamma^{\rm PV}(\q,t_0)\sim \bigO(1)$ as $\q\to0$ at $p=p_c$, while $G(\q,t_0)^{-1}\propto \q^2$. This in turn implies that there is no change in the exponents from their lowest order values, $\eta=0$ and $D_{\rm p}=D_{\rm sc}=2$. Note that here we disregard the possibility that the sum of an infinite number of finite terms might diverge, which might invalidate the conclusion.

For $d\leq6$, this perturbative argument breaks down as the corrections become arbitrarily large as $t_0\to 0$, in particular in the region $t_0<g_0^{4/(6-d)}$ (the Ginzburg criterion). In order to handle this, the use of RG techniques becomes essential. These techniques effectively re-sum and redefine the expansion. There are several formulations of the RG. These may be divided into two classes. One class of particularly powerful techniques is the field-theorists' RG, in which the aim initially is to take $\Lambda\to\infty$ (or $a\to0$) at fixed separations or momenta, in such a way that the limits of the correlation (or connectedness) functions exist, thus recovering a true continuum theory. This is called renormalization of the theory. Subsequently, the renormalized theory is used to set up the RG, and calculate exponents for $d\leq d_c$. The leading alternative is the Wilsonian RG, in which the cutoff is kept finite. The Wilsonian RG is more difficult to use for higher numbers of loops. Both approaches lead to equivalent results for physical quantities such as exponents for $d\leq d_c$, where $d_c=6$ for percolation and MSF$(p_c)$. In this paper we will follow the approach of the field theorists.

For $d>6$, we can see from above that the effective expansion parameter is $g_0^2\Lambda^{d-6}$. As $\Lambda\to\infty$, it is then necessary to make $g_0\to0$ such that $g_0^2\Lambda^{d-6}$ does not diverge. In fact the situation is even worse than this would suggest: there simply is no rational way to define the limit $\Lambda\to\infty$ so that the connectedness functions at fixed $\q$ and (for example) $p=p_c$ have finite limits, without introducing an infinite number of parameters. This is referred to as {\em non-renormalizability} of the perturbation expansion. But by keeping the cutoff $\Lambda$ finite, and using the Wilsonian point of view, we can see that the exponents in this region take their simple zero-loop values, as indicated above. Accordingly, we concentrate on $d\leq 6$ from here on in this article.

\subsection{Renormalizability of the theory}
\label{s_rg_justify}

In this section we outline our proof that the MSF path vertex theory for $d\leq 6$ may be consistently renormalized. The full technical details are in Appendix \ref{app_RGibility}. In the interests of making this article more accessible to readers without a field theory background, we take a somewhat pedagogical approach in discussing the renormalizability and the RG calculation in the remainder of this paper. Of course, we do not have the space here for a full description; the interested reader is directed towards any of the standard textbooks such as \cite{RGbasic,feynint_3}.

We saw above that the Feynman integrals associated to diagrams for certain vertex functions are superficially divergent as $\Lambda\to\infty$ in six dimensions. In fact, closer inspection reveals that subintegrals (integrals over a subset of the loop momenta $\k$, holding the others fixed) may also be superficially divergent, and this can occur even when in integrals that are superficially convergent as a whole, showing that they do not converge after all. However, the superficially divergent integrals (or subintegrals) are associated only with (sub-)diagrams that are, topologically at any rate, of the form of the vertex functions $\Gamma^{(2)}$, $\Gamma^{(3)}$, $\Gamma^{(2,{\rm PV})}$, or $\Gamma^{(2,1)}$. These correspond respectively to the self-energy, cubic coupling, MSF path, and mass-insertion vertex functions. The first two of these suggest a possible way to eliminate the divergences: add the divergent terms to the mass-squared $t_0$ and the coupling $g_0$, respectively, and define renormalized quantities $t$ and $g$, and then insist that these are the ``physical'' or measurable parameters at long length scales. There is also a subleading divergence in the self-energy of order $\q^2$, which perhaps can be removed similarly by rescaling the field, and hence the Green function (this effect also enters the definition of $t$). The divergences in $\Gamma^{(2,{\rm PV})}$ and $\Gamma^{(2,1)}$ can be handled similarly.

This procedure works for conventional field theories at their critical dimension; one such case is the theory of percolation at six dimensions. It is important to recognize why it can work. That is because every occurrence of, for example, a self-energy subdiagram within another diagram occurs with a $d$ weight that can be factored as the $d$ weight for the subdiagram, times that for the quotient diagram, in which the subdiagram is contracted to a single vertex. Further the integrations over wavevectors in a Feynman integral have the property that the integral for a subdiagram always has the same form, independent of the larger diagram of which it is a part. Then the subintegral for any self-energy subdiagram has precisely the same divergence wherever it occurs, independent of the larger diagram of which it is a part. Meanwhile the quotient diagram has the form of a lower order diagram. This enables us to write the leading divergence as a correction to the bare mass-squared $t_0$ that is context-independent, and therefore meaningful. The same has to be true for the other divergent subdiagrams (or ``renormalization parts''). For percolation, the factorization of the $d$ weights can be easily seen in the $Q\to1$ Potts model formulation, in which the $d$-weights arise from contracting together tensors, and then the factorization for subdiagrams is automatic (and similarly for other local field theories). It is not immediately obvious that this will hold for our MSF theory, because: (i) The $d$ weights are replaced by $\dmsf$ weights, which depend on an ordering $\order$, may not factor in the fashion required, and in fact for some orderings do not factor; (ii) The $\Omsf$ operator and summation over orderings raise similar questions.

In Appendix \ref{app_RGibility} we undertake a careful study of these questions. We find that the degree of divergence of a subdiagram for a renormalization part is the same as it would be for the corresponding subdiagram in percolation for certain orderings, and for these the $\dmsf$-weight exhibits the desired factorization properties. Indeed, for the self-energy and cubic coupling renormalization parts, the divergent part has exactly the same coefficient as for percolation. This holds also for part of the subleading divergence in the self-energy case, but there is also another subleading part in that case which does not have these properties. That part is problematic, as the program above provides no apparent way to remove these divergences. However, we eventually find that all such terms cancel, not for a single diagram, but in the sum of diagrams of a given order. In the remainder of this discussion, we will take that for granted, and so continue as if there are no such divergences.

These observations then allow us to absorb all the divergences into the quantities mentioned above. More formally, this amounts to subtracting off the superficially-divergent contribution for each subdiagram of a diagram (including that, if any, for the diagram as a whole). After doing so, we should prove that the remaining integrals are actually finite. Here again, we cannot simply appeal to the usual field theories, as we have modified the Feynman integrands, and so the proofs must be reconsidered. We complete the proof using the Schwinger parametric integral formulation \cite{param_1,param_2,bd,param_div_1,param_div_2}, and a theorem by Berg\`{e}re and Lam \cite{param_div_3}. This then completes the proof of renormalizability of our perturbation expansion.

The renormalization procedure removes the divergent parts of the original Feynman integrals. It does not uniquely fix a finite part that may also be subtracted. This part may be determined by giving some renormalization conditions obeyed by the renormalized vertex functions. A convenient choice for the following is to define the values (and a first derivative) of these functions at zero renormalized mass-squared, $t=0$ (corresponding to $p=p_c$), and a non-zero wavevector of magnitude $\kappa$. For dimensional reasons, one or other of $t$ and $\kappa$ must enter. However, we also mention the scheme of dimensional regularization and minimal subtraction, in which such renormalization conditions are not used. We adopt the present scheme in order to keep things relatively transparent.

The RG is now introduced by obtaining an equation, the RG equation, describing how the renormalized vertex functions behave under a change in $\kappa$. As $\kappa$ decreases, the effective coupling $g$ changes, and may reach a non-zero $\kappa$-independent fixed point. This is then used to calculate the exponents for scale-covariant behavior of the vertex functions or correlation functions. The fixed point is at $g^2$ of order $\ve=6-d$, and so the expansion in powers of $g$ is traded for one in powers of $\ve$. This expansion is essential to obtain useful finite results for $d<d_c=6$. In this way we will obtain the exponents to order $\ve$ via a one-loop RG calculation.


\subsection{RG analysis at one-loop order}
\label{s_rg_calc}
In the preceding sections and appendix \ref{app_RGibility}, we have proved that the diagrammatic expansion for the MSF path theory is renormalizable. These proofs were technical, but having established this fact, we are free to make use of standard RG methods such as those discussed in Ref.\ \cite{feynint_3}. We continue to take a rather pedagogical approach in this section.

As we explained in sections \ref{s_ginzburg} and \ref{app_RGibility}, renormalizability of a theory means that we may absorb the strong $\Lambda$-dependence of all correlation functions into a finite number of parameters and the overall scale of the correlations, at the cost of introducing another scale $\kappa$. In the scheme we use, in which the renormalization conditions are at zero renormalized mass-squared $t=0$ and non-zero wavevector of order $\kappa$, the precise statement is that functions $g$, $Z_\phi$, $Z_{\phi^2}$, $Z_{\rm PV}$ exist such that (here we append subscripts $0$ to denote the ``bare'' vertex functions as constructed above, with cutoff $\lambda$)
\bea
&&Z_\phi^{N/2}(g_0,\kappa,\Lambda)Z_{\phi^2}^L
(g_0,\kappa,\Lambda)
\Gamma_0^{(N,L)}(\{\q_i\},\{\q_j\};g_0,t_0,\Lambda)\nonumber\\
&&\qquad\qquad=\Gamma_R^{(N,L)}(\{\q_i\},\{\q_j\};g,\kappa),\\
&&Z_\phi(g_0,\kappa,\Lambda)Z_{\rm PV}(g_0,\kappa,\Lambda)
\Gamma_0^{(2,\rm PV)}(\{\q_i\};g_0,t_0,\Lambda)\nonumber\\ &&\qquad\qquad=\Gamma_R^{(2,\rm PV)}(\{\q_i\};g,\kappa),\label{eq_renormgamm}
\eea
where $\Gamma_R$ are independent of $\Lambda$ as $\Lambda\to\infty$ with $g$, $t=0$ fixed, up to corrections vanishing in this limit. (Here $L$ is the number of insertions of $\phi^2$, and $\{\q_j\}$ is the set of corresponding wavevectors.) We will also now introduce dimensionless versions of the bare and renormalized couplings $g_0$, $g$:
\be
u_0^2 \equiv \frac{g_0^2 \kappa^{-\ve}}{(4 \pi )^{d/2}}, \quad
u^2 \equiv \frac{g^2 \kappa^{-\ve}}{(4 \pi )^{d/2}}.
\ee
We introduced an angular factor $(4\pi)^{d/2}$ in the above definitions for later convenience to simplify expressions. For the vertex functions not containing the path vertex, the functions and their renormalization is exactly as percolation, and this is also true of the following calculations; we include some details anyway to provide checks on the calculation.

The RG equations are obtained from the observation that the bare functions $\Gamma_0$ are independent of $\kappa$ when written in terms of $g_0$, so $\kappa\partial \Gamma_0/\partial \kappa=0$ at fixed $g_0$, $t_0$, $\Lambda$. Using the definition of $\Gamma_R$ we obtain
\bea
\label{eq_rg}
\left( \kappa \frac{\partial}{\partial \kappa} + \beta(u) \frac{\partial}{\partial u} -\frac{N}{2} \gamma_\phi (u) +L\gamma_{\phi^2}(u)\right)&&\nonumber\\
{}\times\Gamma^{(N,L)}_R (\{ \q_i \},\{\q_j\} ,u,\kappa)&&\nonumber\\ = 0,\qquad&&
\eea
and
\bea
\label{eq_rg_PV}
\left( \kappa \frac{\partial}{\partial \kappa} + \beta(u) \frac{\partial}{\partial u} -\gamma_\phi (u) +\gamma_{\rm PV}(u)\right)&&\nonumber\\
{}\times\Gamma^{(2,{\rm PV})}_R (\{ \q_i \},u,\kappa)&&\nonumber\\ = 0,\qquad&&
\eea
In each of these equations the first and second partial derivatives are at fixed $u$ and fixed $\kappa$, respectively.
The RG $\beta$ and $\gamma$ functions appearing in equations \eqref{eq_rg} \eqref{eq_rg_PV} are defined as
\bea
\beta(u)& =& \left. \kappa \frac{\partial u}{\partial \kappa} \right\vert_{g_0,\Lambda}, \\
\gamma_\phi (u) &=& \left.  \kappa \frac{ \partial \log Z_\phi}{\partial \kappa} \right\vert_{g_0,\Lambda},\\
\gamma_{\phi^2} (u) &=& -\left.  \kappa \frac{ \partial \log Z_{\phi^2}}{\partial \kappa} \right\vert_{g_0,\Lambda},\\
\gamma_{\rm PV} (u) &=& -\left.  \kappa \frac{ \partial \log Z_{\rm PV}}{\partial \kappa} \right\vert_{g_0,\Lambda},
\eea
and are finite as $\Lambda\to\infty$ \cite{feynint_3}. Hence in the limit they are independent of $\Lambda$, and so also of $\kappa$, because $u$, $\beta$, and all $\gamma$'s are dimensionless; they are simply power series in $u$.

We then impose the following renormalization conditions, which are those we reached in Appendix \ref{app_RGibility}, but written now with $t=0$ and $u$ in place of $g$. These serve to fix the dependence of the parameters on one another:
\be
\label{eq_rg_norms}
\begin{split}
\Gamma^{(2)}_R (\q = 0, u,\kappa) &= t = 0, \\
\left. \frac{d}{d \q^2} \Gamma^{(2)}_R (\q, u,\kappa) \right\vert_{|\q| = \kappa} &=1, \\
\left. \Gamma^{(3)}_R (\{\q_i\}, u,\kappa) \right\vert_{\rm SP} &=g, \\
\left. \Gamma^{(2,1)}_R( \{\q_i\}, u,\kappa) \right\vert_{\rm SP} &= 1, \\
\left. \Gamma^{\rm PV}_R(\{\q_i\}, u,\kappa) \right\vert_{\rm SP} &= 1.
\end{split}
\ee
Here $\rm SP$ denotes a symmetry point of the external momenta $\q_1$, $\q_2$, $\q_3$, at which $|\q_i|^2=\kappa^2$, as defined in Appendix \ref{app_RGibility} (though the precise definition is unimportant).
These conditions are now used to determine $\beta$ and the $\gamma$'s from the perturbation theory expansion in $g_0$ of the 1PI vertex functions $\Gamma_0$ with a fixed cutoff $\Lambda$. The expressions make sense provided $g_0$ is sufficiently small. We will calculate to one-loop order, which means that only the one-loop diagrams for the renormalization parts need to be calculated. This will give results for exponents to first order in $\ve=6-d>0$. (More generally, computing to $\bigO(\ve^{\cal L})$ in the $\ve$-expansion requires computing all the renormalization parts with $\cal L$ or fewer loops.)

Then the instances of equation \eqref{eq_renormgamm} with which we need to deal are, to $\bigO(g_0^2)$,
\be
\label{eq_rg_propverts}
\begin{split}
\Gamma^{(2)}_R (\q,u,\kappa) &=Z_\phi \left( (\q^2 + t_0) - \Sigma(\q) \right), \\
\Sigma(\q,g_0) &= d_2 g_0^2 I_2(\q); \\
\Gamma^{(3)}_R (u,\kappa)\vert_{\rm SP} &= Z_\phi^{3/2} \left(\left. g_0 + d_3 g_0^3 I_3\right\vert_{\rm SP} \right); \\
\Gamma^{(2,1)}_R (u,\kappa)\vert_{\rm SP} &= Z_{\phi} Z_{\phi^2} \left(\left. 1 + d_2 g_0^2 I_3\right\vert_{\rm SP} \right); \\
\Gamma^{\rm PV}_R (u,\kappa)\vert_{\rm SP} &= Z_{\phi} Z_{\rm PV} \left( 1 \vphantom{\sum_{\order \in S_3}}\right.\\
&{}\hphantom{=}\left.+g_0^2 \sum_{\order \in S_3} \dmsf(\bigtriangleup| \order) \Omsf(\order)I_3(\order)\vert_{\rm SP} \right).
\end{split}
\ee
Here $d_2, d_3$ are the percolation $d$-weights for these one-loop diagrams, $d_2=-1$, $d_3=-2$, and $d_{\rm PV}$ will be evaluated in a moment from the weights $d(\cG)$ and the operator $\dmsf\Omsf$; the values will be substituted only at the end of the calculation. The negative sign in the equation for $\Gamma^{(2)}_R$ arises because of Dyson's equation.

\begin{figure}[h]
\includegraphics[width=3.1in]{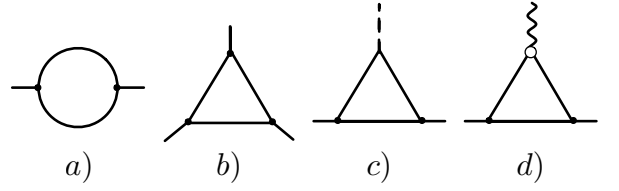}%
\caption{The 1PI one-loop diagrams for vertex functions that are ultraviolet divergent in six dimensions: a) self-energy, b) cubic interaction vertex function, c) mass-insertion vertex function, d) MSF path vertex function. \label{f_renormparts}}
\end{figure}

In equation \eqref{eq_rg_propverts}, $I_2$ and $I_3$ are Feynman loop integrals which we now evaluate. To leading non-trivial order, we can evaluate them at six dimensions, retaining only the terms that diverge quadratically or logarithmically as $\Lambda \to \infty$. These terms may be extracted by any of the standard techniques for evaluating Feynman integrals, including those in Appendix \ref{app_RGibility}; we refer the reader to \cite{feynint_1,feynint_2,feynint_3} in particular. Denoting this approximation by $\simeq$, we find
\begin{align}
I_2(\q) &=  \frac{1}{2} \int^\Lambda \! \frac{d^d \k}{(2 \pi)^d} \, \frac{1}{\k^2 (\k+\q)^2}, \nonumber \\
&\simeq \frac{1}{2} \frac{1}{(4 \pi)^{d/2}} \left( \Lambda^2 - \frac{\q^2}{6} \log \frac{\Lambda^2}{\q^2} \right).
\end{align}
The factor of $1/2$ appearing in $I_2(\q)$ is a diagrammatic symmetry factor, ${\cal A}=2$ (in the other diagrams, ${\cal A}$ or ${\cal A}'=1$). Note that the bare propagators appearing in these integrals should have mass-squared $t_0=t_c$ determined so that $t=0$, but because $t_c$ is $\bigO(g_0^2)$ we may consistently neglect its presence here. The other integral is
\begin{align}
\left. I_{3} \right\vert_{\rm SP} &=  \left. \int^\Lambda \! \frac{d^d \k}{(2 \pi)^d} \, \frac{1}{\k^2 (\k+\q_1)^2 (\k - \q_2)^2} \right\vert_{SP},  \nonumber \\
 &\simeq \frac{1}{2}\frac{1}{(4 \pi)^{d/2}}  \log \frac{\Lambda^2}{\kappa^2}.
\end{align}
For the path vertex function,  as we saw above, $\Omsf(\order, t_c) I_3\vert_{\rm SP}$ is independent of the ordering $\pi$. (Unfortunately, this property does not hold to higher orders in perturbation theory for $\Gamma^{\rm PV}$.) The result for a single ordering is
\be
\Omsf(\order, t_c) I_3\vert_{\rm SP}=\frac{1}{3!}I_3\vert_{\rm SP}.
\ee
Then for brevity we define $d_{\rm PV}$ such that
\be
\label{eq_dsv_def}
 \sum_{\order \in S_3} \dmsf(\bigtriangleup| \order)  \Omsf(\order,t_c) I_3\vert_{\rm SP} = d_{\rm PV}  I_{3}\vert_{\rm SP}.
 \ee
Hence $d_{\rm PV}=\frac{1}{6}\sum_\order \dmsf(\triangle|\order)=-2/3$, as we saw above.

We may now solve by requiring that the renormalized proper vertices defined on the right-hand side of \eqref{eq_rg_propverts} satisfy the normalization conditions \eqref{eq_rg_norms} up to terms of $\bigO(u_0^4)$. We obtain to this order
\bea
t_c &=& {\textstyle\frac{1}{2}} d_2u_0^2 \Lambda^2 ,\\
u&=&\kappa^{-\ve/2} g_0 \left(1 +(d_3-{\textstyle\frac{1}{4}}d_2)
u_0^2  \log \frac{\Lambda}{\kappa} \right),\\
Z_{\phi} &=&1- {\textstyle\frac{1}{6}} d_2 u_0^2 \log \frac{\Lambda}{\kappa} ,\\
Z_{\rm PV}  &=& 1 +( {\textstyle\frac{1}{6}}d_2- d_{\rm PV})u_0^2 \log \frac{\Lambda}{\kappa}.
\eea
Finally, to this order $Z_{\phi^2}$ is the same as $Z_{\rm PV}$ except that $d_2$ replaces $d_{\rm PV}$. Note that the $Z$'s are functions only of the dimensionless variables $u_0$ and $\kappa/\Lambda$, and we can set $d=6$ (so $\ve=0$ and $g_0^2=u_0^2/(4\pi)^3$)) in the one-loop terms, but not in zero-loop  terms.

The RG functions are
\bea
\beta(u)& = &-{\textstyle\frac{1}{2}}\ve u + \left({\textstyle \frac{1}{4}}d_2 - d_3 \right) u^3\nonumber\\&&{}\quad + \bigO(u^5, \ve u^3),\\
\gamma_\phi (u) &=& {\textstyle\frac{1}{6} }d_2u^2+\bigO(u^4, \ve u^2),\\
\gamma_{\rm PV} (u)&=&\left({\textstyle\frac{1}{6}} d_2-d_{\rm PV}\right) u^2+\bigO(u^4, \ve u^2).
\eea

The fixed points of the RG are the values of $u=u^*$ at which $\beta(u)=0$. Clearly, one fixed point is at $u=0$, but is unstable to the introduction of the cubic coupling for $d<6$, because $u$ grows as $\kappa$ {\em decreases}, corresponding to the behavior at {\em larger} length scales. In six dimensions, $u$ approaches zero logarithmically as $\kappa$ decreases, because the coefficient of $u^3$ is $d_2/4-d_3=7/4$ which is positive. Below six dimensions, there is another zero of $\beta$ which results from the competition between the two terms in $\beta$, at
\be
(u^\ast)^2 = \frac{2 \ve}{d_2 -4 d_3} + \bigO(\ve^2).
\ee
Note that $(u^\ast)^2$ is positive for $d<6$ ($\ve>0$). For $d>6$, this fixed point is not relevant to percolation or MSTs.

The values of the $\gamma$'s at the fixed point value of $u$ give the ``anomalous dimensions'' of the various operators (except that in the case of $\phi$, the anomalous dimension is $\gamma_\phi/2$). These are the difference of the total dimensions $x$ of the operators from their canonical or engineering dimensions, which are the zero-loop values discussed earlier \cite{feynint_3}. The co-dimension $d-x$ gives the fractal dimension of the associated geometric object (set of points). The most interesting dimension for us is that of the path vertex, which is $x_{\rm PV}=d-2+\gamma_{\rm PV}$. The codimension yields the fractal dimension $D_{\rm p}=2-\gamma_{\rm PV}$ of the path on the MSF$(p_c)$, as discussed in section \ref{s_fracdim}. We find
\be
\gamma_{\rm PV} (u^\ast) \equiv 2-D_{\rm p}  = \frac{(d_2-6 d_{\rm PV} ) \ve}{3(d_2 - 4 d_3)}  + \bigO(\ve^2),
\ee
that is
\be
D_{\rm p}= 2 - \frac{ \ve}{7}+ \bigO(\ve^2).
\ee
This is the main quantitative result of this paper.

The other $\gamma$'s produce fractal dimensions related to properties of percolation, which also apply to MSF$(p)$. First,
\bea
\gamma_{\phi} (u^\ast) \equiv \eta &=& \frac{d_2 \ve}{3(d_2 - 4 d_3)}  + \bigO(\ve^2)\nonumber\\
&=& - \frac{\ve}{21} + \bigO(\ve^2).
\eea
Hence for the fractal dimension $D_{\rm perc}$ of the critical percolation clusters, which is $D_{\rm perc}=d-x_\phi$, we find
\be
D_{\rm perc}= (d+2-\eta)/2=4-\frac{10\ve}{21}+ \bigO(\ve^2).
\ee

The other $\gamma$ is $\gamma_{\phi^2}$, for which the value at the fixed point can be obtained from the formula for $\gamma_{\rm PV}$ by replacing $d_{\rm PV}$ by $d_2$, that is
\be
\gamma_{\phi^2} (u^\ast) \equiv 2-D_{\rm sc}   = \frac{5\ve}{21}  + \bigO(\ve^2),
\ee
and so for the fractal dimension of the set of singly-connected edges (see section \ref{s_fracdim}) we find
\be
D_{\rm sc} =\nu^{-1}= 2- \frac{5 \ve}{21}+ \bigO(\ve^2) .
\ee
The values we have obtained for both exponents $\eta$ and $\nu$ agree with those in the literature on percolation, to order $\ve$ \cite{pl_perc_ft,amit_percFT,bonfim}, which provides a check on our calculation.

The comparison of $D_{\rm p}$ with $D_{\rm sc}$ raises some questions of inequalities obeyed by $D_{\rm p}$. There are also some other fractal dimensions defined for paths on critical percolation clusters which have been studied. These include $D_{\rm min}$, the fractal dimension of the shortest path on the cluster between the given points, and $D_{\rm max}$, the fractal dimension of the longest (self-avoiding) path between them \cite{stah}. Here the length of the path is the number of edges of the lattice that it traverses. Then the inequalities are fairly obvious: first, because all these paths must pass through the singly-connected edges, $D_{\rm sc}$ is the smallest of all, and the remaining inequalities
\be
D_{\rm sc}\leq D_{\rm min}\leq D_{\rm p}\leq D_{\rm max},
\ee
follow from the definitions. To order $\ve$, one has $D_{\rm min}=2-\ve/6$ and $D_{\rm max}=2-\ve/42$ \cite{dmindmax}, and all the inequalities are obeyed {\em strictly} by the results to this order. $D_{\rm p}$ is close but not equal to $D_{\rm min}$.


\section{Conclusion}
\label{s_conclude}

The results of this paper fall into three main parts. First, we constructed an exact expansion for the Kruskal process, or spanning forest MSF$(p)$, in a series in powers of $p$, which terminates for a finite graph, and is analogous to a low-density expansion for percolation, or a high-temperature expansion in a statistical mechanical model. The expansion is for the probability that the path on the MSF from $\bfx$ to $\bfy$ passes through a vertex $\bfz$. Second, this expansion was used to obtain a continuum formulation (with cut-off) for $p\leq p_c$ (where $p_c$ is the percolation threshold) in terms of Feynman diagrams (the region $p>p_c$ presents additional technical problems, and we will not discuss these further here). This expansion was then shown to be renormalizable to all orders in perturbation theory, so that the limit of infinite momentum-space cutoff (or zero lattice spacing) can be taken. Third, the renormalized perturbation expansion was used to calculate the fractal dimension of any path on MSF$(p)$ at $p=p_c$,
to first order in $\ve=6-d$, for $d\leq 6$: $D_{\rm p}= 2- \varepsilon/7+\bigO(\ve^2)$. For $d>6$, $D_{\rm p}=2$. If the ``superhighways'' idea is correct, then the same $D_{\rm p}$ also applies to the region $p>p_c$, in which we expect the path dimension to be independent of $p$ on large enough length scales.

It is important to realize that it was by no means obvious at the outset that such a field-theoretic renormalization process would be possible. The problem is not obviously given by a local field theory, and our expansion is not based on an action principle (at least, not in any apparent way). Optimization is generally a non-local process as it involves making comparisons among (sums of) costs globally; however, this is also true when one wishes to minimize a Hamiltonian, even if its parameters (corresponding to costs) multiply local interaction terms. For minimum spanning trees, the definition of the allowed or ``feasible'' configurations (i.e.\ spanning trees) is not local either. It was not obvious that the expansion would be renormalizable like that of a local field theory. Indeed, in the end our procedure worked thanks to unexpected and non-local cancellations of some subleading divergences (see Appendix \ref{app_canc_proof}), for which we are unaware of any analogs in local field theories. Undoubtedly the underlying reasons for this success with MSTs should be found in the applicability of Kruskal's greedy algorithm and its connection with percolation.

The calculations can be extended in various ways. The exponents can be calculated to higher orders in $\ve$, with increasing effort required for each additional order. The path vertex function, and not only its scaling dimension, can in principle also be studied, as can more general correlation functions with path vertices and mass-insertion vertices. In six dimensions, there are logarithmic corrections to the simple scaling with $D_{\rm p}=2$ that holds for dimensions bigger than six, and these are calculable.

Independently of these applications of the renormalized perturbation expansion, the exact lattice low-density expansion could be studied in low orders (say, the first thirty terms) in any dimension $d$, as is conventionally done with high-temperature series. This would provide another way to obtain scaling dimensions for correlation functions. Such techniques are frequently very accurate.

A further question is the Borel summability of the perturbation expansion, or of the $\ve$ expansion for the exponents. If an asymptotic expansion of a function is Borel summable, then it uniquely determines that function. If a few terms of the expansion are available, and it is believed to be Borel summable, then an improved estimate for the quantity of interest, such as an exponent, for a non-zero value of the parameter (say $\ve=1$) can be made, and for critical exponents these values may be very accurate (comparable with high-temperature series methods). For percolation at threshold, the asymptotic high-order behavior of the perturbation expansion has been shown in Ref.\ \cite{houghton} to have the form that is a necessary condition for the expansion to be Borel summable. These results also apply to our theory, but again we also need a similar result for the path vertex function. It would be interesting to find a technique to estimate the high order behavior of our expansion.

In conclusion, the introduction of the Kruskal process and geometric object MSF$(p)$, based on an optimization problem, provides a rich area for study not unlike conventional critical phenomena. At $p=p_c$, many techniques can be applied to it. It illuminates numerical work on such problems as optimal paths and transport in random media.

\begin{acknowledgments}
This work was supported by NSF grant no.\ DMR--0706195.
\end{acknowledgments}


\appendix

\section{Low-density expansion in percolation}
\label{app_essam_vs_potts}

This Appendix summarizes various results concerning the low-density diagrammatic expansion for percolation. Some of these results have appeared elsewhere in the literature \cite{essam_dombgreen, essam_perc_review}, but we reproduce them here in order to introduce notation and terminology, and because the derivation in Appendix \ref{s_more_dmst} below closely follows that given here.

In section \ref{s_essam_2pt} we define the low-density graphical expansion for two-point connectendess functions, which is generalized to the case of $n$-point functions in section \ref{s_essam_npt}. These sections summarize results given in \cite{essam_dombgreen, essam_perc_review}. Finally, in section \ref{s_essam_potts} we prove that the expansion is the same as that obtained from the conventional description of bond percolation via the low density (high temperature) series for the $Q$-state Potts model, in the limit $Q \to 1$. The principal results are the graphical expansions \eqref{eq_perc_c}, \eqref{eq_perc_nc2} with diagrammatic weights given in simplest form in \eqref{eq_pottsweight_2pt}, \eqref{eq_pottsweight_npt}.

\subsection{Essam's construction}
\label{s_essam_2pt}

Essam's expansion for percolation \cite{essam_dombgreen, essam_perc_review} is based on the principle of inclusion and exclusion from combinatorics \cite{vanlint_wilson}. As this may not be familiar to all readers we summarize it here. We start with a set of events $\{X_i\}$ indexed by $i$ in an index set $I$. In order to calculate probabilities later we introduce the indicator function
\bea
\indic[X_i] &\equiv &\left\{\begin{array}{cl} 1 &\hbox{$X_i$ true,}\\ 0&\hbox{$X_i$ false.}\end{array}\right.
\eea
The principle of inclusion-exclusion is the expansion
\bea
\label{eq_inclexcl_or}
\indic \biggl[ \bigvee_{i \in I} X_i \biggr]& =& \sum_{i\in I} \indic[X_i]-\sum_{i,j\in I:i<j}\indic[X_i\wedge X_j]+\ldots\nonumber\\
&=&\sum_{\emptyset \neq I' \subseteq I} (-1)^{|I'| +1} \indic \biggl[ \bigwedge_{j \in I'} X_j \biggr],
\eea
where $\vee$ denotes a logical OR and $\wedge$ denotes a logical AND. An analogous series may be obtained for the conjunction of all the events, by using De Morgan's law $\neg ( \vee_i X_i ) = \wedge_i ( \neg X_i)$, where $\neg$ denotes logical NOT. This yields
\bea
\label{eq_inclexcl_and}
\indic \biggl[ \bigwedge_{i \in I} X_i \biggr] &=& 1-\indic \biggl[ \neg\biggl( \bigwedge_{i \in I}\neg X_i \biggr)\biggr] \\
&=&\sum_{I' \subseteq I} (-1)^{|I'|} \indic \biggl[ \bigwedge_{j \in I'} \neg X_j \biggr].
\eea
Note that in this case $I'$ may be the empty set, so the first term of this series is 1.

We apply this to bond percolation at a parameter value $p$ by first investigating the two-point connectedness function, defined as
\be
\label{eq_perc_cdef}
C_{\bfx, \bfy} (p)
= \langle \icxy \rangle.
\ee
where $\icxy$ stands for $\icxy=\indic[ \bfx, \bfy \text{ connected by edges of cost}\leq  p ]$ and the angle brackets denote an average with respect to all realizations of the edge costs. Defining $\gamxy$ to be the set of all self-avoiding walks  on the lattice between $\bfx$ and $\bfy$, we may write
\be
\label{eq_perc_inclexcl1}
\icxy = \indic \biggl[ \bigvee\limits_{ \gamma \in \gamxy} (\gamma \leq p) \biggr].
\ee
where we define the event
\be
(\gamma \leq p)  \equiv \left(  \max_{e \in \gamma} \p_e \leq p \right) ;
\ee
i.e, we require all edges $e$ on the path $\gamma$ to be present by the time the parameter is raised to the value $p$. Using equation \eqref{eq_inclexcl_or} to expand the right-hand side of \eqref{eq_perc_inclexcl1} by inclusion-exclusion yields
\begin{multline}
\label{eq_perc_inclexcl2}
\icxy =  \sum\limits_{\emptyset \neq \Gamma' \subseteq \gamxy} (-1)^{|\Gamma'|+1} \indic \biggl[ \bigwedge\limits_{\gamma' \in \Gamma'} (\gamma' \leq p) \biggr].
\end{multline}

We obtain an expansion in terms of graphs from \eqref{eq_perc_inclexcl2} by grouping together all terms that test the same set of edges on the lattice; the terms in the series are now indexed by graphs $G$, each obtained as the union of some set of paths $\Gamma'\subseteq \gamxy$ (possibly from more than one such $\Gamma'$). We say that such a set $\Gamma'$ covers (the edges of) $G$. Because the paths are self-avoiding walks, all the graphs $G$ generated from such unions must be vertex-irreducible: removing any vertex from the graph must leave at least one of the points $\bfx$, $\bfy$ in each connected component. Let the set of all such graphs with the marked vertices $\bfx$, $\bfy$ be $\gxy$.

Letting $\gamxy(G)$ for $G \in \gxy$ denote the set of paths on $G$ connecting the root points $\bfx, \bfy$, equation \eqref{eq_perc_inclexcl2} can be rewritten
\begin{multline}
\label{eq_perc_c}
\icxy = \\
\sum_{G \in \gxy} \sum_{\Gamma' \subseteq \gamxy(G)} (-1)^{|\Gamma'|+1} \indic [ \Gamma' \text{ covers } G] \indic [ G \leq p].
\end{multline}
The average over the costs can be performed immediately. Referring back to the definition \eqref{eq_perc_cdef}, we obtain the graphical expansion
\be
\label{eq_perc_c2}
C_{\bfx, \bfy} (p) = \sum_{G \in \gxy} d(G) \Pr[ G \leq p],
\ee
by introducing
\begin{multline}
\label{eq_dweight_def}
d(G \in \gxy) \equiv
\sum_{\Gamma' \subseteq \gamxy(G)} (-1)^{|\Gamma'|+1} \indic[\Gamma' \text{ covers } G],
\end{multline}
which is independent of the parameter $p$, and is referred to as the $d$-weight of the graph $G$.
This expression for the $d$-weight can be seen to possess the topological invariance property mentioned in section \ref{s_essam_inclexcl}: the insertion of any number of vertices of degree two (or, equivalently, replacing edges of $G$ with paths of edges) does not change the set $\gamxy(G)$ of paths connecting the root points, or the value of $ \indic[\Gamma' \text{ covers } G]$ for any of the subsets $\Gamma'$.

The definition of $d(G)$ may be extended to cover the case where $G$ is any two-rooted graph as follows: if $G$ consists of more than one connected component, there is no way to cover all its edges with paths connecting the roots, so $d(G) = 0$. Note that, because the covering criterion is defined in terms of the edge set only, addition of isolated vertices does not change a graph's $d$-weight. Similarly, if $G$ is not vertex-irreducible, by definition some edges --- the ``tadpoles'' or ``dangling ends'' --- cannot be covered by a self-avoiding path, since backtracking is forbidden, so again $d(G) = 0$. Since $d(G)$ vanishes for these additional cases, the sum in \eqref{eq_perc_c2} may be extended to all two-rooted subgraphs of the underlying lattice.

We may make further progress if we remark that the preceding derivation also applies to connectedness functions on an arbitrary graph $\Lambda$ instead of the whole lattice; the sum in \eqref{eq_dweight_def} is then over appropriate subgraphs of $\Lambda$. We denote this connectedness function by $C_{\bfx, \bfy} (\Lambda, p)$. Equation \eqref{eq_perc_c2} generalizes to
\be
\label{eq_essam_gconn}
C_{\bfx, \bfy} (\Lambda, p) = \sum_{E' \subseteq E(\Lambda)} d(G_{E'}) \Pr [ G_{E'} < p ],
\ee
where $G_{E'}$ is the subgraph of $\Lambda$ consisting of all vertices of $\Lambda$ and a subset $E' \subseteq E=E(\Lambda)$ of its edges. Evaluating \eqref{eq_essam_gconn} at $p = 1$ yields
\be
\indic [E(\Lambda) \text{ connects } \bfx, \bfy]=C_{\bfx, \bfy} (\Lambda, 1) = \sum_{E' \subseteq E(\Lambda)} d(G_{E'}).
\ee
In the definition of $d(G)$ for $G=G_{E'}$, defined by a subset $E'$ of the edges of $\Lambda$, we may note that vertices of $\Lambda$ incident on no edges can be deleted without changing $d(G)$.
Now because $\Lambda$ is an arbitrary graph, and the sum in \eqref{eq_essam_gconn} is over over all subsets of $E(\Lambda)$, we may easily invert this sum by M\"{o}bius inversion \cite{vanlint_wilson}, which for the present case is related to inclusion-exclusion. We obtain
\be
\label{eq_pottsweight_2pt}
d(G) = \sum_{E' \subseteq E(G)} (-1)^{|E(G)|-|E'|} \indic [E' \text{ connects } \bfx, \bfy].
\ee
This form is equivalent to \eqref{eq_dweight_def}, but easier to work with as it does not require a sum over the set of paths on $G$.

We point out that in this argument the sum over paths was only used to arrive at the form \eqref{eq_perc_c2} for arbitrary $\Lambda$. Once this is known, the expressions \eqref{eq_pottsweight_2pt} for the coefficients were obtained by M\"obius inversion with no further reference to paths. This suggests that a shorter derivation may exist.

\subsection{Extension to $n$-point connectedness functions}
\label{s_essam_npt}

The expansion \eqref{eq_perc_c} generalizes readily to $n$-point connectivity functions; the criterion is simply that $n$ root points $ \bfx_1, \cdots, \bfx_n $ are connected if and only if there exists at least one path from $\bfx_1$ to each $\bfx_i$, $2 \geq i \geq n$, where we select $\bfx_1$ arbitrarily. Note that in enumerating the set of paths from $x_1$ to $x_i$, we must include those paths that pass through other root points. Using inclusion-exclusion \eqref{eq_inclexcl_or} and equation \eqref{eq_perc_inclexcl1} again, we may write the indicator function for this event as
\begin{multline}
\indic[\bfx_1, \cdots, \bfx_n \text{ connected at } p] =
\prod_{i=2}^n \, \indic_c ( \bfx_1, \bfx_i) \vert_{p} \\
= \prod_{i=2}^n \, \sum_{\emptyset \subset \Gamma'_i \subset \Gamma_{\bfx_1, \bfx_i}} (-1)^{|\Gamma'_i|+1} \indic \biggl[ \bigwedge\limits_{\gamma \in \Gamma'_i} (\gamma \leq p) \biggr].
\end{multline}
Repeating the previous derivation and grouping together terms that test the same set of edges, we obtain the diagrammatic expansion
\be
\label{eq_perc_nc2}
C_{\bfx_1, \cdots, \bfx_n } (p) = \sum_{G \in G_{\bfx_1, \cdots, \bfx_n }} d(G) \Pr[ G \leq p],
\ee
where the $n$-point $d$-weight is
\begin{multline}
\label{eq_ndweight_def}
d( G \in  G_{\bfx_1, \cdots, \bfx_n }) \equiv \\
 \prod_{i = 2}^n \, \sum_{\emptyset \subset \Gamma'_i \subset \Gamma_{\bfx_1, \bfx_i}(G)} (-1)^{|\Gamma'_i|+1} \indic [ \cup_i \Gamma'_i \text{ covers } G].
\end{multline}
Again, the fact that $d(G)$ may be computed in terms of sets of paths covering $G$ establishes that it is a topological invariant, unchanged by adding vertices of degree two to $G$.

The argument following \eqref{eq_dweight_def} also carries though, since the above definition of the $d$-weight may be extended to arbitrary graphs and we may perform M\"{o}bius inversion on the connectedness function evaluated on an arbitrary $n$-point graph, obtaining
\begin{multline}
\label{eq_pottsweight_npt}
d(G \in G_{\bfx_1, \cdots \bfx_n}) = \sum_{E' \subseteq E(G)} (-1)^{|E(G)|-|E'|} \\ \times \indic \left[ E' \text{ connects }  \bfx_1, \cdots \bfx_n \right].
\end{multline}
Our final results, equation \eqref{eq_perc_nc2} with \eqref{eq_pottsweight_npt}, constitute a complete low-density expansion for all connectedness properties of percolation clusters.

\subsection{Equivalence with the Potts model}
\label{s_essam_potts}

The development of the field theory for the Potts model is described in detail elsewhere \cite{zia_potts, amit_percFT,pl_perc_ft,luben_review} and we will recall only the parts of the derivation that are relevant to our discussion here. The $Q$-state Potts model on a graph $\Lambda$ \cite{wu_potts} has, associated with each vertex $\bfx$ of $\Lambda$, a degree of freedom $\alpha(\bfx)$ which may take on any of $Q$ discrete states (``colors''). The Hamiltonian for this model in the absence is
\be
\label{eq_hpotts}
H = -J\sum_{\langle \bfx, \bfx' \rangle} (\delta_{\alpha(\bfx), \alpha(\bfx')}-1),
\ee
where the sum is over edges indexed by the two incident vertices $\bfx$, $\bfx' $.
The partition function can be expanded in the form \cite{fk}
\begin{align}
Z&=\sum_{\{\alpha(\bfx):\bfx\in V(\Lambda)\}}e^{-\beta H}\\
&=\sum_{\{\alpha\}}\prod_{\langle\bfx,\bfx'\rangle}\left[(1-e^{-\beta J})\delta_{\alpha(\bfx), \alpha(\bfx')}+e^{-\beta J}\right]\\
&=\sum_{E'\subseteq E} p^{|E'|}(1-p)^{|E|-|E'|}Q^{N_c(G_{E'})},
\end{align}
where $p=1-e^{-\beta J}$ and again $E=E(\Lambda)$. When $Q\to 1$, the partition function becomes $Z=1$, and the expansion corresponds to the sum of probabilities for the sets $E'$ of occupied edges in bond percolation with independent probabilities $p$ for occupying each edge. The $Q$-state Potts model partition function, viewed as a function of $Q$ and $p$, is also (essentially) the Tutte polynomial \cite{tutte}.

The states at each vertex can be represented by an overcomplete set of $Q$ vectors $\vec{e}^\alpha$, $\alpha = 1$, \ldots, $Q$, in a $Q-1$ dimensional space. These vectors are obtained by projecting the position vectors of a regular $Q$-simplex in $Q$-dimensional space onto the subspace orthogonal to the vector $(1,1,\ldots,1)$. More concretely, if we let the coordinates of these vectors with respect to some basis be $e^\alpha_i$, $i = 1, \ldots, Q-1$, the set of vectors may be uniquely defined up to relabeling and change of basis by requiring that
\begin{align}
\label{eq_pottsvec1}
\sum_{\alpha = 1}^Q e^\alpha_i &= 0, \\
\label{eq_pottsvec2}
\sum_{\alpha = 1}^Q e^\alpha_i e^\alpha_j &=  Q \delta_{ij}, \\
\label{eq_pottsvec3}
\text{and }\sum_{i=1}^{Q-1} e^\alpha_i e^\beta_i &= Q \delta_{\alpha \beta} -1.
\end{align}
In equations \eqref{eq_pottsvec1} -- \eqref{eq_pottsvec3}, we have normalized the vectors following the convention used in \cite{amit_percFT, luben_review, bonfim}. Note that \cite{zia_potts} and \cite{pl_perc_ft} adopt a different normalization.

To obtain the two-point connectedness function, we introduce factors $e_{i_1}^{\alpha(\bfx_1)}$, $e_{i_2}^{\alpha(\bfx_2)}$ into the sum. If they are not in the same connected component in the expansion, the sum over all $\alpha$'s gives zero by \eqref{eq_pottsvec1}. That is,
\begin{align}
C_{i_1,i_2}&(\bfx_1,\bfx_2)\equiv\nonumber\\
&\sum_{\{\alpha\}}e_{i_1}^{\alpha(\bfx_1)}
e_{i_2}^{\alpha(\bfx_2)}\prod_{\langle\bfx,\bfx'\rangle}\left[
p\delta_{\alpha(\bfx), \alpha(\bfx')}+(1-p)\right]\\
=&\;\delta_{i_1,i_2}\sum_{E'\subseteq E}\indic \left[E' \text{ connects }  \bfx_1, \bfx_2 \right] \nonumber\\
&\qquad{}\times p^{|E'|}(1-p)^{|E|-|E'|}Q^{N_c(G_{E'})},
\end{align}
where we also used \eqref{eq_pottsvec2}. After removing the factor
$\delta_{i_1,i_2}$ and setting $Q=1$, this is equal to   $C_{\bfx, \bfy} (\Lambda, p)$.

Now we rewrite
\be
p\delta_{\alpha(\bfx), \alpha(\bfx')}+(1-p)=p(\delta_{\alpha(\bfx), \alpha(\bfx')}-1)+1.\ee
(Although this does not explicitly involve the $e_i^\alpha$'s, this choice is motivated by the form of eq.\ \eqref{eq_pottsvec3} as $Q\to 1$; note that there are many similar expressions that become equal to this for $Q=1$.) We expand the Potts correlation function $C_{i_1,i_2}(\bfx_1,\bfx_2)$ using this decomposition for each edge, and then once more for $\delta_{\alpha(\bfx), \alpha(\bfx')}-1$ on each edge. This yields
\begin{align}
C_{i_1,i_2}(\bfx_1,\bfx_2)=&\delta_{i_1,i_2}\sum_{E'\subseteq E(\Lambda)}p^{|E'|}d_Q(G_{E'}),
\end{align}
where
\begin{align}
d_Q(G_{E'})=&\sum_{E''\subseteq E'}\indic\left[E''  \text{ connects }  \bfx_1, \bfx_2 \right]\nonumber\\
&\times (-1)^{|E'|-|E''|}Q^{N_c(E'')}.
\end{align}
Removing $\delta_{i_1,i_2}$ and setting $Q=1$, we recover the expressions eq.\ \eqref{eq_essam_gconn} and \eqref{eq_pottsweight_2pt}. The derivation can be readily generalized, at least to the $3$-point connectedness function. Hence Essam's diagrammatic expansion is identical term-by-term with the low-density expansion of the Potts model in the $Q \to 1$ limit.


\section{Properties of MST paths}
\label{app_pathcompare}
In this Appendix we present proofs of properties obeyed by paths on the MST. These properties are used in Appendix \ref{s_more_dmst} below to construct a diagrammatic expansion for the MSF path vertex.

\subsection{MST paths as geodesics}
\label{app_mstpath_geo}

We first define a {\em minimax path} between two given vertices on the finite graph $\Lambda$: a (self-avoiding) path is a minimax path for the pair of (distinct) vertices $\bfx$, $\bfy$ if among all paths from $\bfx$ to $\bfy$ it has the lowest value of the most costly edge (among all edges on the path). That is, it is a minimum (over the set of paths from $\bfx$ to $\bfy$) of the maximum (over edges on the path) of the cost of the edge. We note immediately that in general there is more than one minimax path for the given vertices, even though they must all share the same most costly edge (we assume that no two edges have equal cost). We say that a path $\gamma$ is a {\em geodesic} if, for all vertices $\bfw$, $\bfz$ lying on $\gamma$, the subset of $\gamma$ which connects $\bfw$, $\bfz$ is a minimax path from $\bfw$ to $\bfz$. A geodesic passing through $\bfx$ and $\bfy$ is necessarily a minimax path for $\bfx$ and $\bfy$. Applying the definition of geodesic for the case where $\bfw$, $\bfz$ are adjacent vertices connected by a single edge, we see that we may equivalently define geodesics as those paths all of whose edges are minimax paths connecting the vertices to which they are incident. We note that a geodesic cannot be a cycle, so it must have endpoints. We may also remark that the geodesic path is the correct strong disorder limit of the optimal path \cite{optpath}, that which minimizes the total cost of all edges on the path with fixed endpoints.

It is not always well-appreciated in the literature that a minimax path is not unique, see for example \cite{dd,bu_diso_strength,wu_superhwy_1,wu_superhwy_2}, which frequently refer in the singular to ``the'' minimax path between two points. These sources really mean the geodesic path, which we will now prove is unique.

We now prove that there is a unique geodesic between any two given vertices, say $\bfx$, $\bfy$ (provided they are on the same connected component of the underlying graph), provided that all edge costs are distinct. Specifically, we let $\gamma$ be a geodesic and we will show that no other path $\olgam \neq \gamma$ which shares the same endpoints may also be a geodesic. We noted above that any minimax path from $\bfx$ to $\bfy$, such as $\olgam$, must pass through the same most costly edge $e_1$. However, at this stage it is not clear that they all do so in the same direction. But if we consider the endpoint of $e_1$ that is encountered first on walking along $\gamma$ from $\bfx$ to $\bfy$, say $\bfw$, then any minimax path, such as $\olgam$, from $\bfx$ to that endpoint must pass through the same edge $e_2$, which is the most costly on the subpath (but clearly less costly than $e_1$). (If $\bfw=\bfx$, then we can start from $\bfy$ instead, and if $e_1$ has endpoints $\bfx$ and $\bfy$ then we are done.) Note that this shows that $\olgam$ traverses the edge $e_1$ in the same direction as $\gamma$, because otherwise the most costly edge on the minimax from $\bfx$ to $\bfw$ would be $e_1$. Using induction on the number of steps on $\gamma$, we find that $\olgam$ must be the same as $\gamma$.


By elementary properties of MSTs, all paths on the MST are geodesics.
Likewise, all geodesics are contained in the MST, because in particular each of their edges is minimax for its two incident vertices, which is a property of the MST. Indeed, the MST of a graph is the union of all of its minimax edges.

It is amusing to realize that the MST has the ultrametric property (the content of this paragraph will not be used elsewhere in the paper). Let us assume that the costs are non-negative (if not, we can add a positive constant to all of them). Then we can obtain a notion of distance, or metric, $\partial$ between any two vertices on the graph $\Lambda$, by defining $\partial(\bfx,\bfy)$ to be the largest (or minimax) cost on a minimax path from $\bfx$ to $\bfy$, with $\partial(\bfx,\bfx)=0$ if $\bfx=\bfy$. By definition, a metric should  be finite and non-negative, symmetric ($\partial(\bfx,\bfy)=\partial(\bfy,\bfx)$), equal to zero if and only if $\bfx=\bfy$, and obey the triangle inequality. The first three properties are clear, while it is easy to see that $\partial$ obeys the stronger property that, for any $\bfx$, $\bfy$, $\bfz$,
\be%
\partial(\bfx,\bfy)\leq \max\left(\partial(\bfx,\bfz),\partial(\bfz,\bfy)\right).
\label{ultrametric}
\ee
These four properties imply that $\partial$ is an ultrametric. Note that the ultrametric inequality eq.\ \eqref{ultrametric} implies the triangle inequality. For ordinary metric spaces, one defines geodesics to be paths of shortest ``length'' using the metric, and this motivates our terminology above. Further, if $\bfx$, $\bfy$, $\bfz$ are three distinct points, the ultrametric property implies that if $\partial(\bfx,\bfy)\leq \partial(\bfx,\bfz)$ and $\partial(\bfy,\bfz)$, then $\partial(\bfx,\bfz)=\partial(\bfy,\bfz)$.  It is well known that an ultrametric space with a finite number of points can be viewed as a tree, which we imagine depicted with the points as the {\em leaves} located on a hyperplane, other vertices to one side of the hyperplane, connected by straight lines, and the ultrametric represented by the height above (in the direction orthogonal to the hyperplane) the leaves to which one must go in walking from one leaf to another along the tree. In the present case, this essentially corresponds to the MST. The tree is trivalent (except at the leaves) with probability one. The trivalent vertices represent the edges on the MST, with their height as their cost. In fact, if we consider the subforest of the tree consisting of the vertices at height less than or equal to some bound, then this represents the MSF$(p)$.

\subsection{Identifying MST paths through binary comparisons}
\label{app_mstpath_compare}

Let $\gmst(\bfx, \bfy)$ be the geodesic from $\bfx$ to $\bfy$, or equivalently the path on the MST.
The geodesic property of $\gmst (\bfx, \bfy)$ allows it to be selected from the set $\gamxy$ of all paths connecting $\bfx, \bfy$ by means of repeated comparisons using a binary ordering relation $\prec$, defined as follows. Let $\gamma$ and $\gamma'$ be two paths in $\gamxy$. Let $e_1, e_2, \cdots e_n$ be the first, second, ... $n$-th most expensive edges on $\gamma$, and likewise for $e'_1, e'_2, \cdots e'_n$ on $\gamma'$. We say $\gamma \prec \gamma'$ if and only if there exists some $j$ such that $\p_{e_j} < \p_{e'_j}$ and $\p_{e_i} = \p_{e'_i}$ for all $i < j$: in other words, we compare the most expensive edges whose costs are not identical. We will prove that
\be
\gmst (\bfx, \bfy) = \min_{\gamma \in \gamxy} \gamma;
\ee
in other words, $\gmst (\bfx, \bfy)$ is the minimal element of the set $\gamxy$ under the ordering defined by $\prec$.


Again, we assume all edge costs to be distinct, which implies that $e_i = e'_i \iff \p_{e_i} = \p_{e'_i}$ and $\neg (\gamma \preceq \gamma') \iff \gamma \succ \gamma'$: i.e., the relation $\prec$ defines a \emph{total} order on the set of all paths between fixed endpoints. Under this assumption, for any two paths $\gamma$, $\gamma' \in \gamxy$ we have
\be
\label{eq_pathcomparedef}
\gamma \prec \gamma' \iff \max_{e \in \gamma - \gamma \cap \gamma'} \p_e < \max_{e' \in \gamma' - \gamma \cap \gamma'} \p_{e'}.
\ee
Let $\gamma$ be a path satisfying
\be
\label{eq_gam_is_min}
\gamma = \min_{\gamma' \in \gamxy} \gamma'.
\ee
In particular, $\gamma$ is less than all paths in $\Gamma(\gamma; \bfx, \bfy) \subset \gamxy$, the set of paths in $\gamxy$ having no edges in common with $\gamma$. By \eqref{eq_pathcomparedef}, this means $\prec$ compares only the most expensive edges on the paths, so if $\gamma \prec \gamma' \in \Gamma(\gamma; \bfx, \bfy)$ then $\gamma$ must be a minimax path for $\bfx$ and $\bfy$.

Similarly, for any subpath $\olgam \subseteq \gamma$ with endpoints $\bfw$, $\bfz$, we define $\Gamma(\gamma; \bfw, \bfz) \subset \gamxy$ as the set of those paths that coincide with $\gamma$ from $\bfx$ to $\bfw$ and from $\bfy$ to $\bfz$, and have no edges in common with $\olgam$; in other words, the set of those $\gamma'$ such that $\gamma - \gamma \cap \gamma' = \olgam$. Because $\gamma \prec \gamma' \in \Gamma(\gamma; \bfw, \bfz)$, $\olgam$ is a minimax path from $\bfw$ to $\bfz$. This holds for all choices of $\bfw$ and $\bfz$, and hence $\gamma$ is a geodesic. It is unique and can be identified as a path $\gamma=\gamma_{\rm MST}(\bfx,\bfy)$ on the MST by the results above.

We finally note that identifying MST paths through the definition \eqref{eq_gam_is_min} is most convenient for the purposes of our diagrammatic expansion \eqref{eq_mst_ctop}. It would be very  inefficient computationally, since we make many unnecessary comparisons with paths that are not in any of the sets $\Gamma(\gamma; \bfw, \bfz)$. On the other hand, the geodesic characterization of MST paths is less directly useful for our purposes, since it requires keeping track of the locations of the most expensive edges. The geodesic properties of MST paths are very useful computationally: they are essential in constructions of linear-time algorithms for MST path verification \cite{MSTverify} which were used in \cite{MSTrandom} to give a randomized algorithm which constructs the entire MST in linear time.


\section{Derivation of low-density expansion for MSF paths}
\label{s_more_dmst}

In this Appendix we derive the exact low-density expansion for the probability $\cxyz(p)$ that the points $\bfx, \bfy$ are connected by a path on the MSF$(p)$ which passes through $\bfz$, on a finite graph $\Lambda$.

First, we can formally define an indicator function
\begin{multline}
\icxyz=\indic [(\bfx, \bfy \text{ connected at } p)
\wedge \\
(\gmst (\bfx, \bfy) \text{ passes through } \bfz)]\label{eq_mst_cdef}
\end{multline}
and then
\be
\cxyz(p)
= \langle \icxyz \rangle.
\ee
In the Kruskal process, edges are never removed from MSF$(p)$ as $p$ is increased, so if a path connecting two points on the MSF at a parameter value $p$ exists, it must be identical to the unique path connecting those points on the completed MST. In Appendix \ref{app_mstpath_compare} above we use the geodesic properties of the MST path to arrive at the definition
\be
\gmst (\bfx, \bfy) = \min_{\gamma \in \gamxy} \gamma,
\ee
where $\min$ denotes the minimal element under the relation $\prec$ defined in \eqref{eq_pathcomparedef}. This lets us write the indicator function in \eqref{eq_mst_cdef} as a sum over all paths $\gamma \in \gamxy$, in the form
\begin{multline}
\label{eq_mst_inclexcl2}
\icxyz = \\
\sum_{\gamma \in \gamxy} \indic [\gamma \leq p] \icz [ \gamma ] \indic \biggl[ \bigwedge_{\gamma' \in \gamxy} ( \gamma \preceq \gamma') \biggr] ,
\end{multline}
where
\be
\icz[\gamma] \equiv \indic [ \gamma \text{ passes through } \bfz],
\ee
where again $\gamxy$ is the set of all paths on $\Lambda$ with endpoints $\bfx$ and $\bfy$. For the time being, we will suppress the dependence of all expressions on the underlying graph $\Lambda$.

Expanding the indicator function $\indic \left[ \bigwedge_{\gamma' \in \gamxy} ( \gamma \preceq \gamma') \right]$ by inclusion-exclusion gives
\begin{multline}
\label{eq_mst_inclexcl3}
\icxyz = \sum_{\gamma \in \gamxy}  \indic [\gamma \leq p] \icz[ \gamma] \\
\times \sum_{\Gamma' \subseteq \gamxy}  (-1)^{|\Gamma'|}  \indic \biggl[ \bigwedge_{\gamma' \in \Gamma'} \neg ( \gamma \preceq \gamma') \biggr] .
\end{multline}
Because the uniqueness of the edge costs implies $\neg ( \gamma \preceq \gamma') \iff (\gamma \succ \gamma')$, we may restrict the sum over subsets of $\gamxy$ to those not containing $\gamma$ itself. We now reorganize the double sum by grouping together all terms that test the same set of edges, as was done for equation \eqref{eq_perc_c} for percolation. For each term in \eqref{eq_mst_inclexcl3}, the edges in $\gamma \cup \Gamma'$ form a graph $G$ in the set $ \gxy$ of all vertex-irreducible graphs with root vertices $\bfx, \bfy$. When we regroup the sum in terms of these graphs, we obtain a sum over sets $\Gamma'$ of paths from $\bfx$ to $\bfy$ which cover $G$ (these sets are the previous $\gamma\cup \Gamma'$ redefined as $\Gamma'$, so contain the chosen path $\gamma$), similar to what was obtained to percolation. Unlike the percolation case, we still have the outermost sum in \eqref{eq_mst_inclexcl3}, which becomes the innermost sum over elements $\gamma$ of $\Gamma'$. Thus the expansion becomes
\begin{multline}
\label{eq_mst_inclexcl4}
\icxyz = \\
\sum_{G \in \gxy} \indic[G \leq p] \sum_{\Gamma' \subseteq \gamxy(G)} (-1)^{|\Gamma'|+1} \indic[ \Gamma' \text{ covers } G] \\
\times \sum_{\gamma \in \Gamma'} \indic \biggl[ \bigwedge_{\gamma' \in \Gamma' - \gamma} ( \gamma \succ \gamma') \biggr] \icz [ \gamma].
\end{multline}
As in the derivation of \eqref{eq_perc_c2}, we may factor out the dependence on the parameter $p$ as
\be
\label{eq_mst_inclexcl5}
\icxyz = \\
\sum_{G \in \gxy} \dmsf(G) \indic[G \leq p],
\ee
where we have introduced $\dmsf(G)$, the analogue of Essam's $d$-weight \eqref{eq_dweight_def} for MSF paths:
\begin{multline}
\label{eq_mst_ddef1}
 \dmsf(G)\equiv \sum_{\Gamma' \subseteq \gamxy(G)}(-1)^{|\Gamma'|+1}  \indic[ \Gamma' \text{ covers } G] \\
\times \sum_{\gamma \in \Gamma'} \indic \biggl[ \bigwedge_{\gamma' \in \Gamma' - \gamma} ( \gamma \succ \gamma') \biggr] \icz[ \gamma].
\end{multline}
$\dmsf(G)$ depends implicitly on $\bfx$, $\bfy$, and $\bfz$, and on the costs of the edges of $G$.

In the analogous statement \eqref{eq_dweight_def} for percolation, we found $d(G)$ was independent of edge costs. Here to evaluate $\dmsf(G)$ we need  to compare paths which cover $G$ using the relation $\succ$. From the definition \eqref{eq_pathcomparedef}, a necessary and sufficient set of information to do this is the relative ordering of the edge costs of $G$. As discussed in Section \ref{s_spec_vert}, we will introduce an ordering by indexing the set $E$ of edges of $\Lambda$ arbitrarily and defining an ordering of their costs to be given by a permutation $\order \in S_{|E|}$ on the set of $|E|$ elements, via
\be
\p_{\order(i)} < \p_{\order(j)} \iff i< j.
\ee
Then the induced ordering on a subset $E'$ of $E$ is written $\order_{E'}$.
With this notation, we see that $\dmsf$ is a function of the graph $G$ and edge cost ordering $\order_{E(G)}$, so we write $\dmsf(G | \order_{E(G)})$ (it still depends implicitly on $\bfx$, $\bfy$, and $\bfz$).

The second sum in \eqref{eq_mst_ddef1} detects whether the maximal path in $\Gamma'$ passes through the point $\bfz$, so for a fixed edge cost ordering $\order$ we may write
\begin{multline}
\label{eq_mst_ddef2}
 \dmsf(G| \order_{E(G)})=\\
  \sum_{\Gamma' \subseteq \gamxy(G)}(-1)^{|\Gamma'|+1}  \indic[ \Gamma' \text{ covers } G] \icz [ \max_{\gamma \in \Gamma'} \gamma].
\end{multline}
Note that, as a consequence of our use of inclusion-exclusion, this result is mildly counterintuitive: we are attempting to calculate the probability that the MSF path passes through $\bfz$, and by definition the MSF path (if it exists at $p$) is the minimum out of all paths in $\gamxy$. However, for each graph in the expansion of this probability, the relevant event is that the \emph{maximal} path of the covering $\Gamma'$ passes through $\bfz$.

We may now take the expectation value of $\icxyz$ over all realizations of the edge costs in order to obtain the analogue of \eqref{eq_perc_c2},
\begin{multline}
\label{eq_mst_clatt}
\cxyz(p) = \sum_{G \in \gxy}  \sum_{\order_{E(G)} \in S_{|E(G)|}}\dmsf(G | \order_{E(G)}) \\
\times\Pr[\order_{E(G)} \wedge(G \leq p)].
\end{multline}
We note that the ordering $\pi_{E(G)}$ and the event that all edges of $G$ be less than $p$ are independent, so the last probability factorizes. Because the edge costs are iid, all orderings of the edge costs are equally probable and $\Pr[\order] = 1/|E(G)|!$.

We may find an alternative expression for $\dmsf$ in terms of a sum over edge subsets instead of sets of covering paths, analogous to our derivation of \eqref{eq_pottsweight_2pt} from \eqref{eq_dweight_def}. The argument proceeds the same way: we first make explicit the dependence of MSF path connectedness functions on the graph $\Lambda$ in the set $\gxyz$ of graphs containing the three root points $\bfx$, $\bfy$, and $\bfz$, writing it $\cxyz(\Lambda, p)$. Because \eqref{eq_mst_ddef2} contains a factor of $ \indic[ \Gamma' \text{ covers } G] $ in the summand, we also have $\dmsf(G|\order) =0$ for disconnected or vertex-reducible graphs. The sum in \eqref{eq_mst_clatt} may therefore be extended to {\em all} subgraphs of $G$ as
\begin{multline}
\label{eq_dmst_def0}
\cxyz(\Lambda, p) =  \sum_{\order \in S_{|E(\Lambda)|}} \sum_{E' \subseteq E(\Lambda)}\\
\times  \dmsf(G_{E'} | \order_{E'}) \Pr[G_{E'} \leq p] \Pr[ \order_{E'}].
\end{multline}
Because $\dmsf(G|\order_{E(G)})$ is dependent on $\order$, we must work under the sum over edge cost orderings in performing the M\"{o}bius inversion step. We therefore work with the conditional quantity
\be
\cxyz(G, p | \order_{E(G)}) \equiv \sum_{E' \subseteq E(G)} \dmsf(G_{E'} | \order_{E'}) \Pr[G_{E'} \leq p]
\ee
appearing as a summand in \eqref{eq_dmst_def0}. Evaluating this at $p = 1$ yields
\begin{multline}
\indic[E(G) \text{ connects } \bfx, \bfy] \indic_c^{(\bfz)}[\gmst(G | \order_{E(G)})] = \\
\sum_{E' \subseteq E} \dmsf(G_{E'}  |\order_{E'}),
\end{multline}
where $\gmst(G | \order)$ is the path connecting the root points $\bfx, \bfy$ on the minimum spanning tree of $G$ obtained under the edge cost ordering $\order$. M\"{o}bius inversion of this sum gives
\begin{multline}
\label{eq_dmst_def_copy}
\dmsf(G  |\order) = \sum_{E' \subseteq E(G)} (-1)^{|E(G)| - |E'|} \\
\times \indic[E' \text{ connects } \bfx, \bfy]  \indic_c^{(\bfz)}[\gmst(G_{E'} | \order_{E'})].
\end{multline}
This definition of $\dmsf$ is more convenient than \eqref{eq_mst_ddef2} for the proofs of appendix \ref{app_rgproof_dweights}. It is also, in principle, more convenient for computation, since for large graphs the size of the set $\gamxy(G)$ of self-avoiding walks grows faster than $|E(G)|$, hence the sum in \eqref{eq_dmst_def_copy} is more easily performed than that in \eqref{eq_mst_ddef2}.


\section{Renormalizability of the MSF perturbation expansion}
\label{app_RGibility}

In this section we give the proofs outlined in Section \ref{s_rg_justify}, which establish that our perturbation expansion for MSF paths is renormalizable. Recall that diagrams of this theory with no path vertex are identical to those of percolation theory and hence pose no problem, while we construct diagrams involving the MSF path vertex by the substitution \eqref{eq_msf_from_perc}:
\begin{multline}
\label{eq_msf_from_perc_copy2}
d(\cG) I (\cG) \mapsto \\
\sum_{\order \in S_{|E(\cG)|}} \dmsf(\cG| \order) \Omsf (\order,t_0) I(\cG,\{t_\epsilon\}),
\end{multline}
where the integrals $I(\cG)$ and $I(\cG,\{t_\epsilon\})$ are identical Feynman integrals with only cubic interaction vertices, containing the factor $g_0$ for each such interaction, but in the latter integral the mass-squared $t_0$ is generalized to a distinct parameter $t_\epsilon$ for each edge $\epsilon$ of the graph $\cG$. In this Appendix, we will drop the prime from the orderings $\order'$ throughout; orderings $\order_{E(\cG)}$ are nonetheless the induced orderings on the set of highest costs $L_\epsilon$ of the set $E(\cG)$ of edges $\epsilon \in E(\cG)$ of a topological graph (Feynman diagram) or subgraph $\cG$. $\dmsf$ and $\Omsf$ were defined in \eqref{eq_dmst_def}, \eqref{eq_mst_op} respectively, and as defined both depend on the structure of the entire graph $\cG$. In particular, it is not entirely evident from the definition \eqref{eq_dmst_def} how $\dmsf(\cG |\order)$ could be computed from knowledge of its values on subgraphs of $\cG$.

The Appendix is structured to give proofs of the following results.
We begin in section \ref{s_app_defs} by introducing terminology common to all sections of this Appendix and explaining the parametric formulation of Feynman integrals. In section \ref{omsfeffect} we obtain the effect of the $\Omsf$ operator, which is very simple in the parametric formulation: it introduces a simple product factor $F_\order$ into the integrand, which depends on the choice of an ordering $\order$ for the costs on the graph.

In section \ref{app_rgproof_bounds}, we prove that the superficially-divergent subintegrations (as the cutoff $\Lambda\to\infty$) associated to a connected subgraph $\cH$ come only from a subset $S' \subset S_{|E(\cG)|}$ of all possible orderings on the edges of $\cG$.
Specifically, if $\cH$ is a three-point subgraph or a 2-point subgraph containing the path vertex, then $S' = \order_{[0]}$, in which all costs in the subgraph are cheaper than all those outside. Similarly, if $\cH$ is a two-point subgraph (i.e.\ a self-energy), the only superficial divergences are for orderings $S' = \order_{[0]} \cup \order_{[1]}$, in which at most one edge in $\cH$ has cost higher than one or more outside $\cH$. Moreover, for these orderings, with one class of exceptions the divergences in self-energy or cubic coupling (3-point) subgraphs are the same as those in the corresponding percolation diagrams, up to the $d$-weights. These results generalize easily to diagrams with several superficially-divergent subdiagrams, if these are pairwise either disjoint or one inside another.

Having identified the important orderings, we consider in section \ref{app_rgproof_dweights} the behavior of the $\dmsf$ weights for these orderings. We show that the weights obey nice factorization properties for connected subdiagrams with two or three external points for orderings in class $\order_{[0]}$, and also (in a different, more general form) for self-energy subdiagrams $\cH$ with orderings in which one or more edges in $\cH$ is more costly than at least one outside $\cH$. The factorization has the form
\be
\label{eq_dmst_factors2}
\dmsf(\cG|\order_{E(G)}) = d(\cH) \dmsf( \cG/\cH| \order_{E(\cG/\cH)})
\ee
if the path vertex is not in $\cH$ (note the appearance of a $d$-weight from percolation), and
\be
\label{eq_dmst_factors1}
\dmsf(\cG|\order_{E(\cG})) = \dmsf(\cH|\order_{E(\cH)}) \dmsf( \cG/\cH| \order_{E(\cG/\cH)})
\ee
if the path vertex is in $\cH$. The precise definitions, in particular for $\order_{E(\cG/\cH)}$, will be given in section \ref{app_rgproof_dweights}. Here and below we use notation $\cG/\cH$ to denote the diagram obtained by contracting the subgraph $\cH$ to a single vertex (which may be of degree 2, producing a harmless extension of the class of diagrams to be considered).

In section \ref{app_renorm_proof} we come to the heart of the proof. We use a theorem of Berg\`{e}re and Lam \cite{param_div_3} to show that the Feynman integral for each diagram in our perturbation expansion can be rendered absolutely convergent by a procedure of subtracting all the superficially divergent parts of the integrand. Furthermore, utilizing the results of preceding sections, all the terms that have to be subtracted for divergent subdiagrams (including those containing the path vertex) are the same as those for a corresponding full diagram, with the exception of one class of terms as mentioned above, which is dealt with in section \ref{app_canc_proof}. Subject to the latter result, this means that all divergences are dealt with by renormalizing parameters and the overall scale of the vertex functions, as in a renormalizable field theory.

Finally, in section \ref{app_canc_proof} we prove that the class of exceptional subleading divergences in the self-energy subdiagrams cancel in the sum over all diagrams of a given order. This completes the proof of renormalizability to all orders in the perturbation expansion.

\subsection{Definitions}
\label{s_app_defs}

We begin by considering an arbitrary Feynman integral associated with
a diagram $\cG$ appearing in the perturbative expansion of a correlation function in, for example, the field theory of percolation.

\begin{figure}[h]
\includegraphics[width=3.1in]{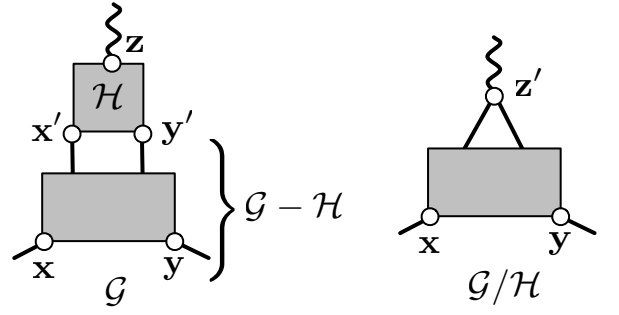}%
\caption{Depiction of an arbitrary graph contributing to the MSF path connectedness function. The various root vertices used in the definition of this function are labeled with open circles. \label{f_graphparts4a}}
\end{figure}

First, we recall the expression for a Feynman integral associated with $\cG$ given in eq.\ \eqref{eq_cont_perc}.
Let $V, E$ be the vertex and edge sets of $\cG$, and let ${\cal N}_{\epsilon,v}$ be its incidence matrix under an arbitrary orientation of its internal edges; i.e.
\begin{align}
\begin{split}
{\cal N}_{\epsilon,v} &= 1 \text{ if $v$ is the head of $\epsilon$,} \\
{} &= -1 \text{ if $v$ is the tail of $\epsilon$, and } \\
{} & = 0 \text{ otherwise.}
\end{split}
\end{align}
Neglecting the cut off for a moment, the integral with which we are concerned is
\begin{multline}
\label{eq_percint_def1}
I(\cG) =\int \! \left(\prod_{\epsilon \in E} \frac{d^d \q_\epsilon}
{(2 \pi)^d}\right)\cdot \left(\prod_{\epsilon \in E}\frac{1}{\q^2_\epsilon + t_\epsilon} \right) \\
\times \prod_{v \in V} (2 \pi)^d \delta^d \left( (\k_\text{ext})_v - \sum_{\epsilon \in E} {\cal N}_{\epsilon,v} \q_e \right).
\end{multline}
Here $(\k_\text{ext})_v$ is the net external momentum incident on the vertex $v$. Since we will replace the percolation $d$-weight with the appropriate $\dmsf$-weight, we neglect the factor $d(\cG)$ and also $g_0$ to the power of the number of internal cubic vertices (or other couplings for interactions of different degree that may be present more generally).

We make progress by expressing the Feynman integral \eqref{eq_percint_def1} in terms of integrals over the Schwinger parameters $\alpha$. This makes use of the identity $1/X = \int_0^\infty \! d\alpha \; e^{-\alpha X}$ to rewrite part of the integrand as
\be
\label{eq_alpha_ident}
\prod_{\epsilon \in E} \frac{1}{\q_\epsilon^2+t_\epsilon} = \prod_{\epsilon \in E} \int_0^\infty \! d\alpha_\epsilon \; e^{-\alpha_\epsilon
(\q_\epsilon^2+t_\epsilon)}.
\ee
(For brevity, we let $A_\cG$ denote the set of parameters $\alpha_\epsilon$ introduced above.) The total of $d|V|$ $\delta$-functions can be rewritten using the identity $2\pi\, \delta(k)=\int d\lambda\, e^{i\lambda k}$ for each. Integrals over the internal momenta $\q_\epsilon$ are now Gaussian and can be performed, and then the $\lambda$-integrations become Gaussian and can be performed, except for one which produces a $\delta$-function expressing conservation of the total momentum, $(2\pi)^d\delta^d(\sum_{v\in V}(\k_{\text{ext}})_v)$. Omitting this $\delta$-function, we have
\be
\label{eq_percint_def2}
\begin{split}
I(\cG) =
\frac{1}{(4 \pi)^{{\cal L}d/2}}\int_0^\infty \! \prod_{\epsilon \in E}  d \alpha_\epsilon \cdot P^{-d/2}(A_\cG) \\
\times \exp \left( -\k^T_\text{ext} {\Delta}^{-1} \k_\text{ext}- \sum_{\epsilon \in E} \alpha_\epsilon t_\epsilon \right).
\end{split}
\ee
Here $\Delta$ is a $|V|\times |V|$ matrix, which is a Laplacian on $\cG$, defined by
\be
\Delta(A_\cG)_{v_1, v_2} = \sum_{\epsilon \in E} {\cal N}_{\epsilon, v_1} \frac{1}{\alpha_\epsilon} {\cal N}_{\epsilon, v_2},
\ee
and $P(A_\cG)$ is defined as
\be
P(A_\cG) =\left( \prod_{\epsilon \in E} \alpha_\epsilon \right) {\det}' \Delta (A_\cG),
\ee
in which the determinant ${\det}'$ is that of $\Delta$ with one row and column removed, so as to remove the zero mode.  $\k_\text{ext}$ is viewed as a $|V|$-component vector, and ${\cal L}={\cal L}(\cG)$ is the cyclomatic number of $\cG$, the number of {\em independent} loops (cycles) of $\cG$. $P(A_\cG)$ is a homogeneous polynomial of degree $\cal L$. These expressions are quite general and may be obtained for the diagrams of any field theory; for a further discussion consult \cite{param_1,param_2}. It is interesting that $P(A_\cG)$ and also $\Delta^{-1}$ can be related to weighted sums over spanning trees on $\cG$ \cite{param_1,param_2} by the Kirchoff matrix-tree theorem \cite{vanlint_wilson}; it is not clear to us whether this fact is deeply involved in the renormalizability of the theory of MSF$(p)$.

For many diagrams, the integral $I(\cG)$ as written in \eqref{eq_percint_def1} or \eqref{eq_percint_def2} is ultraviolet divergent and must be regularized, which is done by restricting the momentum integrations in \eqref{eq_percint_def1} to the region $|\q_\epsilon| < \Lambda$. We implement this in \eqref{eq_percint_def2} by taking the range of integration of each of the $\alpha_\epsilon$ to be $[\Lambda^{-2},\infty)$, which exponentially suppresses contributions from $|\q_\epsilon| \gg \Lambda$.

We make use of the parametric representation for Feynman integrals for two reasons. First, it greatly simplifies the study of renormalization of the expansion, as in the field theories in \cite{param_div_1,param_div_2,param_div_3,param_2}.


The second reason we employ the parametric representation is that, as we show in the next Subsection, the action of $\Omsf$ takes a particularly simple form. Application of $\Omsf$ directly to \eqref{eq_percint_def1} results in intractable integrals over the $\{t_\epsilon\}$ for diagrams beyond one-loop order, while we are able to obtain its action on an arbitrary graph in closed form in equation \eqref{eq_alpha_multi}.

\subsection{Effect of $\Omsf$ operator}
\label{omsfeffect}

The preceding discussion applied to the diagrams from the field theory for percolation. To investigate how things change when we calculate MSF diagrams, we specify a given total ordering $\order$ of the masses of $\cG$, such that
\be
i < j  \iff t_{\order(i)} > t_{\order(j)}.
\ee
The diagrammatic contribution to the MST theory is obtained by summing over all total orderings of edge costs consistent with the placement of the path vertex, according to \eqref{eq_dmst_def_copy}. To find the contribution from one ordering $\order$, we apply the operator $\Omsf(\order,t)$ defined in \eqref{eq_mst_op} to both sides of  \eqref{eq_alpha_ident}, obtaining
\be
\label{eq_alpha_multi}
\begin{split}
&\idotsint\limits_{\substack{\infty > t_{\order(1)} > \cdots \\ \cdots > t_{\order(|E|)} > t}} \prod_{\epsilon \in E} dt_\epsilon \,   \frac{d}{dt_\epsilon} \frac{1}{\q_\epsilon^2+t_\epsilon} \\
&{}= \int_{\Lambda^{-2}}^\infty \prod_{i=1}^{|E|} \frac{\alpha_{\order(i)} \, d\alpha_{\order(i)} }{\sum_{j=1}^i \alpha_{\order(j)} }
e^{-\sum_{\epsilon \in E} \alpha_\epsilon
(\q_\epsilon^2+t)}.
\end{split}
\ee
The integrand on the right-hand side is that appearing on the right-hand side of \eqref{eq_alpha_ident}, multiplied by a factor
\be
\label{eq_falpha_def}
F_\order (A_\cG) \equiv \prod_{i=1}^{|E|} \frac{\alpha_{\order(i)} }{\sum_{j=1}^i \alpha_{\order(j)} } ,
\ee
(We note that $F_\order(A_\cG)$ is of the same form as we obtained on the lattice in equation \eqref{eq_multi_l_int0'}. This is another manifestation of the well-known equivalence between scalar field theory and a system of random walkers.) Thus finally our prescription for evaluating the contribution of each diagram is that it is given by the parametric Feynman integral as for percolation, but with the factor
\be
\sum_\order\dmsf(\cG|\order)F_\order(A_\cG)
\ee
inserted inside the $\alpha$ integrals, replacing the $d(\cG)$ weight for the percolation theory.

The factor $F_\order(A_\cG)$ obeys $0\leq F_\order(A_\cG)\leq 1$ for any $A_\cG \in [0,\infty)^{|E|}$, and has the property that it reduces to one as we go towards the limit in which
\be
\alpha_{\order(1)}\ll\alpha_{\order(2)}\ll\cdots\ll\alpha_{\order(|E|)}.
\ee
It tends to suppress orderings which do not obey the version of these inequalities in which all $\ll$'s are replaced by $<$'s. Thus it acts to replace the strict inequalities on the $t_\epsilon$'s by corresponding but softer conditions on the $\alpha_\epsilon$'s.
This result makes intuitive sense: high-momentum (small $\alpha$) propagators correspond to lattice walks consisting of relatively few edges. In the Kruskal process, we expect the shortest paths to be completed first, at the lowest value of $p$, corresponding to a larger mass-squared $t_0$.

It will be useful to simplify the $F_\order$ factors as much as possible, by performing (or partially performing) the sums of $\dmsf(\cG|\order)F_\order$ over orderings $\order$  as much as possible before performing the integrals. We now give some basic formulas that are a step in this direction.
First, we obtain another proof of \eqref{eq_sum_all_orderings} from the fact that
\be
\sum_{\order \in S_{|E|}} F_\order (A_\cG) = 1.
\label{eq_Fsumid}
\ee
A more general fact that will be useful is that if we consider a subset of edges $E'\subseteq E$ of $E$ and orderings $\pi$ such that the masses on edges of $E'$ are greater than all those in $E-E'$, and sum over all such orderings that fix an ordering on $E-E'$ (such orderings can be written as $\order=\sigma\circ\order_0$ for $\order_0$ any one such ordering and $\sigma$ a permutation in $S_{E'}\subset S_{E}$), then:
\be
\sum_{\sigma\in S_{E'}}F_{\sigma\circ\order_0}(A_\cG)=\prod_{i=|E'|+1}^{|E|} \frac{\alpha_{\order_0(i)} }{\sum_{j=1}^i \alpha_{\order_0(j)}},\label{eq_Fsumid'}
\ee
in which the right-hand side is independent of the choice of $\order_0$. This follows by using eq.\ \eqref{eq_Fsumid} applied to the restricted sum over orderings. Indeed, as the derivation of this identity only used the sum over a smaller set, this can be used in a proof by induction (on the size $|E|$) of eq.\ \eqref{eq_Fsumid} itself. The induction step, of taking $|E'|=|E|-1$ and summing the right-hand side eq.\ \eqref{eq_Fsumid'} over cosets $S_{|E|}/S_{|E|-1}$ is simple.

\subsection{Estimating MSF Feynman integrals}
\label{app_rgproof_bounds}

In this Subsection, we describe how the divergent behavior of a given diagram of the MSF path theory differs from that of the diagram from percolation theory from which it was obtained, and obtain some basic statements about the form of the divergences for each ordering.

In the absence of the $F_\order$ factor, the parametric form of the Feynman integrals may in general suffer from divergences associated with the region $\alpha\to0$ for some or all $\alpha$'s. These take the place of the possibly more familiar divergences at large $\k$ in the original momentum space integrals over $\q_\epsilon$; recall that the latter integrals have already been done, after exchanging orders of integration. For a 1PI graph $\cG$, the superficial degree of divergence of $I(\cG)$ is obtained easily from the momentum-space form by counting the total number of powers of all $\q_\epsilon$'s and integrations $\int d^d\q_\epsilon$, and is given by
\be
\omega(\cG) = d{\cal L}(\cG) - 2 |E(\cG)|.
\ee
This formula holds for any field theory of scalar fields interacting via non-derivative couplings. The same result is easily obtained in the parametric representation also \cite{param_div_1,param_div_2,param_div_3,param_2}. It may be obtained more formally by rescaling the $\alpha_\epsilon\to \rho^2\alpha_\epsilon$ for all edges of $\cG$, with $\rho\to 0^+$. The formula may also be applied to the subintegral associated with a connected 1PI subdiagram $\cH$ of $\cG$ (strictly, a subdiagram is a subset of the vertices of $\cG$, together with all edges that connect these vertices); this will be denoted $\omega(\cH)$. In this case, it is obtained from the behavior as the subset of $\alpha_\epsilon$ associated with edges of $\cH$ are scaled to zero by a common factor. Notice that the superficial degree of divergence for a subgraph $\cH$ might be larger than that for $\cG$. A graph or subgraph is said to be superficially divergent if its superficial degree of divergence is positive or zero, and superficially convergent if its superficial degree of divergence is negative. (A graph with $\omega=0$ may diverge more slowly than any power of $\Lambda$, for example logarithmically, or may be convergent.) It is a theorem that if $\omega$ is negative for $\cG$ and for all its subgraphs, then the associated Feynman (parametric) integral is absolutely convergent.

For the theory with cubic interactions that we consider here, the only connected 1PI graphs that are superficially divergent at $d=6$ dimensions are (a) any self-energy diagram (with two external points), because all have $\omega=2$, (b) any vertex correction diagram, that is a graph with three external points, because all have $\omega=0$, and c) a self-energy graph with a $\phi^2$ insertion, which have the same form as the vertex diagrams in b). Here in a) and b) an external point means that a line that ``leaves'' the graph (joined to it by a cubic vertex like the others) was removed to leave the 1PI part. The graphs containing $\phi^2$ are relevant to the path vertex that we wish to consider in this paper. Other graphs are superficially convergent.

Turning to our theory for MSF$(p)$, the parametric form of the Feynman integral for a given ordering $\order$ is simply modified by the insertion of the factor $F_\order(A_\cG)$.
Because $F_\order(A_\cG)$ is bounded, it follows that the superficial divergence of any diagram or subdiagram of the MSF theory is no worse than the corresponding diagram of percolation theory from which it was obtained. More formally, $F_\order$ is a homogeneous rational function of degree zero, and so the superficial degree of divergence for $\cG$ is again $\omega(\cG)$.

However, for a subgraph $\cH$ of $\cG$,  $F_\order (A_\cG)$  may reduce the superficial degree of divergence below $\omega(\cH)$. Recall that for a subdiagram, we consider the limit as $\alpha_\epsilon$ for $\epsilon\in E(\cH)$ go to zero simultaneously, by scaling them with a common factor $\rho^2$, leaving $\alpha_\epsilon$ for $\epsilon\in \cG-\cH$ unchanged.
Considering each of the $|E(\cH)|$ factors in $F_\order(A_\cG)$ that have numerator $\alpha_\epsilon$ for an edge $\epsilon\in E(\cH)$ appearing in \eqref{eq_falpha_def} in this limit, we see that in this limit $\rho\to0$,
\be
F_{\order} (A_\cG)=\bigO\left(\rho^{2n_\order (\cH,\cG)}\right),
\ee
where we define $n_\order (\cH,\cG)$
to be the number of masses $t_\epsilon$ for $\epsilon \in E(\cH)$ that are less than at least one of the masses in $E(\cG)$ under the ordering $\order$. (Clearly $n_\order(\cH,\cG)=0$ for $\cH=\cG$.) For a fixed 1PI connected subgraph $\cH$ of $\cG$, this provides a useful partitioning of orderings into sets $\order_{[m]}$, $m=0$, $1$, \ldots:
\be
\order_{[m]}=\{\order: n_\order (\cH,\cG) = m\}.
\ee
Thus the orderings $\order_{[0]}$ (which will prove most important in what follows), for which $F_\order=\bigO(1)$ as $\rho\to0$, are those where all of the masses on the edges of the subgraph $\cH$ are larger than those in $\cG-\cH$, that is all the costs in $\cH$ are lower.

We may add this result to the superficial degree of divergence to obtain the overall superficial degree of divergence of a connected 1PI subdiagram $\cH$ of a connected 1PI diagram $\cG$ under the ordering $\order$:
\be
\label{eq_mst_degofdiv}
\omega_{\rm MSF}(\cH,\cG|\order) = \omega(\cH)- 2n_\order(\cH,\cG).
\ee
This implies that it is only for class $\order_{[0]}$ that the superficial degree of divergence of the subgraph $\cH$ is unchanged by $F_\order$. For subgraphs with $\omega(\cH)=0$ (i.e.\ the vertex and path vertex diagrams), orderings other than those in $\order_{[0]}$ give convergent subintegrals. For the self-energy subgraphs, with $\omega=2$, orderings in $\order_{[1]}$ lower the superficial degree of divergence to $0$, and these are additional divergences with which we will have to deal. Moreover, in all cases there are subleading terms in the behavior of $F_\order$ as $\rho\to0$ for a subgraph, and while these terms are superficially convergent in most cases, the first subleading term also has zero superficial degree of divergence in the case of the self-energy subdiagrams.

For further analysis, it is helpful to consider the sum $\sum_\order \dmsf(\cG|\order)F_\order(A_\cG)$ and to attempt to simplify it as much as possible, so that the evaluation of the parametric integrals reduces to those for percolation as much as possible. Indeed, by the ``contribution of a diagram'' in general we mean the weighted sum over orderings. In order to consider divergent subintegrals for subdiagrams, it is useful to have factorization properties of the weights $\dmsf$. It is to this that we turn next.

\subsection{Factorization properties of MSF diagrammatic weights}
\label{app_rgproof_dweights}

In this section we demonstrate that the diagrammatic weights $\dmsf$ possess enough factorization properties for our proof of the renormalizability of the perturbation expansion to go through.
Let us first recall that for the $d$-weights in percolation, the weight for a diagram $\cG$ containing a 2- or 3-point subdiagram $\cH$ factors into the weight for $\cH$ times that for the ``quotient graph'' $\cG/\cH$ in which the subgraph $\cH$ is shrunk to a single vertex (formally, its vertices are identified, and its edges are deleted):
$d(\cG)=d(\cH)d(\cG/\cH)$. This is immediate in the Potts model formulation in which the $d$-weights originate from contracting together tensors, due to $S_Q$ permutation symmetry (apart from the problem of giving a formal definition of the $Q\to 1$ limit). It can also be derived from the combinatorial definitions described in Section \ref{app_essam_vs_potts} (this is shown in the case of some 2-point subdiagrams in Ref.\ \cite{essam_dombgreen}). It is important for the proof of renormalizability, as the contributions of such subgraphs in Feynman integrals will be treated as ``correcting'' or ``renormalizing'' the parameters attached to 2- and 3-point vertices in the graphical expansion. We require some similar properties in the expansion for MSF$(p_c)$.

Recall that the weights can be defined as in eq.\ \eqref{eq_dmst_def_copy} [for $E=E(\cG)$],
\begin{multline}
\label{eq_dmst_def_copy2}
\dmsf(\cG  |\order) = \sum_{E' \subseteq E} (-1)^{|E| - |E'|} \\
\times \indic[E' \text{ connects } \bfx, \bfy]  \indic_c^{(\bfz)}[\gmst(\cG_{E'} | \order_{E'})].
\end{multline}
This differs from the diagrammatic weight for percolation \eqref{eq_pottsweight_2pt} only in the presence of the additional indicator function $\indic_c^{(\bfz)}[\gmst(\cG_{E'} | \order_{E'})]$.
A graph that is not 1PI can be decomposed into (connected) 1PI subdiagrams lying on a chain of single edges and such 1PI parts that form a path from $\bfx$ to $\bfy$, and possibly other 1PI parts. That is, $\cG$ may be constructed as a tree $\cG_0$ (with $\bfx$, $\bfy$, $\bfz$ marked) which is then decorated by replacing its vertices $v$ with subgraphs $\cH_v$. As the MSF path must pass through a chain of  1PI parts, it follows that for those 1PI subdiagrams that do not contain the vertex at $\bfz$, the indicator $\indic_c^{(\bfz)}$ is independent of the path through such a 1PI subdiagram, and accordingly the $\dmsf$ weight factors into a product of weights for the single edges and for the 1PI parts. Moreover the $\dmsf$ factor for each such 1PI subdiagram  reduces to $d$ in percolation for that subdiagram (for a single edge, the weight is 1). Likewise, for a vertex-reducible subdiagram or ``tadpole'', such as a 1PI part connected to the rest by a single edge, the $\dmsf$-weight is the same as in percolation and vanishes. Similarly the $\dmsf$ weight for a diagram that contains a disconnected subdiagram vanishes. Hence from here on we need consider only connected, vertex-irreducible 1PI diagrams $\cG$ that contain the path vertex at $\bfz$, as well as root points that we can relabel as $\bfx$, $\bfy$.

For MSF$(p)$, the weights $\dmsf$ depend on the ordering $\order$ of the costs of the edges of the topological graph $\cG$, as well as on $\cG$. In this section, we will denote these costs by the original symbol $\ell_\epsilon$ for edge $\epsilon\in \cG$ (these costs in fact stand for the maximum, earlier denoted $L_\epsilon$, of the chain of edges that are the image of $\epsilon$ under an embedding of $\cG$ in the lattice). In terms of the costs, the ordering $\order$ is defined by
\be
i < j  \iff \ell_{\order(i)}< \ell_{\order(j)}.
\ee
(We use the costs, rather than the mass-squared's for which the inequalities are reversed, because the authors find that this aids their intuition about MSTs.)
In seeking a factorization similar to that for the $d$-weights in percolation, there are two issues. Because the $\dmsf$-weights depend on a choice of ordering, one issue is whether some factorization holds at all for each ordering, and a second is, if there is some factorization, what ordering would be used for the quotient $\cG/\cH$.
What we obtain below may not be the most general possible result. Instead we obtain statements for two (overlapping) sets of conditions, and these are sufficient for our purposes.

Motivated by the considerations of which orderings produce ultraviolet-divergent Feynman integrals associated with a subgraph, we first show that for orderings $\order$ such that all edges in a connected subgraph $\cH$ have lower cost than all others in $\cG$, where $\cH$ is a 2- or 3-point subgraph, and in the 2-point case the vertex $\bfz$ can also be present, factorization holds:
\be
\dmsf(\cG|\order)=\dmsf(\cH|\order_\cH)\dmsf(\cG/\cH|\order_{\cG-\cH}).
\label{dmsf_factor}
\ee
Here the right-hand side involves the ordering $\order_\cH$, which is $\order$ restricted to $\cH$, and $\order_{\cG-\cH}$ which is $\order$ restricted to $\cG-\cH$. (For graphs $\cG$, subgraphs $\cH$, and quotients $\cG/\cH$, we will allow abuses of notation like $\order_{E(\cH)}=\order_\cH$.) Further, in the case in which $\cH$ does not contain $\bfz$, we already know that $\dmsf(\cH|\order_\cH)=d(\cH)$. We recall that these orderings are those in class $\order_{[0]}$, which produce the {\em leading} divergence for the 2- or 3-point subgraphs. Further, the factorization generalizes to the case in which there are several disjoint such subgraphs, and the costs in the union of the sets of edges of the subgraphs are lower than those in the remainder of $\cG$ (regardless of the relative orderings among the edges in the subgraphs). In this case, each disjoint subgraph carries a weight as for the single subgraph considered above. Then, because the 2-point (or self-energy) subgraph (that does not contain $\bfz$) also has subleading divergences that occur when its costs do not obey the preceding conditions, we also derive a more general result for such a subgraph for any ordering. These results can be combined to handle a large class of orderings and subgraph structures.


First we show that, if $\cH$ is a 2- or 3-point subgraph, then in the sum over subsets of edges $E'\subseteq E$ in $\dmsf$ we can replace
\begin{multline}
\label{eq_connected_factors}
\indic[ E' \text{ connects }\bfx, \bfy] = \indic[E'(\cH) \text{ connects } \{ \bfx_i \}] \\
\times \indic[ E'(\cG/\cH) \text{ connects } \bfx, \bfy],
\end{multline}
where $\bfx, \bfy$ are the root points of $\cG$, and $ \{ \bfx_i \}$ are the root points of $\cH$, because other terms cancel. To see this, first notice that if for a 2- or 3-point subgraph $\cH$, the ``diluted'' edge set of $\cH$, $E'(\cH)=E'\cap E(\cH)$ [and similarly for $E'(\cG/\cH)$], does not connect all the root vertices, then there is at least one root vertex not connected to any of them (this does not hold for a subgraph with more than three root points). Choose one of these, and without loss of generality suppose it is $\bfx_1$. In $\cG$ there is a single edge incident on $\bfx_1$ that is not in $\cH$ (call it $\epsilon_{\bfx_1}$). The minimum spanning tree path $\gmst(\cG_{E'} | \order_{E'})$ from $\bfx$ to $\bfy$ on $E'$ clearly cannot pass though $\bfx_1$ for such an $E'$, whatever the ordering $\pi$. We can pair off such subsets $E'$ by choosing pairs of $E'$ which are the same subsets except that the edge $\epsilon_{\bfx_1}$ is in one and not in the other. These subsets differ in size by one, and the indicator function $\indic_c^{(\bfz)}[\gmst(\cG_{E'} | \order_{E'})]$ takes the same value for both. Hence these contributions cancel, and the result follows.

\begin{figure}[h]
\includegraphics[width=3.2in]{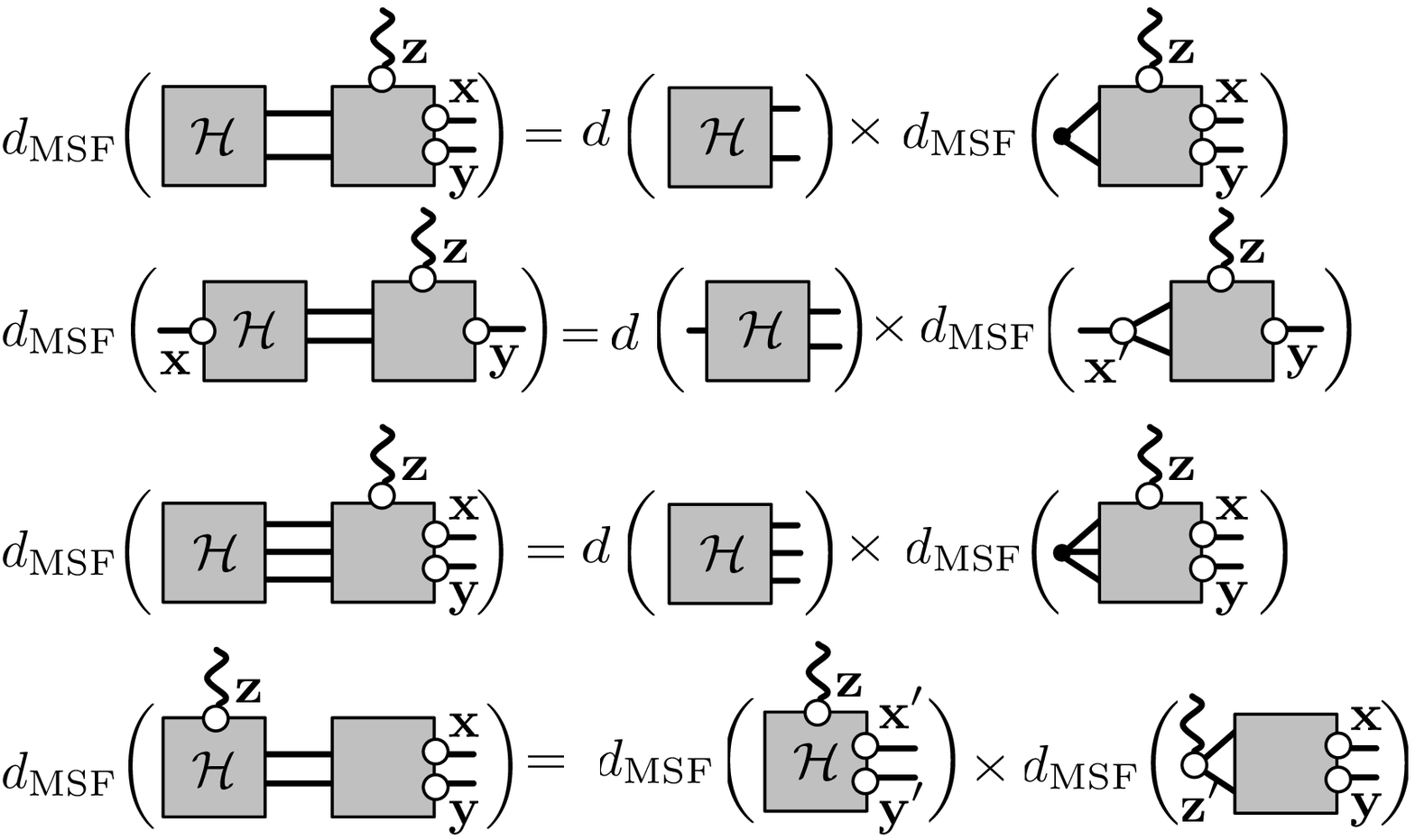}%
\caption{Factorization properties of $\dmsf(\cG|\order)$ for connected subgraphs that are proved in the text for orderings in the class $\order_{[0]}$. \label{f_dmst_factor} }
\end{figure}

Now we turn to the factoring of $ \indic_c^{(\bfz)}[\gmst(\cG_{E'} | \order_{E'})] $; it is here that the form of the ordering enters.
The case in which all edges in the subgraph $\cH$ have costs lower than all those in $\cG-\cH$ is quite simple. First, the same property is inherited in the ordering $\order_{E'}$ restricted to $E'$. As the Kruskal process runs on $E'$, these edges are tested first, and when that is completed the root points of the subgraph are connected (this follows because we have shown that $E'$ connects these vertices). For the remainder of the process, from which the path $\gmst(\cG_{E'} | \order_{E'})$ is obtained, the subgraph $\cH$ [or its diluted version which we denote $\cH_{E'(\cH)}$] can be viewed as collapsed to a single vertex to produce $\cG_{E'}/\cH_{E'(\cH)}$.  It is useful now to distinguish two cases: either $\bfz$ is in $\cH$, or it is not. In the first case, the MSF path must enter $\cH$ to reach $\bfz$, and then leave. This implies that a) on $\cG_{E'}/\cH_{E'(\cH)}$, the image of $\cH$ is the point through which the MSF path must pass, and b) once within $\cH$ the path must pass through $\bfz$. That is, we can write for the indicator function
\begin{multline}
\label{eq_msfindic_factor1}
\indic_c^{(\bfz)}[\gmst(\cG_{E'} | \order_{E'})]=
\indic_c^{(\bfz)}[\gmst(\cH_{E'(\cH)} |\order_{E'(\cH)})] \\ \times\indic_c^{(\cH)}[\gmst(\cG_{E'}/\cH_{E'(\cH)} | \order_{E'-E'(\cH)})]
\end{multline}
The summation over subsets of the edges $E'$ can be written as a sum over subsets $E'(\cH)$ and over $E''=E'-E'(\cH)$, and so the factorization of the $\dmsf$-weights as in eq.\ \eqref{dmsf_factor} follows.  Likewise, in the case where the path vertex is located in $\cG-\cH$, we can simply write
\begin{multline}
\label{eq_msfindic_factor2}
\indic_c^{(\bfz)}[\gmst(\cG_{E'} | \order_{E'})] \\
=\indic_c^{(\bfz)}[\gmst(\cG_{E'}/\cH_{E'(\cH)}  | \order_{E'-E'(H)})],
\end{multline}
and again the form in eq.\ \eqref{dmsf_factor} follows, though now $\dmsf(\cH|\order_\cH)=d(\cH)$.
Together these prove all the relations shown in Fig.\ \ref{f_dmst_factor} for the the stated class of orderings.

For graphs with several disjoint subgraphs of the same type as the single connected subgraphs considered above, the same proof goes through if all edges in all the subgraphs are less costly than those outside. That is, the edges in one of the connected subgraphs need not be all more, nor all less, costly than those in another one of the subgraphs. Thus this result is more general than simply iterating the application of the preceding result, though the final factored form of the $\dmsf$-weight is the same as if it were.

We now turn to a more general argument for the case of a connected 1PI 2-point subgraph $\cH$ and {\em any} ordering $\order$. It holds if the path vertex $\bfz$ is within $\cH$, however for orderings not in the class $\order_{[0]}$ which are already covered by the preceding proof, the corresponding Feynman integrals are convergent, so we will not make of this, and can assume that $\bfz$ is not in $\cH$.

Let $\epsilon_\cH$ be the most costly edge in $\cH$ for the ordering $\order$. Now we consider the evaluation of the $\dmsf$-weight. For each diluted edge-set $E'$, this involves comparison of paths from $\bfx$ to $\bfy$ (which exist because the contribution to $\dmsf$ vanishes if $\bfx$ and $\bfy$ are not connected). We saw above that we can assume that the root points $\bfx_1$, $\bfx_2$ of $\cH$ are connected by $E'(\cH)$. Further, there are paths from $\bfx$ to $\bfy$ through $\cH_{E'(\cH)}$, because otherwise $\cH_{E'(\cH)}$ is either disconnected from both $\bfx$ and $\bfy$, or is part of a tadpole, and in either case the weight vanishes as we saw above. To find the MST path from $\bfx$ to $\bfy$, the task can be broken into subtasks, and one of these is first to find the MST path through $\cH_{E'(\cH)}$ between its root points. If the MST path from $\bfx$ to $\bfy$ passes through $\cH_{E'(\cH)}$, the portion within $\cH_{E'(\cH)}$ must be this MST path. We now show that (within the sum defining $\dmsf$) this path $\gmst(\cH_{E'(\cH)} |\order_{E'(\cH)})$ between $\bfx_1$ and $\bfx_2$ must pass through the most costly edge $\epsilon_\cH$ of $\cH$. For suppose that $\epsilon_\cH\in E'(\cH)$, but the MST path does not traverse it. Then there is another edge set which is the same as $E'$ except that $\epsilon_\cH$ is omitted, and these terms cancel in pairs (note that the MST paths are the same for these edge sets). But the terms with $\epsilon_\cH\in E'(\cH)$ and $\epsilon_\cH$ on $\gmst(\cH_{E'(\cH)} |\order_{E'(\cH)})$ do not cancel in a similar way, as removing $\epsilon_\cH$ from this edge set leaves the root vertices $\bfx_1$ and $\bfx_2$ disconnected, and we know that those edge sets cancel among themselves. The reason the root vertices become disconnected on removing $\epsilon_\cH$ (so $\cH_{E'(\cH)}$ is not 1PI) is that if not, then a less-costly path (in the sense of the ordering $\prec$ in section \ref{app_mstpath_compare}) between the roots would exist.

\begin{figure}[h]
\includegraphics[width=3.1in]{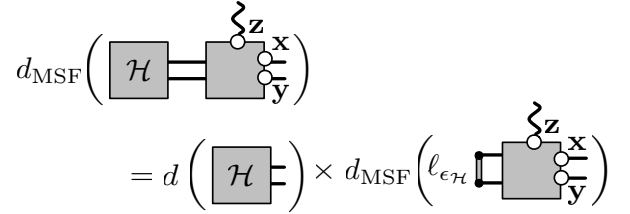}%
\caption{The factorization property of $\dmsf(\cG|\order)$ proved in the text for arbitrary ordering of edge costs. \label{f_dmst_factor_2pt}}
\end{figure}

It follows that in comparing possible MST paths on $\cG_{E'}$, the subgraph $\cH_{E'(\cH)}$ can be replaced by a single edge from $\bfx_1$ to $\bfx_2$ with cost $\ell_{\epsilon_\cH}$. We use this result to define the induced ordering $\order_{\cG/\cH}$ for the quotient graph for such a 2-point subgraph $\cH$; this ordering gives the ordering for any diluted edge set $E'(\cG/\cH)$. Note however that here we are forced to view $\cH$ as replaced by an edge, not a vertex, in the quotient graph. [Further, $\cH$ is bordered by two other edges, and these three edges form a chain, which by the general elementary arguments given earlier can be replaced by a single edge of cost the maximum of the costs of the three edges, for the purposes of finding the MST path $\gmst(\cH_{E'} |\order_{E'})$ from $\bfx$ to $\bfy$.]
We can summarize this whole argument as showing that the indicator function can be written as
\begin{multline}
\label{eq_msfindic_factor3}
\indic_c^{(\bfz)}[\gmst(\cG_{E'} | \order_{E'})]=
\indic_c^{(\epsilon_\cH)}[\gmst(\cH_{E'(\cH)} |\order_{E'(\cH)})] \\ \times\indic_c^{(\bfz)}[\gmst(\cG_{E'}/\cH_{E'(\cH)} | \order_{E'(\cG/\cH)})],
\end{multline}
while the edge subsets $E'(\cG/\cH)$ that have to be summed over are subsets of the set $(E'-E'(\cH))\cup\{\epsilon_\cH\}$ (the latter change cause no difficulty, and again the three edges in a chain can be replaced by one, with the cost as described above). This then shows that the weight factors as
\be
\dmsf(\cG|\order)=\dmsf(\cH|\order_\cH)\dmsf(\cG/\cH|\order_{\cG/\cH}),
\label{dmsf_factor2}
\ee
where the $\dmsf$ for $\cH$ is that for the path from $\bfx_1$ to $\bfx_2$ to pass though $\epsilon_\cH$, while the second simply requires a path on the quotient graph $\cG/\cH$ to pass through $\bfz$. However, the argument already given above for the MST path within $\cH$ on the diluted edge sets $E'(\cH)$ shows that if the former condition is dropped, then the evaluation of the sum is the same. That is
\be
\dmsf(\cH|\order_\cH)=d(\cH)
\ee
for $\dmsf(\cH|\order_\cH)$ with the MST path vertex at the most costly edge of $\cH$ under the ordering $\order_\cH$. Thus we obtain factorization in the same form as before, as desired (see Fig.\ \ref{f_dmst_factor_2pt}). This agrees with the result for an ordering in class $\order_{[0]}$ (because then $\order_{\cG/\cH}=\order_{\cG-\cH}$), but gives the correct generalization to other orderings, for the case of a 2-point subgraph. For other orderings, the highest cost in $\cH$ has to be compared with those in the remainder of $\cG$; we emphasize again this aspect of the definition of $\order_{\cG/\cH}$.

In the present case, the argument can simply be used again if $\cG/\cH$ contains a 2-point subgraph. For 3-point subgraphs, we expect that a more complicated generalization exists, but we have not looked for it.

\subsection{Proof of renormalizability}
\label{app_renorm_proof}

In this section, we assemble the preceding results to describe the divergences of the diagrams or subdiagrams in MSF$(p)$ theory, and compare them with the corresponding ones in the perturbation expansion for percolation. Here by a diagram, we mean the corresponding Feynman integral, including the sum over orderings of the $\dmsf F_\order$ factors inside the parametric integral. The initial results provide the direct motivation for the renormalization of the perturbation series. Then we describe the proof of renormalizability.

We saw in section \ref{app_rgproof_bounds} that for a subdiagram $\cH$ and an ordering in the class $\order_{[0]}$ (or for the whole diagram $\cG$, and any ordering), the superficial degree of divergence $\omega_{\rm MSF}(\cH,\cG|\order)$ is the same as $\omega(\cH)$. We consider only 2- or 3-point subdiagrams, including the 2-point subdiagram that contains the path vertex. For any fixed ordering on the edges $E(\cG)-E(\cH)$ not in $\cH$, we can consider the sum of $\dmsf(\cG|\order)F_\order(A_\cG)$ over all the orderings $\order_\cH$ of edges in $E(\cH)$ such that the ordering of all edges is in $\order_{[0]}$. For each ordering in the sum, we saw in section \ref{app_rgproof_dweights} that the $\dmsf$ weight reduces to $\dmsf(\cH|\order_\cH)\dmsf(\cG/\cH|\order_{\cG-\cH})$ (for the path vertex case) or $d(\cH)\dmsf(\cG/\cH|\order_{\cG-\cH})$ (for the other cases). For the latter cases in which the path vertex is not in $\cH$, the weight $\dmsf(\cG|\order)$ is independent of the ordering $\order_\cH$, and the sum over the latter can be performed using eq.\ \eqref{eq_Fsumid'}, which shows that the part of the $F_\order$ associated with the subdiagram has reduced to unity, as in eq.\ \eqref{eq_Fsumid}. The remaining factor on the right hand side of eq.\ \eqref{eq_Fsumid'} depends on the $\alpha$'s for the subdiagram, but of course not on their ordering. To leading order as all those $\alpha$'s are scaled to zero, the resulting subintegration has exactly the same divergence (not only degree of divergence) as the corresponding subdiagram in the field theory of percolation near criticality [including the $d(\cH)$ weight], and the remaining $F_\order$ factor is that for the quotient graph, $F_{\order_{\cG-\cH}}(A_{\cG/\cH})$. (Note that here we disregarded the possibility that $\cH$ itself contains a subdiagram that is divergent; this will be handled later.) For the path vertex, the leading divergence of the subintegral is not identical to any in the percolation theory, as the sum of $F_\order$ over $\order_\cH$ does not reduce to a factor unity for the subdiagram; the 2-point vertex function with a mass ($\phi^2$) insertion, which it resembles, is different, though it has the same degree of divergence [an example of such an integral was discussed in eq.\ \eqref{eq_pathvertcorr}]. Nonetheless, the weighted sum of $F_\order$ has similar factorization properties.

Motivated by these observations, we aim to prove that our perturbation expansion for the path vertex function can be renormalized in a manner very similar to that for percolation. Indeed, the mass, field, and coupling renormalizations will be exactly the same as in percolation, even when they occur inside a 1PI diagram for the path vertex function (we saw earlier that this is so for the self energy parts outside the 1PI path vertex function, that is connected to this function by a single line). For the path vertex itself, the renormalization works and takes a similar form as that for a mass insertion in percolation, but the coefficients are different. We will prove this to all orders in perturbation theory. First, we will establish that it is possible to perform subtractions as in ordinary field theory Feynman diagrams (e.g.\ for percolation), with the result that our subtracted amplitudes are non-diverging for each graph in every order in perturbation theory. Then we will show that, because of a cancellation of some sub-leading pieces involving self-energy insertions, the subtractions take the same form as in percolation, as indicated above. This then leads almost immediately to the RG equations, and the epsilon expansion for the exponents.

The idea for rendering the Feynman integral associated to a diagram finite is intuitively simple. We identify all the diverging sub-integrations associated to subdiagrams of the types already listed above (called ``renormalization parts''), for which the divergence is related to the behavior of the integrand as a corresponding set of $\alpha$ parameters is scaled to zero, and then subtract away these parts of the integrand. One would hope that the resulting integrand is then convergent, and even absolutely convergent. It is necessary to prove this non-obvious result, which we will do using results from the literature.

The procedure is somewhat complex because a given diagram may contain several diverging subdiagrams. The subdiagrams may themselves contain diverging sub-subdiagram (these are revealed by considering several dilatation parameters $\rho_i$ attached to distinct subdiagrams, which go to zero in some order). These possibilities cause no problems for disjoint subdiagrams (that have no common edges), nor for a nested sub-subdiagram (entirely contained in a subdiagram). The case of subdiagrams that are neither disjoint nor nested, called ``overlapping divergences'', is more difficult, but turns out not to be a problem. One makes subtractions corresponding only to  non-overlapping and nested subdiagrams. The procedure was defined by Bogoliubov and Parasiuk in recursive terms \cite{bphz_1}, finiteness was proved by Hepp \cite{bphz_2}, and a non-recursive definition in terms of ``forests'' was given by Zimmerman \cite{bphz_3}. Together, this formulation is called the BPHZ method. These authors worked in terms of momentum-space integrals. For the later formulation and proofs within the parametric formulation, see Refs.\ \cite{param_div_1,param_div_2,param_div_3} and the review in Ref.\ \cite{param_2}.

We briefly outline the result due to Berg\`{e}re and Lam \cite{param_div_3} that we will use. First it will be useful to introduce the ``generalized Taylor operators'' ${\cal T}^n$ \cite{param_div_2,param_div_3}. For a function $f(x)$ of a positive variable $x$ that behaves as $f(x)\sim a_0x^\nu$ as $x\to0$ ($a_0\neq 0$), such that $x^{-\nu}f(x)$ is infinitely-differentiable on $[0,a)$ ($a>0$), and for our purposes with $\nu$ an integer, (such a function is said to have the Taylor series property) these are defined for any integer $n$ to extract the Laurent-like series of terms:%
\be
{\cal T}^nf(x)=a_0x^\nu+a_1x^{\nu+1} +\ldots+a_{\nu+n}x^{n},
\ee
(where $a_0$, \ldots $a_{\nu+n}$ are complex numbers) with properties ${\cal T}^nf(x)=0$ if $n<\nu$, and $(1-{\cal T}^n)f(x)\sim x^q$ with $q>n$. While the series has the Laurent form, we do not assume $f$ is complex differentiable away from $0$, and the coefficients can be calculated from $f$ at positive $x$ only, by ordinary Taylor expansion of $x^{-\nu}f(x)$ at $x=0$. For a function of several variables $x_1$, $x_2$, \ldots, we may define generalized Taylor operators ${\cal T}_{x_i}^{n_i}$ similarly by acting with one of them at a time, but we must be careful as they do not generally commute.

In the following these operations will be applied acting on some subset $E'$ of $\alpha$'s for a graph $\cG$ by a dilatation parameter $\rho$ as $\rho\to0$, and then setting $\rho=1$ in the result: ${\cal T}_{E'}^nf(A_{\cG})={\cal T}_\rho^nf(\{\rho^2\alpha_i:i\in E'\},\{\alpha_i:i\not\in E'\})|_{\rho=1}$. Thus these extract precisely the leading and subleading terms that we have been discussing, up to order $n$. Here when $f(\{\rho^2\alpha_i:i\in E'\},\{\alpha_i:i\not\in E'\})$ has the Taylor series property as a function of $\rho$, we say it has it with respect to the set $E'$, and it is in this case that the operator ${\cal T}_{E'}^n$ is defined.

We will need some definitions for properties of the functions to which the Theorem applies. We will consider what Berg\`{e}re and Lam \cite{param_div_3} call a ``nest'' of edge subsets, which is a filtration, that is a set ${\cal N}=\{E_1,\ldots,E_r\}$ of edge subsets such that
\be
\emptyset\subset E_1\subset E_2\subset\cdots \subset E_t\subseteq E(\cG)\ee
in which the inclusions are strict except possibly the last. For a function $Z(A_\cG)$, we say that it has the ``simultaneous Taylor series property'' with respect to the filtration $\cal N$ if there is a set of integers $\nu_{E_r}$ such that $(\prod_{r=1}^t\rho_r^{-\nu_{E_r}})Z(\rho^2 A_{\cG})$ has simultaneous Taylor series in the set of $\rho_r$ near, and does not vanish at, $\rho_r=0$ for all $r$; here $\rho^2 A_\cG$ stands for the ordered set $A_\cG$ of $\alpha$'s, but each $\alpha_i$ acquires a factor $\rho_r^2$ for each subset $E_r$ to which $i$ belongs. For example, the function $1/(\alpha_1+\alpha_2)$ has the simultaneous Taylor series property for the filtration $E_1=\{1\}$, $E_2=\{1,2\}$.

Now we can state (a special case of) the theorem of Berg\`{e}re and Lam: if (i) $Z(A_\cG)$ is infinitely differentiable for $0<\alpha_i<\infty$; (ii) $Z(A_\cG)$ and its $\alpha$ derivatives are polynomially bounded when arbitrary subsets of $A_\cG$ are scaled to $\infty$; (iii) $Z(A_\cG)$ has the simultaneous Taylor series property with respect to every filtration ${\cal N}$ of edge subsets, then the integral
\be
I_R=\int_0^\infty\prod_{i=1}^{|E(\cG)|}d\alpha_i\, e^{-\sum_i\alpha_it}R[Z(A_\cG)]\ee
with $t>0$ is absolutely convergent. Here the $R$ operation is the subtraction operator which can be defined as
\be
R=1+\sum_{\cal N}\prod_{E'\in {\cal N}}(-{\cal T}_{E'}^{-2|E'|}),
\ee
where the sum is over all filtrations $\cal N$ of the set $A_\cG$ of $\alpha$'s.

In its general form, the theorem applies to many integrals that are not related to Feynman diagrams in any obvious way. Now we wish to apply it to the Feynman integrals in our perturbation expansion. First we point out that these integrals do satisfy the hypotheses of the theorem. Indeed, the integrands of our integrals contain factors that occur in the field theory of percolation, which for this purpose is no different from a cubic-interaction scalar field theory, times the factor $F_\order$ for some ordering $\order$ (times $\dmsf$ and summed over $\order$, but we need not consider this here; this sum can be exchanged with the integral and then taken under the $R$ operation if desired). The integrand in the cubic theory satisfies the conditions, and it is easy to see that the $F_\order$ factor does not change this.

To go further, we note that when applied to Feynman integrals based on a graph (the graph made no appearance in the statement of the theorem), $R$ can also be expressed in many other ways, one of which is as the sum over forests of renormalization parts \cite{param_div_3}. As we know, a forest is a collection of trees, but here the trees are not spanning trees on our lattice or our graph $\cG$. Instead, a forest is any set of renormalization parts in $\cG$ (which are 1PI connected 2- or 3-point subdiagrams), such that for any two such parts in the set, either one is entirely contained in the other (both for its vertices and its edges), or else they are disjoint. (Often in the literature, a forest is pictured as a set of non-intersecting boxes overlaid on the depiction of the Feynman diagram.) In this form for $R$, the sum over all filtrations is replaced by a sum over all forests, and each edge set $E'$ in the product is that of a single renormalization part belonging to that forest. We note that in the BPHZ formulation, whether in parametric form or not, no divergent integral or cutoff is mentioned. The subtractions are carried out instead on the integrand (which however, before the subtractions are performed, does have the property of diverging more strongly in some limits).

We will apply the Theorem to the integral for a diagram $\cG$, in which the sum over orderings, and $\dmsf$ factors, are taken into the integrand. That is,
\begin{multline}
Z(A_\cG)=\sum_\order \dmsf(\cG|\order)F_\order(A_\cG)\\
\times P^{-d/2}(A_\cG)
\exp ( -\k^T_\text{ext} \Delta^{-1} \k_\text{ext}).
\end{multline}
Our earlier remarks imply that for each renormalization part, the subtractions (in forest form) remove precisely all the superficially divergent pieces and no more. Thus in this form, the $R$ operation is exactly what one might expect it to be from the discussion preceding the statement of the theorem, and the theorem says that these subtractions result in an absolutely convergent integral. For subdiagrams  $\cH$ of a diagram $\cG$, these subtractions are exactly the same as those for a diagram of the same type (number of external points, and presence or absence of the path vertex), with one exception. This is the subleading superficial divergence in the case of $\cH$ a self-energy subdiagram. In the subleading generalized Taylor expansion (that is, ${\cal T}^{-2|E(\cH)|}-{\cal T}^{-2-2|E(\cH)|}$ acting on the dilatation factor for the subdiagram), part of it comes from expanding $F_\order$ to order $\alpha$ (for some $\alpha$ in the subdiagram) times the leading term from the percolation integrand; the terms from $F_\order$ arise from orderings in class $\order_{[1]}$ and from subleading terms in class $\order_{[0]}$. This does not correspond to the subtraction made to any whole diagram, and would thus be difficult to include in the renormalization scheme. Fortunately, these subtractions cancel, not for the given diagram, but between diagrams of the same order that differ only in the placement of the self-energy insertion in the graph. This cancellation result will be proved in section \ref{app_canc_proof} below.

Hence because we are always interested in the sum of all diagrams in each order anyway, the only subtractions that have to be made correspond to those that would be made to $\cG$ when it is a renormalization part. It follows that the subtractions correspond to subtracting the Taylor series in $\k^2$ for the subdiagram, where $\k$ is the wavevector entering the subdiagram, and replacing the original graph by the quotient by the subdiagram, times these Taylor coefficients in place of the subdiagram. For the vertex and path vertex cases, the subtraction is simply at zero wavevector, while for the self-energy
the first order term in $\k^2$ has to be subtracted also. This is easily seen, as the subtractions to $\cG$ itself are just its Taylor expansion in $\k^2$ to the given order \cite{param_div_1,param_div_2,param_div_3,param_2}. If we include the zero-loop parts of the vertex functions, this implies that the renormalized vertex functions in this renormalization scheme obey
\be
\label{eq_rg_norm_k0}
\begin{split}
\Gamma^{(2)}_R ({\bf 0}; g,t) &= t, \\
\left.\frac{d}{d \k^2} \Gamma^{(2)}_R (\k; g,t) \right\vert_{\k=\bf 0} &=1, \\
\Gamma^{(3)}_R ({\bf 0,0,0}; g,t)  &=g, \\
\Gamma^{(2,1)}_R({\bf 0,0,0}; g,t)  &= 1, \\
\Gamma^{(2,\rm PV)}_R({\bf 0,0,0}; g,t)  &= 1.
\end{split}
\ee
In view of the condition on the 2- and 3-point vertex functions at $\k={\bf 0}$, the coupling and mass-squared appearing in the propagators in the expansion can be identified with the {\em renormalized} values, so there is no subscript zero on these quantities.

Now that the renormalized perturbation series defining the $\Gamma_R$ are known to be finite, for example at non-zero wavevectors away from the point $\bf 0$ at which the above conditions are given, we can modify the renormalization scheme. Namely, we can add a finite part (more accurately, a series of finite terms) to each subtracted piece in the definition of the renormalized integrand. These can be chosen in each order to modify the renormalization conditions, and the combinatorics again works out. This changes the renormalization scheme, and for example we can modify the conditions above to specify values at non-zero wavevectors (except for the mass-squared):
\be
\label{eq_rg_norms2}
\begin{split}
\Gamma^{(2)}_R (\q = 0; g,t) &= t, \\
\left. \frac{d}{d \q^2} \Gamma^{(2)}_R (\q; g,t) \right\vert_{|\q| = \kappa} &=1, \\
\left. \Gamma^{(3)}_R (\{\q_i\}; g,t) \right\vert_{\rm SP} &=g, \\
\left. \Gamma^{(2,1)}_R(\{\q_i\}; g,t) \right\vert_{\rm SP} &= 1, \\
\left. \Gamma^{(2,\rm PV)}_R(\{\q_i\}; g,t) \right\vert_{\rm SP} &= 1.
\end{split}
\ee
Here SP $=$ symmetry point denotes a symmetric configuration of external momenta $\q_1$, $\q_2$, $\q_3$, which (by rotational symmetry) we take to be any triple satisfying $\q_i^2=\kappa^2$ for $i=1$, $2$, $3$, $\q_i\cdot\q_j=-\kappa^2/2$ ($i\neq j$). Note that $g$ and $t$ now have a different meaning than before. In this form, we can now set $t=0$ and work directly at the critical point, as in each order in perturbation theory the non-zero wavevector scale $\kappa$ prevents the left-hand-sides from diverging in the infrared (the self-energy $-\Gamma^{(2)}({\bf 0};g,0)$ is not infrared divergent). This renormalization at zero mass-squared is quite convenient technically.

As we mentioned above, the BPHZ subtraction scheme requires no reference to, nor use of, a cutoff. It is possible to develop the RG equations directly from this scheme, working with non-divergent expressions only, and leading for example to the Callan-Symanzik equation when the renormalization scheme at zero wavevector, non-zero $t$ is used  \cite{callan}. However, for calculational purposes, we prefer to write intermediate quantities in terms of expressions that diverge as $\Lambda\to\infty$ as in traditional approaches. The bare vertex functions are given by the original, unsubtracted Feynman integrals with cut-off, including as always the $\dmsf F_\order$ factors. For emphasis, we now write these as $\Gamma_0$'s. They are viewed as functions of the bare coupling $g_0$ and mass-squared $t_0$, as well as the wavevectors and cutoff $\lambda$. Then all the subtractions that define the renormalized amplitudes can be collected into changes of the parameters to $g$ and $t=0$, and changes in the scale of the ``operators'' $\phi$, $\phi^2$ and that described by the path vertex. That is
\bea
&&Z_\phi^{N/2}(g_0,\kappa,\Lambda)Z_{\phi^2}^L
(g_0,\kappa,\Lambda)
\Gamma_0^{(N,L)}(\{\q_i\},\{\q_j\};g_0,t_0,\Lambda)\nonumber\\
&&\qquad\qquad=\Gamma_R^{(N,L)}(\{\q_i\},\{\q_j\};g,\kappa,\Lambda),\\
&&Z_\phi(g_0,\kappa,\Lambda)Z_{\rm PV}(g_0,\kappa,\Lambda)
\Gamma_0^{(2,\rm PV)}(\{\q_i\};g_0,t_0,\Lambda)\nonumber\\ &&\qquad\qquad=\Gamma_R^{(2,\rm PV)}(\{\q_i\};g,\kappa,\Lambda),
\eea
and in the limit $\Lambda\to\infty$ the dependence of all $\Gamma_R$ on $\Lambda$ drops out. These equations require five equations to define the dependence of $Z_\phi$, $Z_{\phi^2}$, $Z_{\rm PV}$, $g_0$, and $t_0$ on $g$, $\kappa$, and $\Lambda$, and these are provided by the five conditions \eqref{eq_rg_norms2}, when these are expanded in perturbation theory in $g_0$. At this point the treatment of our theory has come to closely resemble an ordinary field theory, the main difference being the form of the Feynman rules for calculating $\Gamma^{(2,\rm PV)}$. The most important conclusion of the analysis is that the path vertex is renormalized multiplicatively by $Z_{\rm PV}$. We describe in the main text the derivation of the RG equations, and the calculation of exponents to one-loop order.

\subsection{Cancellation proof for subleading terms}
\label{app_canc_proof}

In this section we present the proof that the particular subleading terms in the Laurent expansion as the $\alpha$'s in a self-energy part
(not containing $\bfz$) go to zero, that do not appear for the self-energy in an external line, actually all cancel among graphs with the same self-energy part inserted in different edges.

First we show that the dependence on the $\alpha$'s in a self-energy of the weighted sum of $F_\order$ factors simplifies. We suppose throughout this section that we consider a fixed graph $\cG_0$
with an ordering $\order_0$, and we then modify this graph to obtain $\cG_i$ by inserting a given self-energy graph $\cH$ on an edge $i$ of $\cG_0$. Thus, the edge $i$ is replaced by two edges $i'$, $i''$, with the self-energy $\cH$ in between. In the parametric integral for the diagram, the parameter $\alpha_i$ in $\cG_0$ is replaced by parameters $\alpha_{i'}$, $\alpha_{i''}$, and there are additional parameters for the edges of $\cH$.

We know from Section \ref{app_rgproof_dweights} that the $\dmsf$ weight for $\cG_i$ is determined by the ordering $\order_{\cG_i/\cH}$ in which the cost replacing the original $\ell_i$ is the largest of $\ell_{i'}$, $\ell_{i''}$ and those in $\cH$, independent of how these are ordered relative to each other. Moreover, the weight factors as
\be
\dmsf(\cG_i|\order)=\dmsf(\cH)\dmsf(\cG_i/\cH|\order_{\cG_i/\cH}).
\ee
Throughout the argument, we will compare cases in which $\order_{\cG_i/\cH}=\order_0$ (in an obvious sense), and is fixed. The sum of the $F_\order$ factors over the orderings of the edges that replace $i$  can be calculated, and this is done most easily by returning to the original calculation of $F_\order$ from the action of the $\Omsf$ operator in section \ref{omsfeffect}. The desired sum has the effect of simplifying the integro-differential operator to the following form, and acting under the parametric integral gives
\be%
\begin{split}
&\idotsint\limits_{D} \prod_{\epsilon \in E(\cG_0)} dt_\epsilon \,   \frac{d}{dt_\epsilon} e^{-\sum_{\epsilon \in E(\cG_i)} \alpha_\epsilon t_\epsilon}\\
&{}=\prod_{j=1}^{E(\cG_0)}\frac{\alpha'_{\order_0(j)} }{\sum_{k=1}^j \alpha'_{\order_0(k)} },
\end{split}
\ee
where (i) $D$ is the usual ($|E(\cG_0)|$-dimensional) integration domain for $\cG_0$ with ordering $\order_0$, defined by $t_{\order_0(1)}>\cdots>t_{\order_0(|E(\cG_0)|}>t_0$, and in the integrand, $t_{i'}$, $t_{i''}$, and the $t$'s associated to edges in $\cH$ are all set equal to $t_i$, and (ii) $\alpha_j'$ are the same as $\alpha_j$ except for $\alpha_i'$, which is the sum of $\alpha_{i'}$, $\alpha_{i''}$ and all the $\alpha$'s in $\cH$. The product on the right-hand side is simply $F_{\order_0}$ for $\cG_0$ but with this substitution; we denote it $F_{\order_0}'(i)$.

A second trick that is commonly used for parametric integrals is also useful: if the integrand only depends on the sum of two parameters, say $\alpha_1$ and $\alpha_2$, then these integrations can be combined into a single integral over $\alpha$ which takes the place of $\alpha_1+\alpha_2$, at the cost of introducing a factor $\alpha$
into the integrand:
\be%
\int d\alpha_1d\alpha_2\, \ldots=\int d\alpha\,\alpha\ldots.
\ee
(This can be generalized to any number of variables, but we do not require that.) It can be shown that, like the weighted sum of $F_\order$ factors, the rest of the parametric integrand only depends on the sum $\alpha_{i'}+\alpha_{i''}$ (this can be shown by some further use of the relation, mentioned earlier, of this integrand to combinatorics of weighted spanning trees, which we do not enter into). Then we use this result to recover an integral over a single $\alpha_i$ in place of those two.

We now consider the generalized Taylor expansion of the integrand with respect to a dilatation parameter $\rho$ applied to the $\alpha$'s in $\cH$. The leading behavior is seen to give simply the integrand for $\cH$ from percolation (with zero external wavevectors), times the integrand and $F_{\order_0}$ factor for the quotient graph, as discussed above. We now turn to the subleading terms of a particular form: those that come from expanding the above factor $F_{\order_0}'(i)$ to first order in $\rho^2$ (or simply in $\alpha$'s in $\cH$), times the leading behavior of the rest of the integrand, as $\rho\to0$. The rest of the integrand factors into that for the quotient graph times that for the subgraph, and we make use of the technique for replacing the $\alpha_{i'}$, $\alpha_{i''}$ by $\alpha_i$. This factor in the integrand is now independent of which edge $i$ of $\cG_0$ was chosen for insertion of $\cH$. The $\dmsf$ factor is also independent of $i$ because of the choices of ordering made earlier. Let us write $\sum{\alpha_{\rm s.e.}}$ for the sum of $\alpha$'s associated to $\cH$. There are two types of terms in the expansion of $F_{\order_0}'(i)$ at first order in $\sum{\alpha_{\rm s.e.}}$ for each $i$: a) those in which $\sum{\alpha_{\rm s.e.}}$ appears in the numerator, in which case it replaces $\alpha_i$, so giving the factor $\sum{\alpha_{\rm s.e.}}/\alpha_i$ times $F_{\order_0}$; b) those in which it comes from expanding a denominator, which must be one of those indexed $j\geq \order_0^{-1}(i)$. This gives a factor $-\sum{\alpha_{\rm s.e.}}/\sum_{k=1}^j\alpha_{\order_0(k)}$ times $F_{\order_0}$. We remember to multiply by $\alpha_i$ (because we replaced two $\alpha$'s by this one), and then sum over the positions $i$ of the self-energy insertion. This gives
\be
\begin{split}
&\sum_i 1-\sum_i \sum_{j:j\geq \order_0^{-1}(i)}\frac{\alpha_i}{\sum_{k=1}^j\alpha_{\order_0(k)}}\\
&{}=\sum_i 1-\sum_j \frac{\sum_{i: \order_0^{-1}(i)\leq j}\alpha_i}{\sum_{k=1}^j\alpha_{\order_0(k)}}\\
&=0,
\end{split}
\ee
times other common factors. This completes the proof.

What we have shown is that this type of subleading term actually cancels in the sum of diagrams in each order. Because the superficial degree of divergence of a self-energy diagram at six dimensions is $2$, these subleading parts are also superficially divergent, and are subtracted for each diagram by the $R$ operation defined earlier. We view the present result as showing that the subtracted terms cancel, and because the sum of the subtracted integrals is finite, these subtracted terms can be dropped and the result is still finite. The cancellation is independent of other subtractions associated with renormalization parts, which might appear either inside $\cH$ or disjoint from it. In particular, this allows the cancellation to be made for any number of self-energy insertions in a graph. The remaining terms subtracted from a self-energy insertion are then exactly those that occur in percolation, for any number of self-energy insertions.

The principle underlying these pleasant cancelations is not entirely clear to us. It seems likely (because they involve the derivative of the self-energy in percolation with respect to $t$) that they are associated with the notion that the $\Omsf$ operator should be renormalized, so that it acts on $t$ rather than $t_0$. This operator has the property that it is invariant under any reparameterization of the variables $t_\epsilon\to T_\epsilon=T(t_\epsilon)$ provided that $T$ is a monotonic function and has no explicit dependence on $\epsilon$ (this invariance of the geometry of MSTs is related to that emphasized in Ref.\ \cite{dd}). This property of $\Omsf$ was not explicitly used anywhere in our construction. We will not attempt to give here a conceptual proof using these ideas.

\bibliographystyle{apsrev}

\end{document}